\documentclass{book}
\usepackage[contents={},opacity=1,scale=1,color=black]{background}
\usepackage[a4paper]{geometry}
\usepackage{tikzpagenodes}
\usepackage{totcount}
\usepackage{helvet}
\usepackage{needspace}
\usepackage{amssymb}
\usepackage{enumerate}
\usepackage{slashed}
\usepackage{floatrow}

\usepackage{xcolor}
\definecolor{ultramarin}{RGB}{0,32,96}
\definecolor{lightgray}{gray}{0.75}
\definecolor{lightgray1}{gray}{0.85}
\definecolor{lightgray2}{gray}{0.8}
\definecolor{3Colour}{RGB}{116, 4, 45}
\definecolor{2Colour}{RGB}{163,17,71}
\definecolor{1Colour}{RGB}{0,100,0}

\usepackage{amsmath}
\usepackage{hyperref}
\usepackage{float}
\usepackage{graphicx}
\usepackage{lipsum}
\usetikzlibrary{calc}
\usepackage{tikzpagenodes}
\usepackage{totcount}

\DeclareMathAlphabet{\mathpzc}{OT1}{pzc}{m}{it}
\newcommand{\lambdabar}{{\mkern0.75mu\mathchar '26\mkern -9.75mu\lambda}}

\usepackage{bbold}
\usepackage{minitoc}
\usepackage{lipsum}
\usepackage[english]{babel}
\usepackage{wrapfig}

\usepackage{lipsum}
\usepackage[absolute,overlay]{textpos}

\newif\ifMaterial

\newlength\LabelSize
\AtBeginDocument{
\regtotcounter{chapter}
\setlength\LabelSize{\dimexpr\textheight/\totvalue{chapter}\relax}
\ifdim\LabelSize>2.5cm\relax
  \global\setlength\LabelSize{2.5cm}
\fi
}

\newcommand\AddLabels{
\Materialtrue
\AddEverypageHook{
\ifMaterial
\ifodd\value{page} 
 \backgroundsetup{
  angle=90,
  position={current page.east|-current page text area.north  east},
  vshift=10pt,
  hshift=-\thechapter*\LabelSize,
  contents={
  \tikz\node[fill=gray!70,anchor=west,text width=\LabelSize,
    align=center,text height=15pt,text depth=10pt,font=\large\sffamily] {\textcolor{white}{Chapter \thechapter}};
  }
 }
 \else
 \backgroundsetup{
  angle=90,
  position={current page.west|-current page text area.north west},
  vshift=-10pt,
  hshift=-\thechapter*\LabelSize,
  contents={
  \tikz\node[fill=gray!70,anchor=west,text width=\LabelSize,
    align=center,text height=15pt,text depth=7pt,font=\large\sffamily] {\textcolor{white}{Chapter \thechapter}};
  }
 }
 \fi
 \BgMaterial
\else\relax\fi}
}

\newcommand\RemoveLabels{\Materialfalse}

\usepackage{titlesec}  

\titleformat{\chapter}[display]
{\fontsize{60}{60} \color{gray!70}	\sffamily \bfseries}
{\huge}
{0pt \flushright {\fontsize{60}{60}\selectfont \color{gray!70} \sffamily  \thechapter }}  {\huge}
 \titlespacing{\chapter}{70pt}{-70pt}{20pt}
  \titlespacing{\chapter}{0pt}{-70pt}{20pt}
 \titlespacing{name=\chapter,numberless}{0pt}{16pt}{15pt}

\usepackage{fancyhdr}
\usepackage{graphicx}
\usepackage{sidecap}

\newcommand{\eqb}{\begin{equation}}
\newcommand{\eqe}{\end{equation}}
\newcommand{\dmb}{\begin{displaymath}}
\newcommand{\dme}{\end{displaymath}}

\newcommand{\eab}{\begin{eqnarray}}
\newcommand{\eae}{\end{eqnarray}}

\newcommand{\be}{\begin{equation}}
\newcommand{\ee}{\end{equation}}

\newcommand{\nn}{\newline}

\newcommand{\CMB}{\textnormal{\tiny \textsc{cmb}}}

\begin{document}

\dominitoc


\begingroup
\thispagestyle{empty}

\centering
\vspace{3cm}

\par\normalfont\fontsize{25}{25}
\hspace{1cm}{\Large Fakultät für Physik und Astronomie\\\vspace{2mm}
\hspace{1cm}Ruprecht-Karls-Universität Heidelberg}

\vspace{14cm}

\hspace{1cm}{\large Masterarbeit\\\vspace{3mm}
\hspace{1cm}im Studiengang Physik\\\vspace{3mm}
\hspace{1cm}vorgelegt von\\\vspace{3mm}
\hspace{1cm}Janning Martin Hermann Meinert\\\vspace{3mm}
\hspace{1cm}aus Hamburg\\\vspace{3mm}
\hspace{1cm}\hspace{6.5cm}2021
}
\endgroup


\newpage
\RemoveLabels
\Materialfalse
\thispagestyle{empty}

{\centering

\vspace{4cm}
{\Huge
\hspace{-1cm}Kondensate ultraleichter Axionen und\\\vspace{3mm}
\hspace{-1cm}eine Verknüpfung leptonischer Skalen\\\vspace{5mm}
\hspace{-1cm}zu dunkler Materie
}
\vspace{10cm}

\hspace{-1cm}{\large Die Masterarbeit wurde von Janning Meinert\\\vspace{3mm}
\hspace{-1cm}ausgeführt am\\\vspace{3mm}
\hspace{-1cm}Institut für Theoretische Physik der Universität Heidelberg\\\vspace{3mm}
\hspace{-1cm}unter der Betreuung von\\\vspace{3mm}
\hspace{-1cm}Herrn PD Dr. Ralf Hofmann\\\vspace{9mm}
\hspace{-1cm}sowie von \\\vspace{3mm}
\hspace{-1cm}Herrn Prof. Dr. Matthias Bartelmann\\\vspace{3mm}
\hspace{-1cm}Institut für Theoretische Physik
} 

}

\newpage
\RemoveLabels
\Materialfalse
\thispagestyle{empty}
\vspace{3cm}

{\centering

\hspace{1cm}{\Large Department of Physics and Astronomy\\\vspace{2mm}
\hspace{1cm}University of Heidelberg}

\vspace{14cm}

\hspace{1cm}{\large Master thesis\\\vspace{3mm}
\hspace{1cm}in Physics\\\vspace{3mm}
\hspace{1cm}submitted by\\\vspace{3mm}
\hspace{1cm}Janning Martin Hermann Meinert\\\vspace{3mm}
\hspace{1cm}born in Hamburg\\\vspace{3mm}
\hspace{1cm}\hspace{6.5cm}2021
}
}


\newpage
\RemoveLabels
\Materialfalse
\thispagestyle{empty}

{\centering

\vspace{3cm}
{\Huge
\hspace{-1cm}Condensates of ultralight axions\\\vspace{3mm}
\hspace{-1cm}and a link of leptonic scales\\\vspace{5mm}
\hspace{-1cm}to dark matter
}
\vspace{10cm}

\hspace{-1cm}{\large This Master thesis has been carried out by Janning Meinert\\\vspace{3mm}
\hspace{-1cm}at the\\\vspace{3mm}
\hspace{-1cm}Institut for Theoretical Physics\\\vspace{3mm}
\hspace{-1cm}under the supervision of\\\vspace{3mm}
\hspace{-1cm}PD Dr. Ralf Hofmann\\\vspace{9mm}
\hspace{-1cm}and\\\vspace{3mm}
\hspace{-1cm}Prof. Dr. Matthias Bartelmann\\\vspace{3mm}
\hspace{-1cm}Institute for Theoretical Physics
} 

}


\newpage
\RemoveLabels
\Materialfalse
\vfill
\thispagestyle{empty}
\vspace{2cm}

Wir bestimmen die Masse ultraleichter Axionen, um die explizite U(1)$_A$ Symmetrie-Brechungs-Skale $\Lambda$ bei einer Peccei-Quinn Skale von der Größenordnung der Planck-Masse zu bestimmen.
Dabei nehmen wir an, dass der dominante Beitrag zur Masse einer Galaxie mit geringer Oberflächenhelligkeit nur von einer Axionspezies im Sinne von fuzzy dark matter (Klumpen) bestimmt wird. Für Rotationskurven-Fits an galaktische Rotationskurven benutzen wir deshalb das Soliton-Navarro-Frenk-White-Modell, welches gemäß der Lösung des Poisson-Schrödinger-Systems einen Kondensat-Kern plus korrelierte Axionen im Halo annimmt. Zusätzlich betrachten wir drei häufig verwendete Massendichteprofile: Navarro-Frenk-White-, Pseudo-Isothermal- und Burkert-Modell. Eine Masse $ m_a $ von $0.675 \times 10^{- 23} \,$ eV wird extrahiert, welche die Ergebnisse von \cite{Bernal:2016gxb} reproduziert. Dies impliziert eine effektive Yang-Mills-Skala von $ \Lambda \sim 287 \, $ eV, die nur um einen Faktor von $ 15 $ kleiner ist als die Yang-Mills-Skala einer SU(2)-Theorie, die zur Beschreibung der ersten Lepton-Familie vorgeschlagen wurde \cite{Hofmann:2017lmu}. Das kosmologische Modell SU(2)$_{\rm CMB}$ legt nahe, dass drei SU(2) Yang-Mills-Theorien, die jeweils für die Entstehung der Lepton-Dubletts (e, $\nu_e$), ($\mu$,$\nu_\mu$) und ($\tau$,$ \nu_\tau$) verantwortlich sind, gleichermaßen zur gegenwärtigen Dichte der dunklen Materie beitragen. Parameter eines isolierten Klumpen, wie der gravitative Bohr-Radius oder die Virialmasse, werden ausschließlich durch die Planck-Masse und die entsprechende Leptonmasse bestimmt. Wenn ein Großteil der in einer Galaxie enthaltenen dunklen Masse durch e-Klumpen repräsentiert ist, könnte möglicherweise eine Mischung aus $\tau-$ und $ \mu$-Klumpen das Vorhandensein massiver kompakter Objekte in galaktischen Zentren erklären, und $\tau$-Klumpen könnten im Zusammenhang mit Kugelsternhaufen und der Halo-Masse stehen. Eine naive Abschätzung eines gravitativen $\tau$-Klumpen Kollaps könnte eine theoretische Erklärung für die Massen-Lücke zwischen stellaren und super massiven schwarzen Löchern liefern.

\vspace{1cm}

We determine the mass of ultralight axions in order to determine the explicit U(1)$_A$ symmetry breaking scale $ \Lambda $ at a Peccei-Quinn scale of the magnitude of the Planck mass. We assume that the dominant contribution to the mass of a galaxy with low surface brightness is only determined by one axionic species in the sense of fuzzy dark matter (lumps). For rotation curve fits to galactic rotation curves, we therefore use the Soliton-Navarro-Frenk-White model, which assumes a condensate core plus correlated axions in the halo according to the solution of the Poisson-Schrödinger system. In addition, we consider three commonly used mass density profiles: Navarro-Frenk-White, pseudo-isothermal and Burkert model. An axion mass $m_a$ of $0.675\times10^{-23}\,$eV is extracted, which reproduces the results of \cite{Bernal:2016gxb}. This implies an effective Yang-Mills scale of $\Lambda \sim 287\,$ eV, which is only a factor of $15$ smaller than the Yang-Mills scale of an SU(2) theory that is used to describe the first lepton family \cite{Hofmann:2017lmu}. The cosmological model SU(2)$_{\rm CMB} $ suggests that three SU(2) Yang-Mills theories, each for the formation of the lepton doublets (e,$\nu_e$), ($\mu$,$\nu_\mu$), and ($\tau$,$\nu_\tau$) are equally responsible for contributing to the current density of dark matter. Parameters of an isolated lump, such as the gravitational Bohr radius or the virial mass, are determined solely by the Planck mass and the corresponding lepton mass. If the dominant constituent of the dark mass contained in a galaxy is represented by e-lumps, a mixture of $ \tau- $ and $ \mu$-lumps could possibly explain the presence of massive compact objects in galactic centers, and $ \tau$-lumps could be related to globular clusters and the halo mass. A naive estimate of gravitational $ \tau$-lump collapse could provide a theoretical explanation for the mass gap between stellar and super massive black holes.


\newpage
\RemoveLabels
\Materialfalse
\thispagestyle{empty}\vspace{2cm}

{\large\noindent Erklärung:\\\vspace{1cm}

 \noindent Ich versichere, dass ich diese Arbeit selbstständig verfasst habe und keine anderen als die angegebenen Quellen und Hilfsmittel benutzt habe.\\\vspace{1cm}
 
\noindent Heidelberg, den 09.06.2021     \hspace{-4cm}\makebox[\linewidth]{\rule{4cm}{0.4pt}}

\vspace{-1,5cm}
\begin{figure}[H]
\end{figure}

}



\Materialfalse
\tableofcontents 


\renewcommand{\headrulewidth}{0pt}

\rhead[]{\rmfamily \thepage}

\lhead[\rmfamily \thepage]{}


\chapter{ Introduction }
 \AddLabels

\vspace{0.5cm}

\begin{minipage}{0.6\textwidth}
\begin{figure}[H]
\includegraphics[width=9cm]{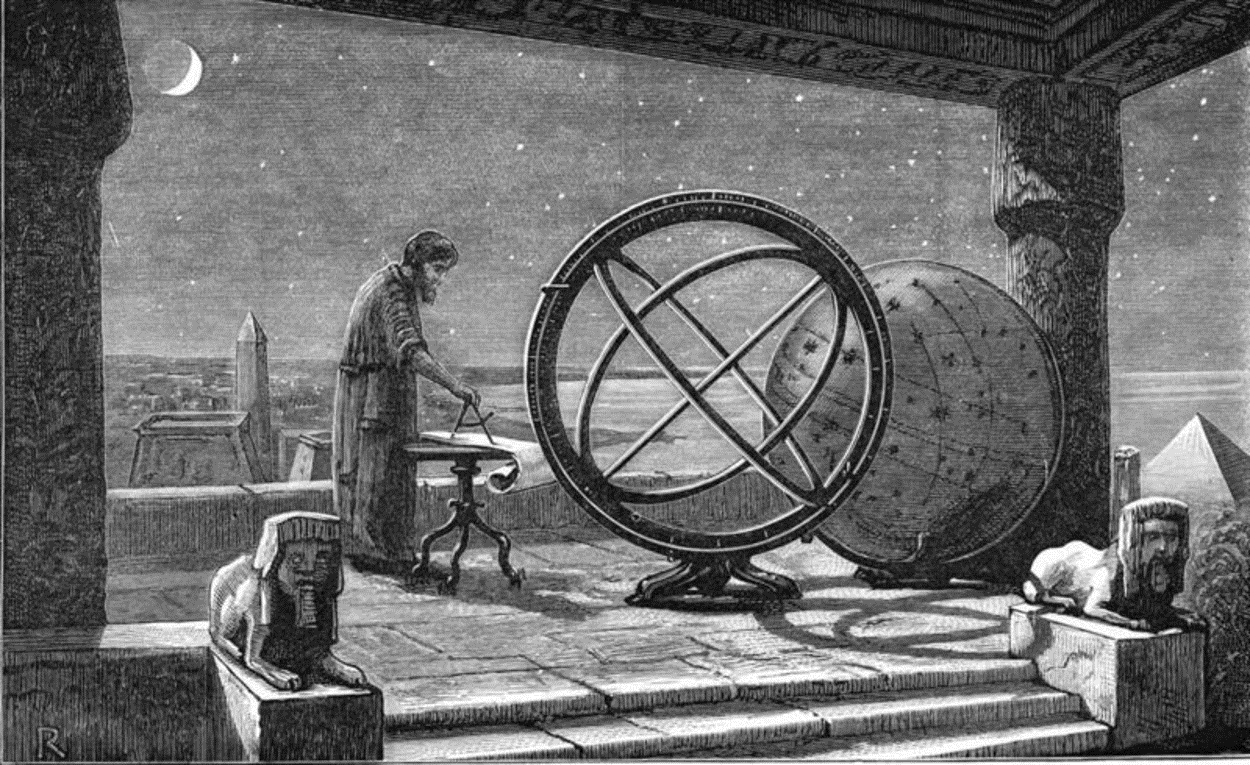}
\caption{Thinking late at night in Egypt, see \cite{EgyptPicture}}
\end{figure}
\end{minipage} \hfill
\begin{minipage}{0.4\textwidth}
\begin{itemize}\sffamily

\item[\textcolor{gray!90}{\textbullet}] \textcolor{gray!90}
{What is the motivation of this thesis?}

\item[\textcolor{gray!90}{\textbullet}] \textcolor{gray!90}
{What are fibre bundles?}

\item[\textcolor{gray!90}{\textbullet}] \textcolor{gray!90}
{What are Lie groups?}

\item[\textcolor{gray!90}{\textbullet}] \textcolor{gray!90}
{How does the chiral anomaly induce a mass?}

\end{itemize}
\end{minipage}\vspace{0.5cm}

\minitoc

\newpage

\section{Introduction} 

Both the Standard Model of particle physics and the Standard Model in cosmology, $\Lambda$CDM are highly successful in their descriptions of nature \cite{altarelli2014higgs,kibble_2015,Planck2014,Planck2016}. Yet, old problems like the nature of dark matter and, more recentlyk dark energy still remain. While both are implemented in the Standard Model in cosmology, they have no representation in the Standard Model of particle physics\footnote{Unless neutrinos turn out to be able to account for dark matter. However, due to thermalisation in the (hot) early Universe it is believed that neutrinos can only contribute to hot dark matter \cite{Primack:2000iq}, which is currently disfavoured by experimental evidence \cite{Hamann_2011}.}. This problem becomes ever more  pressing as new problems in $\Lambda$CDM emerge such as the Hubble constant crisis \cite{Riess:2016jrr}, the discrepancy of cosmologically local and model dependent Hubble constant determinations; the missing baryon problem \cite{Bregman:2007ac}, which refers to a higher baryon density than expected by $\Lambda$CDM and Big Bang nucleosynthesis (BBN) than locally observed; small-scale problems such as the missing satellites problem \cite{Klypin_1999} and the core-cusp problem \cite{Walker_2011}; 
and unexplained anomalies in the Cosmic Microwave Background (CMB) \cite{DelPopolo:2016emo}. 

Many particle theories for explaining dark matter have been considered such as weakly interacting massive particles (WIMPs) \cite{queiroz2017wimp,Arcadi:2017kky}, QCD axions \cite{Di_Luzio_2020,Du:2020eej}, sterile neutrinos \cite{Adhikari_2017,Boyarsky_2019}, and other approaches such as primordial black holes \cite{villanuevadomingo2021brief,PhysRevD.97.081302}, modified gravity \cite{Cembranos:2015svp,Martens:2020lto} and many more. In this thesis we focus on ultralight Planck scales axions \cite{Frieman:1995pm} coupled to an SU(2) theory which describes thermalized photon gases, the so called SU(2)$_{\rm CMB}$ \cite{Hahn:2018dih}, and its implications for the dark sector. The main motivation behind this approach is that SU(2) Yang-Mills thermodynamics incorporate a non-trivial structure of the vacuum \cite{bookHofmann}. This particular aspect of an SU(2) Yang-Mills theory could potentially solve two problems at once:\nn

\noindent i) The vacuum structure of the Standard Model of particle physics is trivial. This leads to a vacuum expectation value (VEV) which is many orders of magnitude higher, $\rho_{\rm QFT} \sim (10^{28} \,\text{eV} )^4$ \cite{Luo:2013hva} than expected by astronomical observations $\rho_{\rm obs.} \sim ( 10^{-3} \,\text{eV} )^4$ \cite{Weinberg:1988cp,Luo:2013hva,Agashe:2014kda}. This discrepancy is known as the cosmological constant problem \cite{Adler:1995vd}. While an SU(2) Yang-Mills theory for the photon has a too small VEV of $\rho_{\rm SU(2)} \sim (6 \times 10^{-5} \,\text{eV} )^4$ \cite{bookHofmann}, coupled to a \emph{Planck} scale axion it is $\rho_{\rm SU(2),M_P} \sim 0.7\times\rho_c\sim (5.92 \times 10^{-3} \,\text{eV} )^4$ \cite{Giacosa:2005nn} which is in relatively good agreement with cosmological observation.\nn

\noindent ii) The U(1) description of the photon has one thermal phase, the photon gas. In contrast, an SU(2) Yang-Mills theory description of the photon has three thermal phases: The deconfined, preconfined and confined phase \cite{bookHofmann}. A transition between the first two phases could explain the Gaussian offset in the CMB line temperature as observed by the Arcade 2 experiment and previous ground based radio experiments \cite{Fixsen_2011}. Electric monopoles\footnote{In a pure gauge theory those monopoles are magnetic, however, here they are interpreted in dual way.} are assumed to emerge in the deconfined phase \cite{Hofmann_2009}. This non-trivial ground state would dynamically break the U(1) symmetry of the photon\footnote{The pure SU(2) description of the photon is only valid for high redshifts. This symmetry is dynamically broken down to U(1) \cite{bookHofmann}, today one can assume that SU(2)$_{\rm CMB} \underset{\mathrm{today}}{=}$ U(1)$_Y$ \cite{Giacosa:2005nn}.} rendering it Meissner massive \cite{Hofmann_2009}.\nn

Moreover, anomalies in the CMB may also be explained by SU(2)$_{\rm CMB}$ \cite{Hofmann:2013rna}. A fit of SU(2) theory to the CMB \cite{Hahn:2018dih} retrieves a Hubble constant of $H_0 = 74.24 \pm 1.46$ (km/s)/Mpc which agrees with the locally obtained values \cite{Riess:2016jrr}, a reionisation redshift $z_{re} \sim 6.23^{+0.41}_{-0.42}$ which agrees with results from studying quasar spectra \cite{Becker_2001,Djorgovski_2001} and a baryon density $\omega_{b,0} = 0.0173 \pm 0.0002$ which might mitigate or resolve the missing baryon problem \cite{Hahn:2018dih}. Compared to $\Lambda$CDM's baryon density of $\omega_{b,0} = 0.0225 \pm 0.00016$ \cite{Planck2014} the SU(2) baryon density is lower. The baryon density of the SU(2) fit lies within the bounds of earlier BBN baryon densities  $\omega_{b,0} =  0.019 \pm 0.0024$ \cite{Tytler_2000}, however, disagrees with more recent BBN results $\omega_{b,0} =  0.022305 \pm 0.000225$ \cite{Cyburt_2016}. Another implication of the CMB fit is a transition between dark energy and dark matter at a redshift of about $z \sim 53$ \cite{Hahn:2018dih}, in order to obtain a sufficiently high dark matter density. This so called process of \emph{depercolation} goes hand in hand with earlier attempts of connecting dark matter and dark energy, i.e. quintessence, compare  \cite{Frieman:1995pm,Wetterich_2002,Barbieri:2005gj,Hall:2005xb,Tsujikawa_2013}.\nn

The main purpose of this thesis is to extract a Planck scale axion mass by fitting rotation curves of low luminosity galaxies from the SPARC library \cite{Lelli:2016zqa}. The first chapter of the thesis will use the mathematically rigorous concept of fibre bundles to introduce the chiral anomaly in an informal way. The second chapter introduces some aspects of SU(2)$_{\rm CMB}$ cosmology. The third chapter will link the chiral anomaly with axions and give a brief introduction to dark matter and rotation curves. In the fourth and following chapters, rotation curve fits will be used in order to extract an ultralight Planck scale axion mass. This mass will then determine an explicit symmetry breaking scale $\Lambda$. We assume that the dominant contribution to the mass of a galaxy with low surface brightness is only determined by one axionic species in the sense of fuzzy dark matter (lumps). Those lumps can be assumed to be a non-thermal and non-relativistic boson condensate which acts as a superfluid \cite{Giacosa:2005nn}. For rotation curve fits to galactic rotation curves, we therefore use the Soliton-Navarro-Frenk-White model, which assumes a condensate core plus correlated axions in the halo according to the solution of the Poisson-Schrödinger system. In addition, we consider three commonly used mass density profiles: Navarro-Frenk-White, pseudo-isothermal and Burkert model. An axion mass $m_a$ of $0.675\times10^{-23}\,$eV is extracted, which reproduces the results of \cite{Bernal:2016gxb}. This implies an effective Yang-Mills scale of $\Lambda \sim 287\,$ eV, which is only a factor of $15$ smaller than the Yang-Mills scale of an SU(2) theory that is used to describe the first lepton family \cite{Hofmann:2017lmu}. In the same way as the photon, the electron is also described with an SU(2) Yang-Mills theory in this thesis, SU(2)$_e$. Therefore, the factor of $15$ can be interpreted as the difference between the confining phase in which the axion condensate exists now, and the deconfining phase of the electron for which the electron scale $\Lambda_e$ was calculated \cite{Hofmann:2017lmu}.\nn

The central result of this thesis is that each leptonic SU(2) Yang-Mills theory, SU(2)$_{e}$, SU(2)$_{\mu}$, and SU(2)$_{\tau}$ is linked to the current dark matter content of the Universe and contributes equally to it. In contrast, the axion field which is associated to SU(2)$_{\rm CMB}$ has not depercolated yet (formed lumps) and is thus spatially homogeneous. It can be interpreted as dark energy \cite{Giacosa:2005nn}, however, both claims include strong assumptions and cannot be proved as of today. One of the core assumptions is that all four axion fields are created at the Planck-scale. This may be motivated by the VEV as mentioned earlier. The conclusion after linking the leptonic sector to three SU(2) Yang-Mills theories is to add those groups to the gauge group SU(2)$_{\rm CMB}$ which might describe thermalized photon propagation. The model now has the following structure: SU(2)$_{\rm CMB} \times$SU(2)$_{e} \times$SU(2)$_{\mu} \times$SU(2)$_{\tau}$. Parameters of an isolated lump, such as the gravitational Bohr radius or the virial mass, are determined solely by the Planck mass and the corresponding lepton mass. If the dominant constituent of the dark matter contained in a galaxy is represented by e-lumps, a mixture of $ \tau- $ and $ \mu$-lumps could possibly explain the presence of massive compact objects in galactic centers, and $ \tau$-lumps could be related to globular clusters. A naive estimate of gravitational $ \tau$-lump collapse could provide a theoretical explanation for the mass gap between stellar and super massive black holes. Unless stated otherwise, we always work in natural units $c=\hbar=k_B=1$. We use Einstein's sum convention and sum over indices where Greek letters go from $\mu = 0,1,2,3$ and Latin letters start at $1$. The Feynman slash notation is used: $\slashed{\partial}= \gamma^\mu \partial_\mu$.\nn

\par\noindent\colorbox{lightgray1}{
    \begin{minipage}{\textwidth}
    \vspace{1mm}
    \textbf{Definition:} Definitions and additional information will be given in a grey box
\vspace{1mm}
    \end{minipage}
    }

\newpage
\section{From fields to fibres}\label{From fields to fibres}

In the last century the cornerstones of present-day fundamental physics have emerged. Prominently among them is the formulation of gauge theories and in particular Yang-Mills theory, which poses a generalisation of Maxwell's electrodynamics to the dynamics governed by non-Abelian gauge groups. The key quantity of interest in gauge theories is the vector potential $A_\mu$ which is easiest to describe in the fiber bundle formalism. Before focusing on gauge groups and in particular on the SU(2) group in the next chapter, the purpose of this chapter is to provide physical intuition to the bundle formalism. For a more and precise formulation of fibre bundles, instantons, gauge and Yang-Mills theories consider reading \cite{Rudolph:2013ofa,Rudolph:2017pug}. For thermal quantum SU(2) and SU(3) Yang-Mills theories and calorons -- finite temperature instantons -- consider \cite{Hofmann:2020wvr}.\nn

\begin{figure}[H]\centering
\includegraphics[width=10cm]{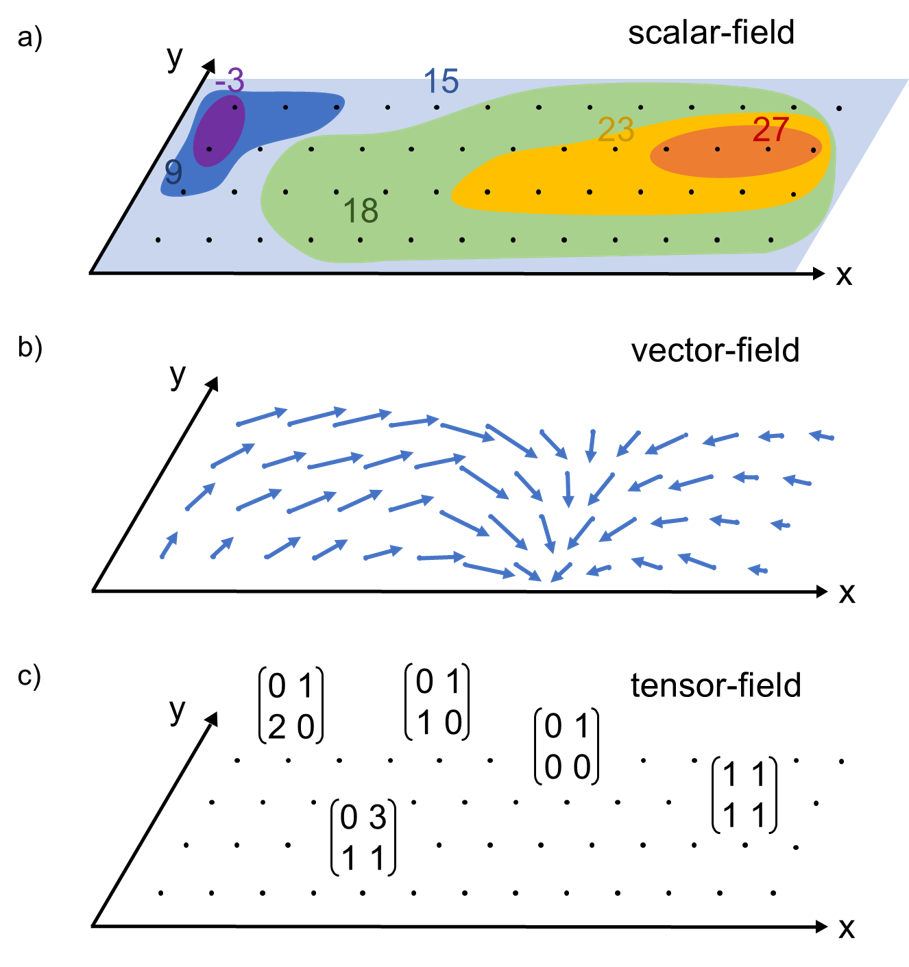}
\caption{An example for a scalar field is a temperature map a). A vector field is usually used to describe airflow, not only the direction but also the velocity of the wind b). Curvature of a Riemannian manifold can be described by a tensor field c).}
\label{fig:fields}
\end{figure}

Michael Faraday was arguable one of the first physicist to use a field description for electricity. The easiest example of a field is a scalar field. Fig.\,\ref{fig:fields} a) shows a temperature map, where each point is associated with a scalar, in this case a temperature.\newline

Another example of a field is a vector field, where each point on the base manifold is associated with a given a vector, see figure \ref{fig:fields} b). A little bit more abstract use of the fields is given in general relativity, where the curvature of spacetime is described by  a tensor field $R_{\mu\nu}$, see c) for the two-dimensional case. \newline

\begin{figure}[H]\centering
\includegraphics[width=11cm]{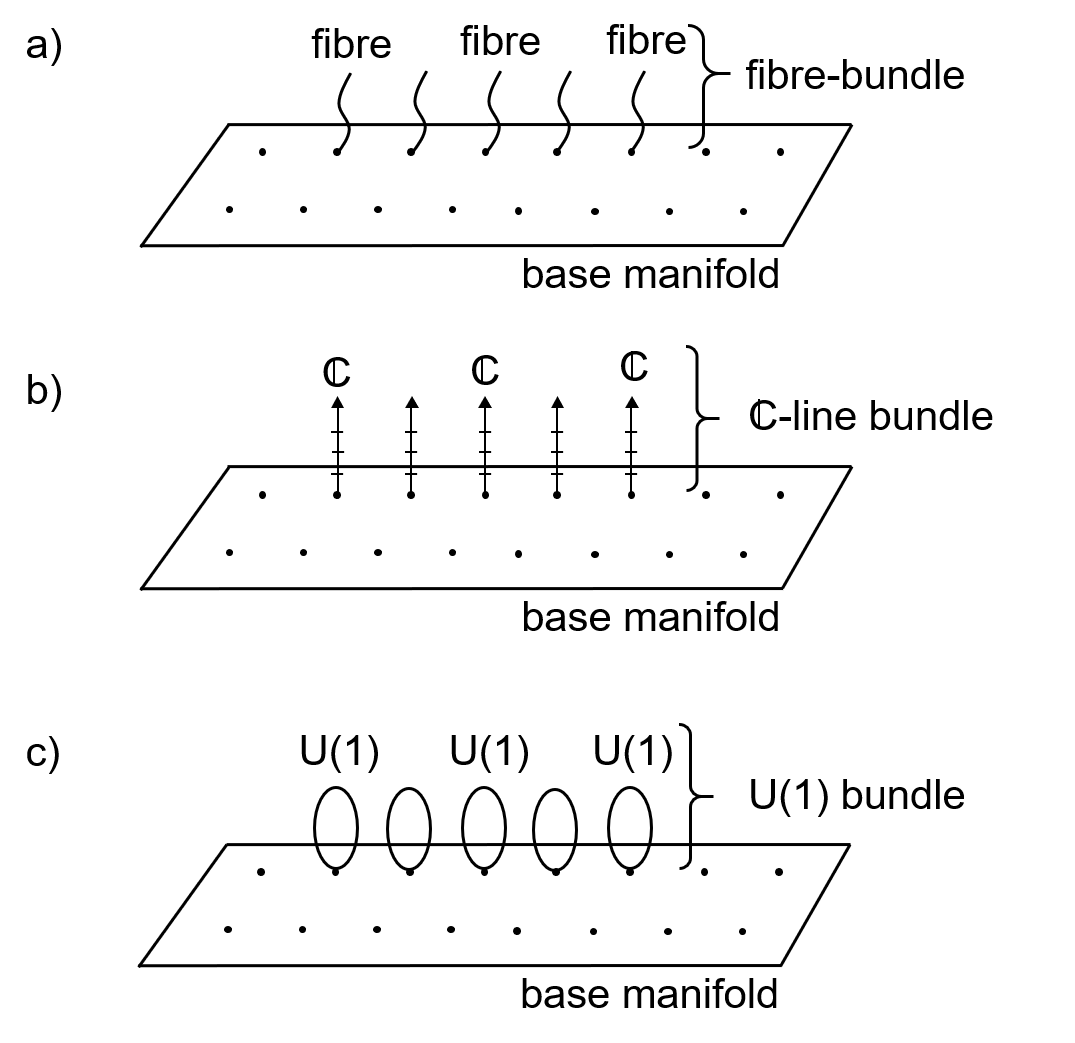}
\caption{Examples of fibre bundles: a) identical fibres with the manifold $F$ are glued to a base manifold $M$,  b) if each point of the base manifold $M$ associates with the complex number line it is called "C-line bundle". c) A physically relevant bundle is a U(1) bundle. In the case of Maxwell's theory a U(1) gauge group describes the freedom to locally change a phase along $F=S^1$.}
\label{fig:bundles}
\end{figure}

The first theory which included a gauge freedom in the definition of its basic field variable $A_\mu$ was Maxwell's formulation of electromagnetism, see Section \ref{sec:Gaugetheories}. Simply put, a gauge field $A_\mu$ has more degrees of freedom than the 
physical fields of the theory. For Lie-groups (Section \ref{sec:Lie Groups}), the symmetry given by the group structure is continuous. 
Before giving examples of gauge fields, it is instructive to introduce fibre bundles first. A fibre is a manifold $E$ which is attached to a point of a base manifold $M$, see figure \ref{fig:bundles}. We will use four-dimensional Minkowski spacetime as a base manifold from now on.\newline

\par\noindent\colorbox{lightgray1}{
    \begin{minipage}{\textwidth}
    \vspace{1mm}
    \textbf{Fibre bundle:} Let $\pi: E \rightarrow M$ be a smooth surjection. Then the triple $(E,M,\pi)$ is called a fibre bundle if there
exists a manifold $F$ such that the following holds. Every $m \in M$ admits an open neighbourhood $U$ and a diffeomorphism $\chi : \pi^{-1}(U) \rightarrow U \times F$ fulfilling ${\rm pr}_U \circ \chi =
\pi$. The manifold $F$ is called the typical fibre of $\pi$. This definition has been taken from \cite[p. 2]{Rudolph:2017pug}.
\vspace{1mm}
    \end{minipage}
    }

\par\noindent\colorbox{lightgray1}{
    \begin{minipage}{\textwidth}
    \vspace{1mm}
    \textbf{Principal bundle:} A principal bundle, is a fiber bundle $\pi: P \rightarrow M$ together with a continuous right action $P \times G \rightarrow P$ such that $G$ preserves the fibers of $P$, where $G$ is any topological group (also called principal $G$-bundle). The tuple $(P, G, M, \Psi, \pi)$ is called a principal bundle if for every $m \in M$ there exists an open neighbourhood $U$ of m and a diffeomorphism $X : \pi^{-1}(U) \rightarrow U \times G$ such that $X$ intertwines $\Psi$ with the $G$-action on $U \times G$ by translations on the factor $G$, and ${\rm pr}_U \circ X (p) = \pi(p)$ for all $p \in \pi^{-1}(U)$. 
    Compare \cite[p. 1]{Rudolph:2017pug} 
    \vspace{1mm}
    \end{minipage}
  }\nn
  \vspace{3mm}

An abstract version of a fibre bundle is given in \ref{fig:bundles} a), it is the collection of multiple, identical fibres of the manifold $R$ which are glued to a base manifold $M$. A concrete example of a fibre bundle is given in \ref{fig:bundles} b), each point is given a complex number line. This is called "C-line bundle". Physically relevant fibre bundles usually use smooth manifold as a fibre, see figure \ref{fig:bundles} c). In fact, the physically most relevant bundles use the manifolds of Lie-groups as fibres. Those bundles are also called principle bundles, because the group structure is represented in the fibre itself. This means that each fiber of the principle bundle is homeomorphic to the group $G$ itself.\newline

\par\noindent\colorbox{lightgray1}{
\begin{minipage}{\textwidth}
\vspace{1mm}
    \textbf{Section:} 
    Let $(P, G, M, \Psi, \pi)$ be a principal bundle. A section of $P$ is a smooth mapping 
    
    $s: M \rightarrow P$ such that $\pi \circ s = \mathbb{1}$. 
    This definition has been taken from \cite[p. 3]{Rudolph:2017pug}.
    \vspace{1mm}
    \end{minipage}%
  }
  \vspace{3mm}

\begin{figure}[H]\centering
\includegraphics[width=9cm]{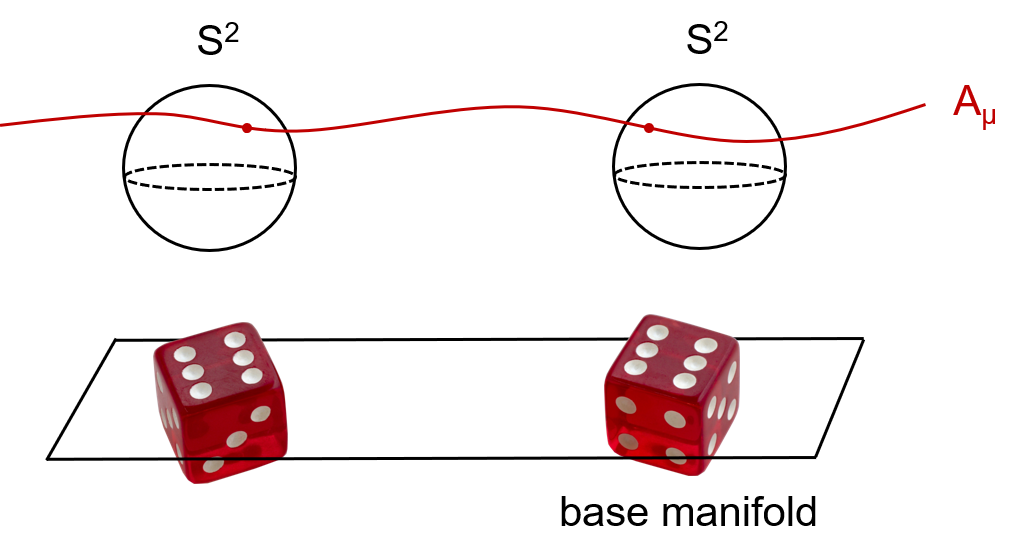}
\caption{
The rotation of a dice can be described by a point on a $S^2$ sphere. The photos of the dices were taken by Brett Jordan.
}
\label{fig:dice}
\end{figure}

Describing fundamental particle physics with gauge theory assumes that the interaction is mediated by a gauge potential $A_\mu$. Geometrically, a gauge potential represents a local connection. This connection lives on a principal fibre bundle over Minkowski spacetime, compare \cite{Rudolph:2017pug}. A connection generalizes infinitesimal motion on a smooth manifold and is thus the generalisation of the directional derivative of objects on a manifold, compare \cite{Bartelmann:2019}. In order to aqcuire some intuition of this formalism of describing nature, imagine a dice. The orientation of the dice can be described with a position $S^2$ sphere, see Fig.\,\ref{fig:dice}. The section of the $S^2$-bundle, from one point in space to another should be smooth. From now on we will refer to the rather mathematical term \emph{section} as connection $A_\mu$, since it is the term more commonly used by physicist.\nn 
 
What does this abstract way of introducing fibres and their sections have to do with physics and gauge fields? Gauge fields are sections through principal fibre bundles. 
For example, the section of a U(1) bundle can be interpreted as the photon field $A_{\mu}$.

\section{Lie groups and Lie algebras}\label{sec:Lie Groups}

In this thesis we will focus on the postulate that the SU(2) Lie group describes the photon propagation rather than a U(1)$_Y \times$ U(1)$_{\rm SU(2)}$ group and its implications for the dark sector. Lie groups and Lie algebras are of special interest in gauge theories, because Lie algebras generate Lie groups and their commutation relation induces the quantisation of gauge fields not only in quantum mechanics but also in quantum field theory (canonical quantisation). Therefore, we will briefly introduce them here.\nn

Two elements of a \textit{non}-abelian group, like SU(2) for example, do not commute. However, both groups have in common that they are continuous groups. A continuous group $G$ of order $n$ can be parametrized by $n$ real parameters. If the set of parameters lives in a differentiable manifold $M^n$, the group G is called a Lie group.\nn 

Every element $g$ of $G$ can be described by the exponential map $e^{i\omega^a t^a}$. Where $t^a$ are the generators and  $\omega_a$ are finite parameters. If the group structure group is defined by a linear trafo, matrix is given by a matrix, as it is the case for SU(2), the exponential map is given by the series, compare \cite{Falquez:2010ve}

\begin{align}
  e^{i\omega^a t^a} \, = \, \sum_{n=0}^{\infty} \frac{i^n}{n!} \, \left( \omega^a t^a \right)^n \,.
  \label{defExpMap}
\end{align}
Elements of $G$ which are close to the identity can be written as
\begin{align}
  g \, = \, \mathbf 1 + i \omega^a t^a\,,
  \label{defInfg}
\end{align}

Let $g_1,g_2 \in G$ be two elements of the group $G$ which are close to the identity. Using the equation above \eqref{defInfg} we can write them as $g_1=\mathbf 1 + i \omega_1^a t^a$
and $g_2=\mathbf 1 + i \omega_2^a t^a$. Multiplying both up to first order gives

\begin{align}
 g_1 g_2 = \mathbf 1 + i (\omega_1^a + \omega_2^a) t^a + O(\omega^2)\,.
  \label{LieComm}
\end{align}

which commutes $[g_1,g_2]=O(\omega^2)$. Now we can demand that any two generators of a Lie group $t^a$ and $t^b$ constitute another generator $t^c$:

\begin{align}
  [t^a, t^b] \, = \, i f^{abc} t^c\,.
  \label{defCommLie}
\end{align}

\begin{figure}[H]\centering
\includegraphics[width=10cm]{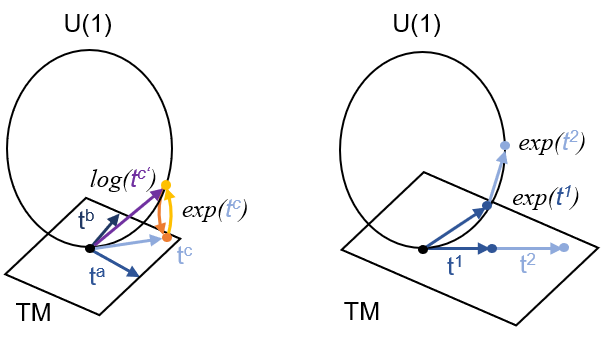}
\caption{
The Lie bracket of two generators $[t^a,t^b]$ corresponds to a tangential manifold $TM$ at the identity \cite{articleLieAlgebra}. With the exponential (yellow) and logarithmic map (orange), any point on the tangential manifold $TM$ can be projected on to the fibre $F=U(1)$ and back to the base manifold.
}
\label{fig:Exponentialmap}
\end{figure}

The Lie bracket guarantees that a composition of two group elements can be represented by the exponential map by virtue of the Baker-Campbell-Hausdorff formula. The reason why one demands in Lie groups G that the commentator of two generators is again a generator is that the exponential representation must also apply to the product of two group elements. A visualisation of the exponential map is given in Fig.\,\ref{fig:Exponentialmap}.\nn

Where $f^{abc} \in \mathbb R$ is the structure constant of the Lie algebra $\mathfrak{g}$ with respect to the basis $t^a$. The Lie algebra $\mathfrak{g}$ of the group $G$ is defined in general 

\begin{align}
  [\cdot, \cdot]:\mathfrak{g}\times\mathfrak{g}\rightarrow\mathfrak{g}, (x,y)\rightarrow [x,y] \,.
\end{align}

The rank $n$ of the Lie algebra $\mathfrak{g}$ is defined as the number of mutually commuting and linear independent generators. In the following we will focus on the non-abelian Lie group SU(2):

\eqb
\label{LagrangianSU(2)}
SU(2) = \{A \in {\rm GL}(2, \mathbb{C}) | A\bar{A}^t =1, {\rm det}\, A =1\}\,,
\eqe

where $GL(n,K)$ is the general linear group of rank $n$ over a body $K$. One can reformulate this statement as

\eqb
SU(2) = \left\{  \begin{pmatrix}
\alpha & \beta \\
-\bar{\beta} & \bar{\alpha} 
\end{pmatrix} : \alpha , \beta \in \mathbb{C}, | \alpha |^2+| \beta |^2=1
\right\}\,.
\eqe

Now $\alpha$ can be expressed as $\alpha=x_1+i x_2$ and $\beta= x_3+i x_4$ with $x_i \in \mathbb{R}$. With this we can identify SU(2) as being diffeomorphic to $S^3 = \{ x_1^2+x_2^2+x_3^2+x_4^2 = 1 \} \subset \mathbb{R}^4$, compare \cite{kirillov}.

\newpage
\section{Gauge fields}\label{sec:Gaugetheories}

The three fundamental forces of the Standard Model, strong, weak and electromagnetism are described with gauge fields. In a gauge theory, the Lagrangian is \textit{locally} invariant under certain transformations. The Lagrangian has spurious degrees of freedoms, which can be cancelled by choosing an explicit \textit{gauge}. Historically, electrodynamics was the first theory which was described with a gauge symmetry.\nn

The dynamics of the photon field $A_\mu$ in the presence of an external source are described by the Maxwell Lagrangian density $\mathcal{L}_{M}$

\eqb
\label{Lagrangian}
\mathcal{L}_{M} \equiv -\frac{1}{4}\,F^{\mu\nu}F_{\mu\nu}-e\,j^\mu A_\mu
\eqe

where $F_{\mu\nu}$ is the field strength or Faraday tensor, $e$ is the coupling constant of electromagnetic interaction, and  $j^\mu$ is the current density. The field strength tensor $F_{\mu\nu}$ can be defined by the action of the commutator $[D_\mu,D_\nu]$ on an complex scalar $\phi$

\eqb
\label{fieldstrength}
F_{\mu\nu} \phi=\frac{i}{e}[D_\mu,D_\nu]\phi\,,
\eqe

where $D_\mu$ is the \textit{covariant} derivative in the direction of $\mu$, which is defined as an ordinary derivative and an additional contribution from the gauge potential $A_\mu$

\eqb
\label{covariant}
D_\mu = \partial_\mu -i e A_\mu\,.
\eqe

The familiar electromagnetic vector fields \textbf{E} and \textbf{B} are recovered from the field strength tensor by the identities

\begin{align}
E^i & \equiv -F^{0i}\,, \\
B^i & \equiv -\frac{1}{2} \epsilon^{ijk} F_{jk}\,,
\end{align}

where $\epsilon^{ijk}$ is the Levi-Civita-Symbol and $F_{jk}$ is the electric part of the Faraday tensor
$F_{\mu\nu}$. The Aharonov-Bohm effect implies that the gauge potential $A_\mu$ is the more fundamental description than the field strength tensor $F_{\mu\nu}$. 


\par\noindent\colorbox{lightgray1}{
    \begin{minipage}{\textwidth}
   \vspace{1mm}
    \textbf{The Aharonov-Bohm effect} can be observed in the interference pattern change of an electron double slit experiment \cite{hiley2013early}. Without an electric current the interference pattern is unaffected (black trajectories of the electron in Fig.\,\ref{fig:AB}) while it is shifted for a shielded current (blue trajectories of the electron in Fig.\,\ref{fig:AB}). A shielded current induces a magnetic field which should classically have no influence on the electrons. The common interpretation is that the vector potential has a more fundamental meaning than its associated force field.\nn
    
    \begin{figure}[H]\centering
\includegraphics[width=8 cm]{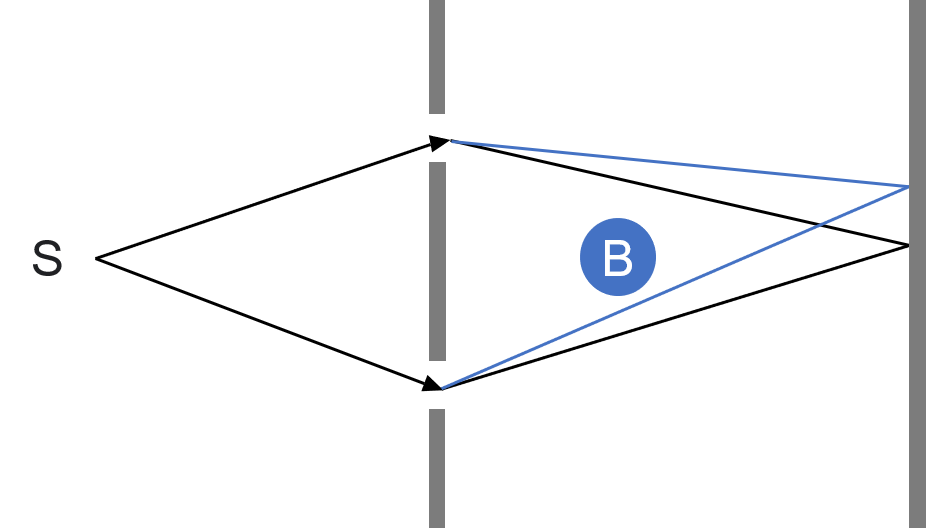}
\caption{ 
The Aharonov-Bohm effect in an electron double slit experiment. Electrons are emitted at a source $S$, pass a screen with two holes and create an interference pattern on a second screen. If a magnetic field $B$ (blue circle) is induced by a shielded current, a deviation from the normal interference pattern (indicated by black trajectories) is observed (indicated by blue trajectories) in the form of a phase change is observed.
}
\label{fig:AB}
\end{figure}
    
\vspace{1mm}
    \end{minipage}
    }
\vspace{3mm}

Putting Eq.\,(\ref{covariant}) into Eq.\,(\ref{fieldstrength}) shows their relationship

\eqb
\label{fieldstrength2}
F_{\mu\nu} = \partial_\mu A_\nu- \partial_\nu A_\mu\,.
\eqe

With this expression it is easier to see that the fieldstrength $F^{\mu\nu}$ and the Maxwell Lagrangian \ref{Lagrangian} are invariant under U(1) transformation $A^\mu \longrightarrow {A^{\mu}}'=A^\mu + \partial^\mu \alpha(x)$:

\begin{align*}
F_{\mu\nu}' & = \partial_\mu A_\nu'- \partial_\nu A_\mu' \\
& = \partial_\mu (A_\nu + \partial_\nu \alpha(x))- \partial_\nu (A_\mu + \partial_\mu \alpha(x)) \\
 & = \partial_\mu A_\nu - \partial_\nu A_\mu 
 +\partial_\mu\partial_\nu \alpha(x)
 -\partial_\nu\partial_\mu \alpha(x)\,.
\end{align*}

The last two terms cancel because the interchangeability of second-order derivatives. 
For the transformed Lagrangian $\mathcal{L}_{M}'$ we therefore have

\begin{align}
{\mathcal{L}_{M}}' & = -\frac{1}{4}\,{F^{\mu\nu}}'{F_{\mu\nu}}'-e\,j^\mu {A_\mu}'\\
& =-\frac{1}{4}\,F^{\mu\nu}F_{\mu\nu}
-e\,j^\mu A^\mu-e\,j^\mu \partial^\mu \alpha(x)
\end{align}

The last term does not affect the action $S[A^\mu]$ as long as \textbf{j} or $\nabla \alpha$ vanish on the boundary of the integral.\nn

\begin{figure}[H]\centering
\includegraphics[width=10cm]{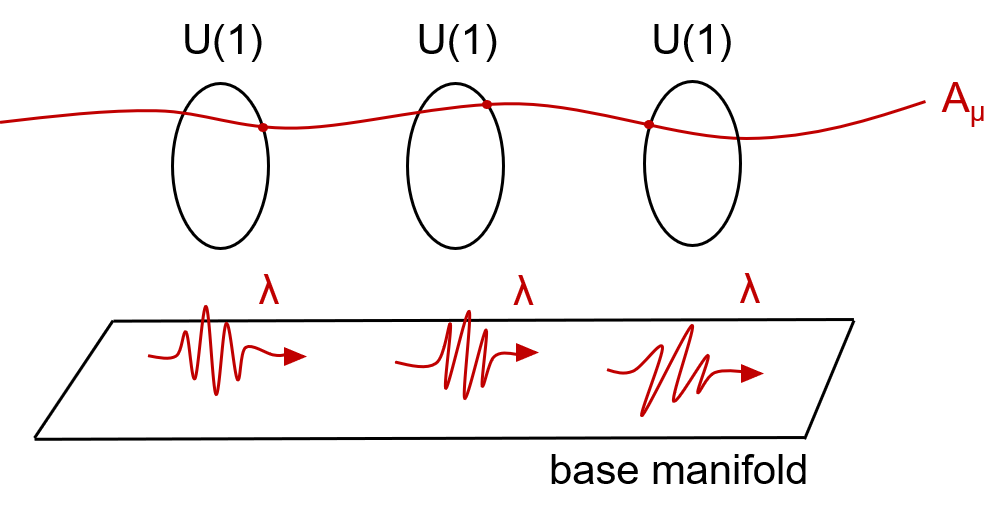}
\caption{The connection of a U(1) bundle is the gauge potential $A_\mu$, it can be interpreted as photon field. The base manifold $M$ should be associated with Minkowskian spacetime, in the case of pure electromagnetism.}
\label{fig:Amu}
\end{figure}

One way of visualization is offered by the bundle formalism as discussed in section \ref{From fields to fibres}, see Fig.\,\ref{fig:Amu}. Here the bundle structure is given by the U(1) group, and the base manifold is chosen to be Minkowski spacetime. In the case of electromagnetism the abundant degrees of freedom are local phases, i.e. polarization. A specific phase can be represented by a point on a $S^1$ circle representing the U(1) group. The gauge potential $A_\mu$, which can be interpreted as photon field is the connection from one U(1) fibre to another. 

\par\noindent\colorbox{lightgray1}{
    \begin{minipage}{\textwidth}
    \vspace{1mm}
    \textbf{Meissner massive photons} form inside a medium, e.g. a superconductor of the type II after a spontaneous symmetry braking of the U(1) symmetry. Under the critical temperature this U(1) symmetry is spontaneously broken because two entangled electrons form Cooper pairs. See figure blow, Fig.\,\ref{fig:superconductorII} for flux tubes (left) which lead to levitating superconductors (right) in a phenomena which is called flux pinning.\nn
    
     \begin{figure}[H]\centering
\includegraphics[width=12cm]{ 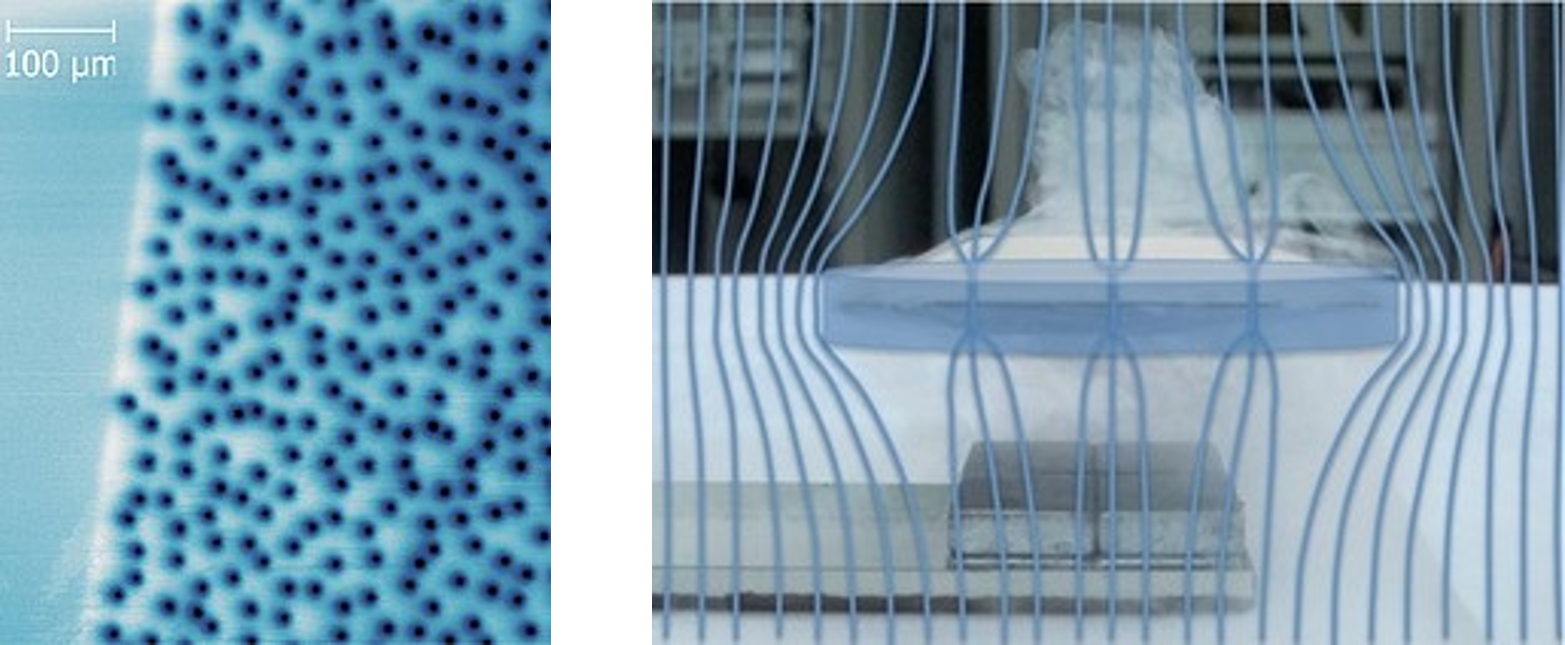}
\caption{ 
left: Images of vortices in a thin YBCO film which was taken by a SQUID scanning Microscop below the critical temperature, compare \cite{Wells_2015}. right: Flux pinning in a superconductor type II. Inside the flux tubes, the photon is massive \cite{FluxtubeImage}. 
}
\label{fig:superconductorII}
\end{figure}
    
\vspace{1mm}
    \end{minipage}
    }
\vspace{3mm}

\newpage

\begin{figure}[H]\centering
\includegraphics[width=15cm]{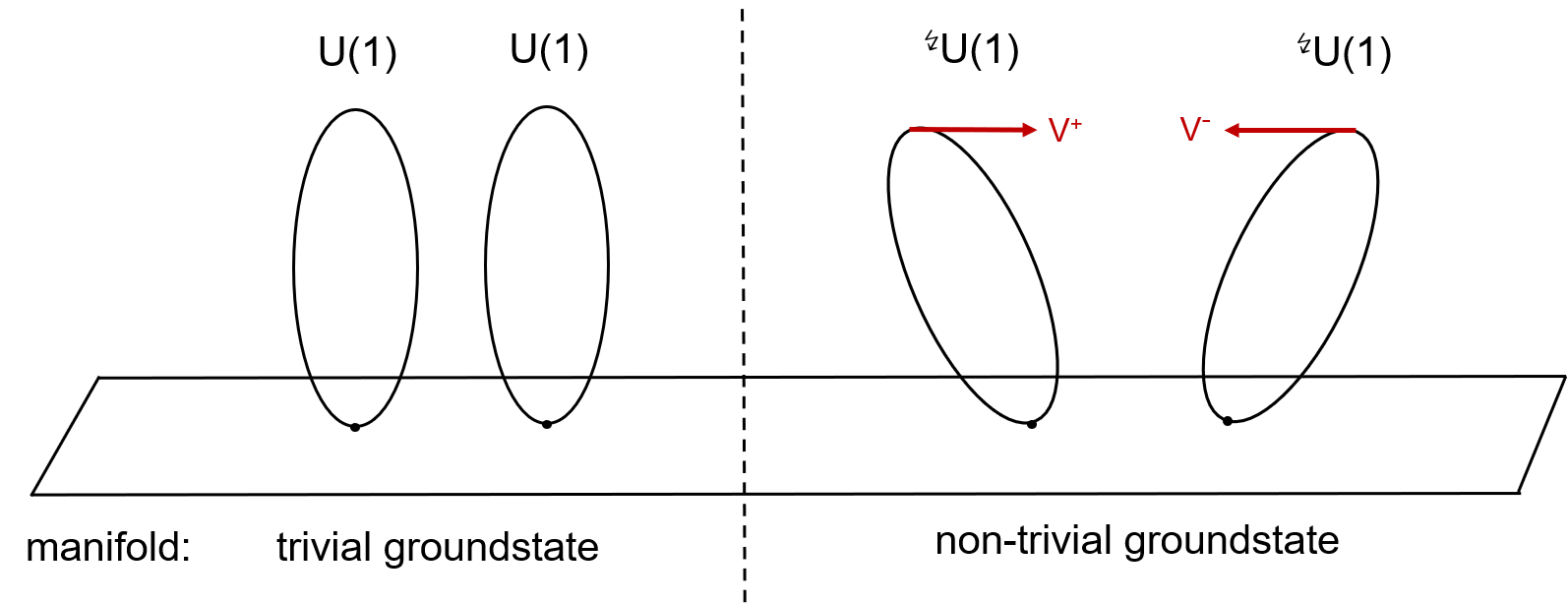}
\caption{
For a trivial ground state, a section / gauge connection through an U(1) bundle describes a photon field which has two degrees of freedom. An additional degree of freedom is introduced if the U(1) symmetry is dynamically broken, e.g. by a medium, Cooper pairs or electric monopoles in a non-trivial ground state. This induces a so called Meissner mass.
}
\label{fig:meissnermass}
\end{figure}

If one associates the gauge field of a local U(1) fibre with the local polarisation, as suggested by Fig.\,\ref{fig:Amu}, it is possible to introduce the so called Meissner massive photons in an intuitive way. The informal argument is that a photon has two polarisations orthogonal to the direction of its propagation. Since the speed of light $c$ is only attained by massless particles, and it is the absolute velocity of the Universe, it must not have an additional polarization while traveling with the velocity $c$. A third polarization, as depicted on the right-hand side of Fig.\,\ref{fig:meissnermass} would allow to temporarily transport information faster and slower than the speed of light. In order to gain a third, longitudinal polarisation the photon needs to slow down. This is equivalent with gaining a small mass. Note however, that a third polarisation breaks the U(1) symmetry dynamically. This argumentation can be turned around, arguing that a small mass induces a third polarisation and breaks the U(1) symmetry. Hopefully, this will help in section \ref{sec:Chiral anomaly} to understand how the chiral anomaly induces masses for axions. In short, the chiral anomaly describes exactly that, a non conservation of the U(1)$_A$-symmetry after quantisation and thus a small induced mass.\nn

In reality, this argument is a little more complicated. According to the Goldstone theorem, a massless boson of spin zero is created by a spontaneous symmetry breaking of a global symmetry \cite{Goldstone:1961eq}. Originally this concept was applied in superconductors by Yoichiro Nambu \cite{Nambu:1960tm}, this is why they are also often referred to as Nambu-Goldstone bosons.

\begin{figure}[H]\centering
\includegraphics[width=12cm]{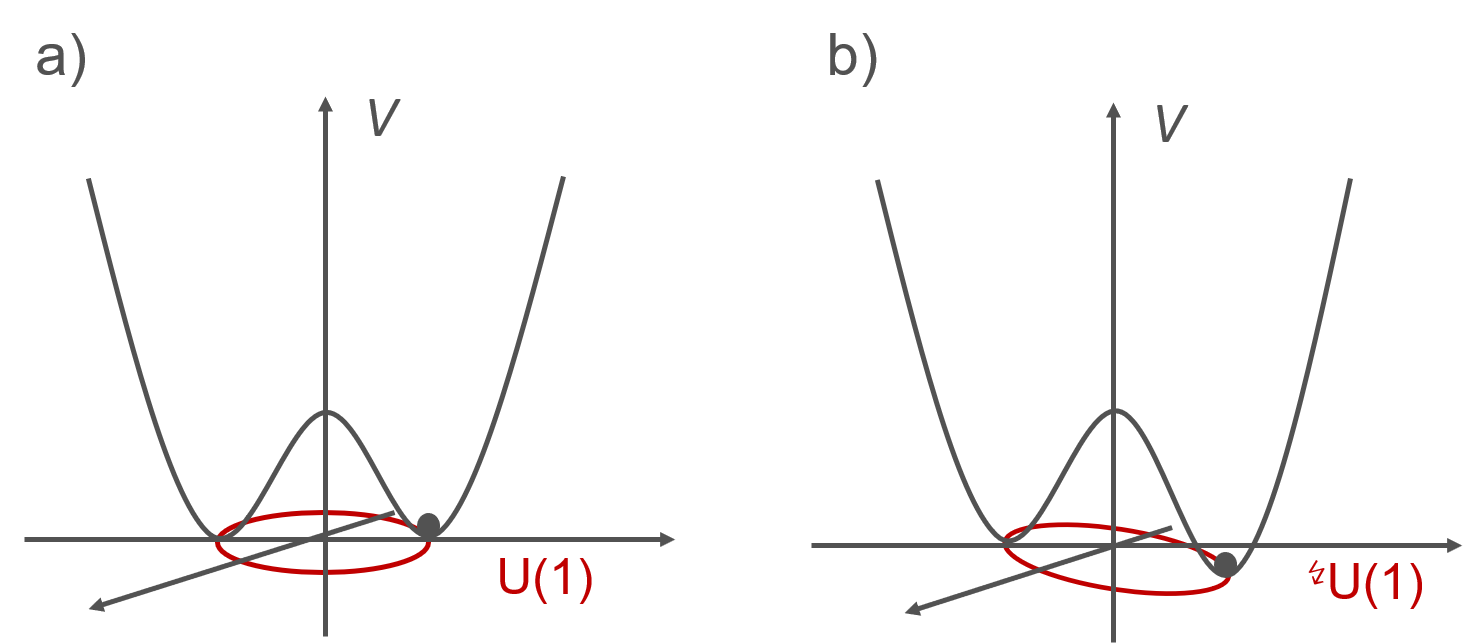}
\caption{
The broken U(1) symmetry in the Higgs mechanism. While a Goldstone boson is created by a spontaneous symmetry breaking a), it is massless and obtains a mass after dynamical symmetry breaking b). Note, that the Higgs potential is three dimensional and only a slice has been shown in one plain, while the U(1) symmetry (red) is shown in an orthogonal plain.
}
\label{fig:higgs}
\end{figure}

However, if a gauge symmetry is broken instead of a global symmetry the would-be Goldstone boson gets "eaten" by the gauge bosons of the broken generators. This particle physicist formulation of eating Goldstone bosons is equivalent to saying that the most natural way of interpreting the additional longitudinal gauge mode is to associate it with a mass. This mechanism is also called the Higgs mechanism and renders the $W^{\pm}$ and the $Z$ boson massive when the electroweak symmetry SU(2)$\times$U(1) is broken down to U(1) \cite{Goldstone:1962es}. A depiction of the dynical symmetry breaking in this context is shown in Fig.\,\ref{fig:higgs}.\nn

\section{Chiral anomaly}\label{sec:Chiral anomaly}

In physics the non-conservation of classical symmetries is often referred to as an \emph{anomaly}. The chiral anomaly is the failure of the chiral symmetry of massless free fermions after quantization, see \cite{Bell:1969ts,Fujikawa:1979ay}. The massless free fermion field $\phi$ is described by the equation of motion

\eqb
D\phi=0\,,
\label{eq:eom}
\eqe

where $\phi$ can be interpreted as a section through the vector bundle\footnote{the vector bundle is actually $\mathbb{Z}$/2-graded, please consider \cite{sardanashvily2009lectures} and \cite{Rabinovich:2017kyj} for more details.} $V \rightarrow M$ over a Riemannian manifold $M$, and where $D$ is a generalized Dirac operator, compare \cite{Rabinovich:2017kyj}.\nn

\begin{figure}[H]\centering
\includegraphics[width=10cm]{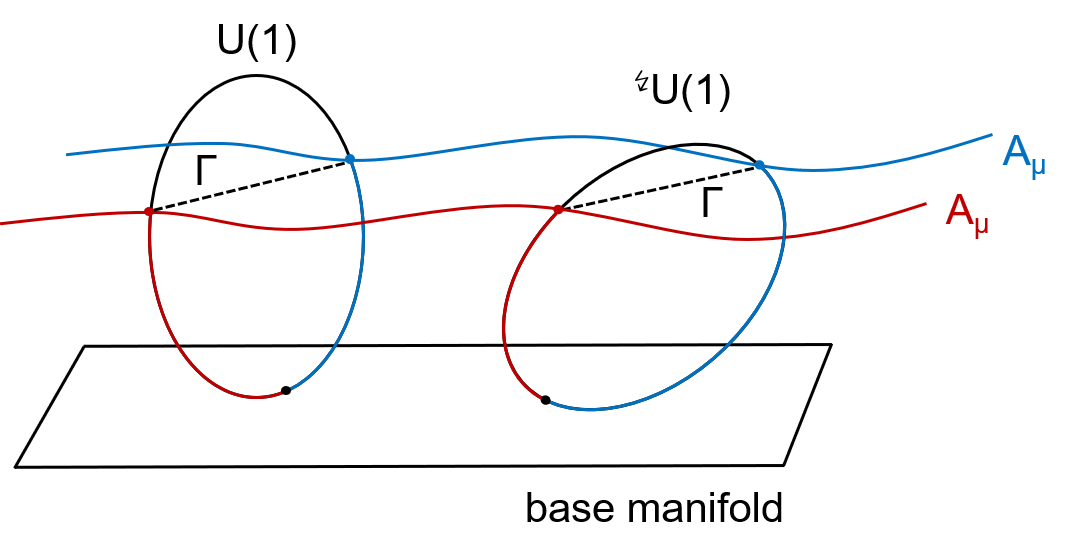}
\caption{
Chirality and helicity are the same for massless (e.g. intact U(1)) particles but differ for massive particles (e.g. broken U(1)). 
The chiral involution operator $\Gamma$ is a measure for equivalence of left- and right-handedness. In the case of massless particles (left) both paths are equivalent. Hence chirality is the same as helicity. After breaking the axial symmetry by making the particle massive (right) it \emph{matters} 
whether the right- or left-handed path is taken.
}
\label{fig:chirality}
\end{figure}

A very brief and abstract introduction into the chiral anomaly can be given by the chiral involution operator $\Gamma$. The operator for the chiral involution $\Gamma$ can be constructed by subtracting the identity on even sections
of $V$ with the identity on odd sections of $V$ \cite{Rabinovich:2017kyj}. In the case of a U(1) bundle, subtracting all sections through even elements of the fibre minus all sections through odd elements is equivalent to the difference between a small left-handed rotation followed by a small right-handed rotation (or vice versa). In this context, the chiral involution operator $\Gamma$ can be thought of as a measure for the equivalence of left- and right-handedness, compare Fig.\,\ref{fig:chirality}. The operator $\Gamma$ preserves the equations of motion because it anti-commutes with $D$ (since $D$ is odd for the $\mathbb{Z}$/2-grading). It then follows because of Eq.\,(\ref{eq:eom}) that

\eqb
D(\Gamma \phi)=0\,.
\eqe

Since $\Gamma$ preserves the equations of motion it generates a U(1) symmetry of the classical theory. This symmetry is called chiral or axial symmetry. While it is a symmetry of classical physics it is broken after quantisation, this is known as \emph{axial anomaly} and induces a mass similarly to the Meissner massive photon as discussed before.\nn

\par\noindent\colorbox{lightgray1}{
    \begin{minipage}{\textwidth}
    \vspace{1mm}
    \textbf{Helicity}, sometimes called "handedness", refers to the sign of the spin projected onto the momentum vector. A negative sign is "left"-handed and a positive sign is called "right"-handed. For massless particles chirality is equivalent to helicity, for massive particles (e.g. after a broken U(1) symmetry) that is not the case.\nn
   
   In general, the \textbf{Chirality} of a particle refers to whether the particle transforms in a right- or left-handed representation of the Poincaré group. In cases where the representation has both left and right handed components, such as a spinor field (section through a spinor bundle) projection operators that treat one handedness at a time can be constructed.
\vspace{1mm}
    \end{minipage}
    }
\vspace{3mm}

Chirality and helicity are the same for massless (e.g. intact U(1)) particles but differ for massive particles. The chiral operator $\Gamma$ is zero in the massless case. The visual clue is the same path lengths by the left handed and right handed paths along U(1) in Fig.\,\ref{fig:chirality}. Similarly, the chiral involution is non zero for a broken U(1) fibre bundle. 
In other words, the chiral symmetry is broken for massive particles.
Of course this statement can be turned around by saying, breaking the chiral symmetry (or axial current) makes a particle massive.
\nn

Another way of illustrating the chiral anomaly is by directly computing that while the vector current $V^a_\mu$ is conserved, the axial current $A^a_\mu$ is not. Therefore, let the massless Lagrangian for the up- and down quark be 

\eqb
\mathcal{L} = i \bar{\Psi} \slashed{\partial} \Psi \,,
\eqe

where the isospinor notation for the fermions, $\Psi$ = (u, d) is used and $\slashed{\partial}$ is defined as $\partial_\mu \gamma^\mu$.\nn

(i) For the vector-current, consider the following transformation: 

\eqb
\Lambda_{\textrm V}: \Psi \longrightarrow e^{-i\frac{\sigma}{2}\Vec{\Theta}}\Psi \simeq (1-i\frac{\sigma}{2}\Vec{\Theta}) \Psi\,,
\eqe

where $\sigma$ refers to the Pauli matrices. The Lagrangian is invariant under this transformation 

\eqb
i \bar{\Psi} \slashed{\partial} \Psi \longrightarrow
i \bar{\Psi} \slashed{\partial} \Psi - i \Vec{\Theta} \left(
\bar{\Psi} i \slashed{\partial} \frac{\sigma}{2} \Psi
-\bar{\Psi} \frac{\sigma}{2} i \slashed{\partial}  \Psi
\right)
=i \bar{\Psi} \slashed{\partial} \Psi\,;
\eqe

the associated Noether current is called vector current and is given by

\eqb
V^a_\mu=\bar{\Psi} \gamma_\mu \frac{\sigma^a}{2} \Psi\,.
\eqe

(ii) For the axial-current, consider the transformation:

\eqb
\Lambda_{\textrm A}: \Psi \longrightarrow e^{-i \gamma_5\frac{\sigma}{2}\Vec{\Theta}}\Psi \simeq (1-i\gamma_5\frac{\sigma}{2}\Vec{\Theta} )\Psi\,,
\eqe

due to the anti-commutation relations of the gamma matrices, the Lagrangian is also invariant under $\Lambda_{\textrm A}$ and the associated axial-current is

\eqb
A^a_\mu=\bar{\Psi} \gamma_\mu \gamma_5 \frac{\sigma^a}{2} \Psi\,.
\eqe


The Lagrangian of massless fermions is invariant under $\Lambda_{\textrm V}$ transformations, as well as $\Lambda_{\textrm A}$ transformation. This is another way of formulating \textit{ciral} symmetry. The associated group structure is SU(2)$_{\textrm V}\times$SU(2)$_{\textrm A}$. If a mass term $m \bar{\Psi}  \Psi$ is added to the Lagrangian we get

\eqb
\mathcal{L} = i \bar{\Psi} \slashed{\partial} \Psi +m \bar{\Psi}  \Psi\,,
\eqe

it remains invariant under the transformation $\Lambda_{\textrm V}$, however, not under the transformation $\Lambda_{\textrm A}$

\eqb
\Lambda_{\textrm A}: m \bar{\Psi}\Psi  \longrightarrow 
m \bar{\Psi}\Psi - 2 i  m \Vec{\Theta}\left(
\bar{\Psi} \frac{\sigma}{2} \gamma_5 \Vec{\Theta} \Psi
\right)\,,
\eqe

and thus the axial current is not conserved for massive quarks, compare \cite{Koch_1997}. The group structure of chiral-symmetry transformations is given by SU(2)$_{\textrm V}\times$SU(2)$_{\textrm A}$. If this symmetry is \textit{spontaneously} broken down to SU(2)$_V$, three massless Goldstone modes are created, the pions $\pi^-,\pi^0,\pi^+$. In reality, those particles are of course massive and hence also called \emph{pseudo}-Goldstone bosons. The chiral anomaly is thoroughly tested and experimentally verified by measuring the lifetime of neutral pions $\pi^0$. A short summary is given in the info-box below and the corresponding Fig.\,\ref{fig:fuji}.\nn

\vspace{3mm}
\par\noindent\colorbox{lightgray1}{
    \begin{minipage}{\textwidth}
    \vspace{1mm}
    \textbf{Neutral pion decay:} Yet another way of introducing the chiral anomaly using Feynman diagrams was originally pursued by Adler \cite{Adler:1969er,Adler:1969gk},  Bell and Jackiw \cite{Bell:1969ts}. One can estimate the lifetime of the neutral pion $\pi^0$ by calculating the matrix element as illustrated below on the left-hand side. The unexpectedly short lifetime of the neutral pion is caused by the chiral anomaly which enables another decay channel into to photons \cite{Larin:2020bhc}. The Feynman diagram of the "decay" of the divergence of the axial current into two photons is shown on the right-hand side below. 
    
    \begin{figure}[H]\centering
\includegraphics[width=10cm]{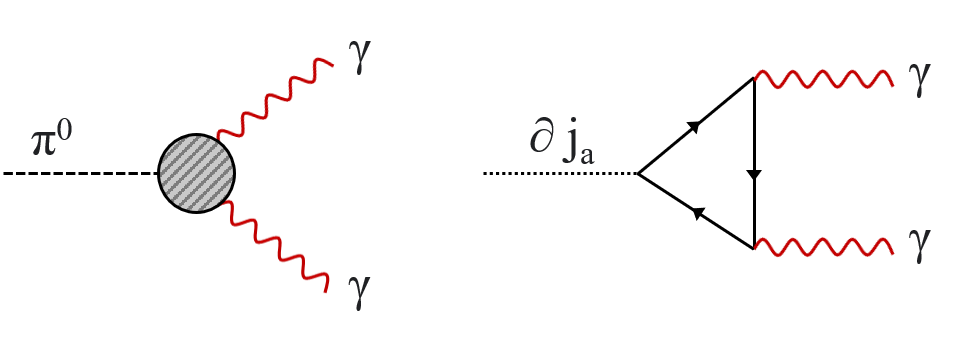}
\caption{Two Feynman diagrams are shown: On the left the decay of the neutral pion into two photons. Its decay is predominantly implemented via the axial current anomaly \cite{Fujikawa:1979ay,Larin:2020bhc}. This diagram is shown in the right.}
\label{fig:fuji}
\end{figure}
\vspace{1mm}
    \end{minipage}
    }
\vspace{3mm}

However, the main application of the axial anomaly is the axion. As it also enables the axion to acquire a mass, see section \ref{axions}.\nn


\chapter{ \texorpdfstring{SU(2)\textsubscript{CMB}}{SU(2)CMB}}\label{chapter-Su(2)}

\vspace{0.1cm}

\begin{minipage}{0.7\textwidth}
\begin{figure}[H]
\includegraphics[width=10cm]{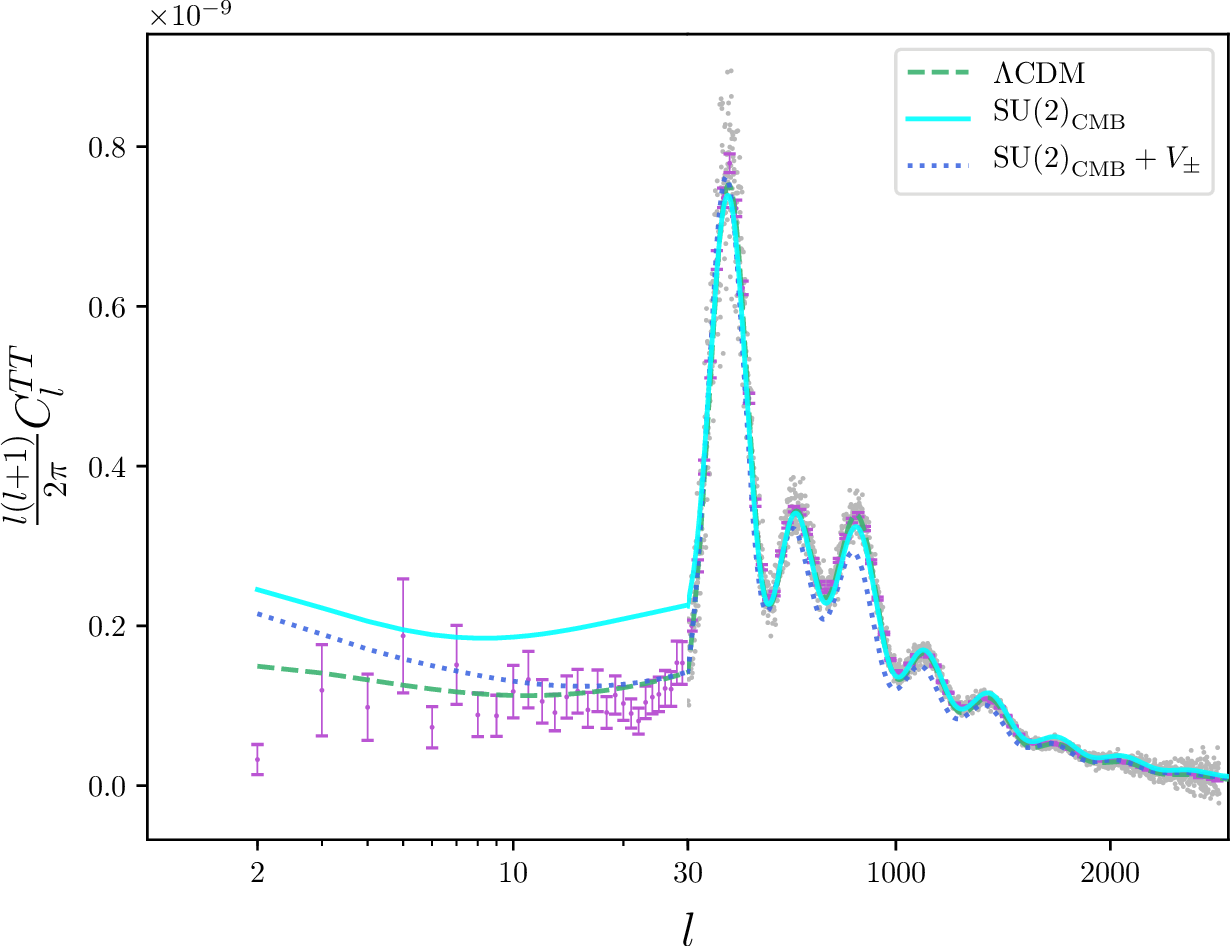}
\caption{
Comparison of two CMB angular power spectrum fits: The SU(2)\textsubscript{CMB} model fit (turquoise) and a $\Lambda$CDM fit (green, dotted). The 2015 Planck data points have been used for the fit, compare \cite{Planck2016}.
}
\end{figure}
\end{minipage} \hfill
\begin{minipage}{0.3\textwidth}
\begin{itemize}\sffamily

\item[\textcolor{gray!90}{\textbullet}] \textcolor{gray!90}{What are the current challenges of $\Lambda$CDM?}

\item[\textcolor{gray!90}{\textbullet}] \textcolor{gray!90}{What is SU(2)\textsubscript{CMB}?}

\item[\textcolor{gray!90}{\textbullet}] \textcolor{gray!90}{How could SU(2)\textsubscript{CMB} mitigate current tensions?}

\item[\textcolor{gray!90}{\textbullet}] \textcolor{gray!90}{What are the implications of SU(2)\textsubscript{CMB} for the dark sector?}
\end{itemize}
\end{minipage}\vspace{1.5cm}

\minitoc

\newpage

\section{Tensions with the \texorpdfstring{$\Lambda$CDM}{LCDM} model}

\begin{figure}[!h]
  \centering
  \includegraphics[width=10cm]{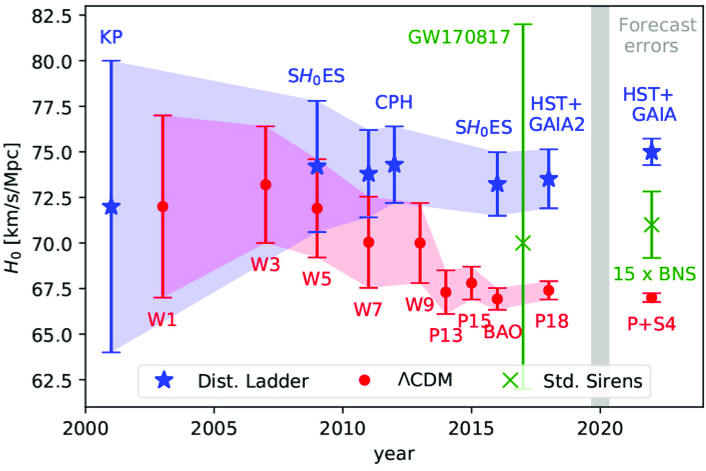}
  \caption{
  Compilation of Hubble Constant predictions and measurements taken from the recent literature:
  This graphic is from \cite{10.3389/fspas.2018.00044} and adopted thereby the graphic from \cite{Beaton:2016nsw};
  Local distance ladder measurements such as S$H_0$ES \cite{Shoes2} are shown in blue, whereas model dependant early-Universe data obtained by WMAP are shown in red.
  The first standard sirens measurement GW170817 is shown in green \cite{Abbott:2017xzu}.
  Forecasts of CMB Stage IV, standard sirens and distance ladder with full GAIA and HST \cite{Casertano_2017} are also depicted.}\label{fig:H0}
\end{figure}

The Cosmological Standard Model is the basis of modern cosmology. To determine all parameters of $\Lambda$CDM at a high accuracy, cosmological distance scales can be calibrated by high-redshift data (inverse distance ladder, global cosmology), coming from precision observations of the Cosmic Microwave Background (CMB) or from large-scale structure surveys probing Baryon Acoustic Oscillations (BAO). Alternatively, low-redshift data (direct distance ladder, local cosmology) can be used by appeal to standard or standardisable candles such as cepheids, TRGB stars, supernovae Ia, and supernovae II. However, depending on whether distances are calibrated with inverse distance ladder data or with local distance ladder anchors (using standard or standardisable candles such as cepheids, TRGB stars, and supernovae Ia), tensions may arise \cite{Verde_2019} 
in some of the  $\Lambda$CDM parameters (e.g., $H_0$ \cite{Aghanim:2018eyx,Shoes1,Shoes2,Wong:2016dpo} and $\sigma_8-\Omega_m$ \cite{DES1,DES2,Tr_ster_2020}. Moreover, it can be shown that those tensions cannot be resolved by a modification of $\Lambda$CDM at low redshift \cite{krishnan2021does}. Figure \ref{fig:H0} shows a time-line of this discrepancy. 
Local distance ladder measurements such as S$H_0$ES \cite{Shoes2} are shown in blue, whereas model dependant early-Universe data obtained by WMAP are shown in red. One measurement based on Sirens is included in green.\nn

\begin{figure}[!h]
  \centering
  \includegraphics[width=15cm]{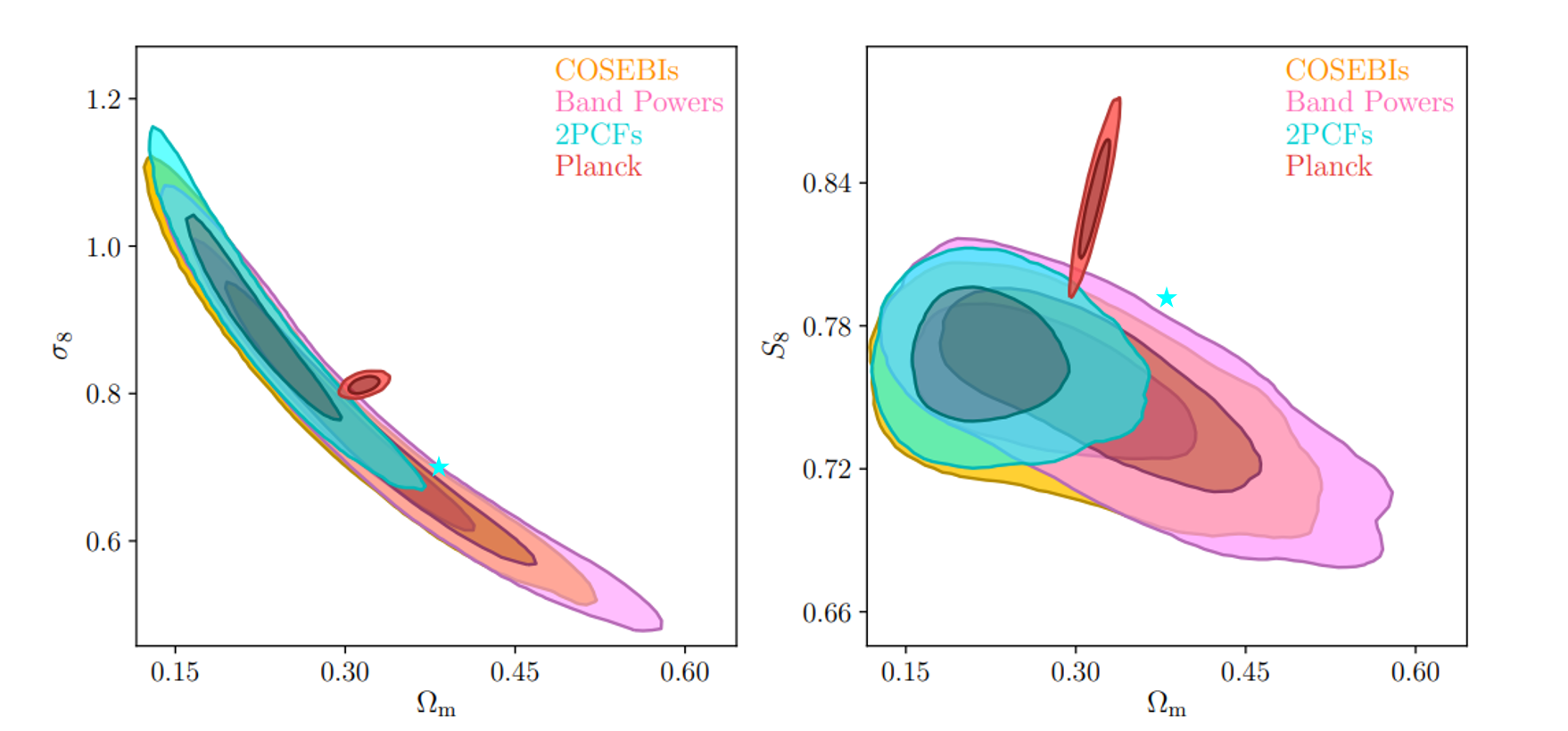}
  \caption{Cosmic shear constraints for the joint distributions of $\sigma_8$ and $\Omega_m$ (left), as well as $S_8=\sigma_8\sqrt{\Omega_m/0.3}$ and $\Omega_m$ (right). The 68\% and 95\% credible regions are shown for COSEBIs (orange), band powers (pink) and the 2PCFs (cyan). Planck (2018, TT, TE, EE+lowE) results are shown in bright red. The turquoise star indicates roughly the values of $\sigma_8$ and $\Omega_m$ according to SU(2)\textsubscript{CMB} in Fig.\,\ref{fig:tabelleDensities}. Note that the stars are drawn in by hand and are \emph{not} plotted. This graphic was adapted from \cite{Asgari_2021}.
  }\label{fig:ShearBanana}
\end{figure}

Another discrepancy may occur in the $\sigma_8$ and $\Omega_m$ relation as shown in Fig.\,\ref{fig:ShearBanana}, the graphic was adapted from \cite{Asgari_2021}. A relatively new tool in Cosmology is Cosmic shear \cite{Kilbinger_2015}. Cosmic shear is created by large-scale structure in the Universe which distort images of distant galaxies due to weak gravitational lensing.
Properties on large scales as well as the geometry of the Universe can be concluded by measuring galaxy shape correlations. Note, however, that in the following study a spatially flat Universe based on $\Lambda$CDM has been adopted.
The so called \textit{Cosmic shear bananas} compare the relationship between $\sigma_8$ and $\Omega_m$ as it can be seen on the left hand side of Fig.\,\ref{fig:ShearBanana} and $S_8$ and the total mass density $\Omega_m$ on the right hand side. Thereby is $S_8$ defined as $S_8=\sigma_8\sqrt{\Omega_m/0.3}$ and $\sigma_8$ is the standard deviation of matter density fluctuations in spheres with a comoving radius of 8 Mpc/h. The Planck data results are shown in bright red. Depending on the redshift calibration, discrepancies up to $3\sigma$ may arise. The turquoise stars in the figure below, Fig.\,\ref{fig:ShearBanana} indicate roughly the values of $\sigma_8$ and $\Omega_m$ according to the SU(2)\textsubscript{CMB} fit. Their values are shown in Fig.\,\ref{fig:tabelleDensities}. Note that the stars in Fig.\,\ref{fig:ShearBanana} are drawn in by hand and are \emph{not} plotted and that the shear model relies on $\Lambda$CDM.\nn

One way of addressing these tensions is to postulate that photon propagation is governed by an SU(2) rather than a U(1) gauge principle. The idea is to describe the CMB as a gas of thermal photons supplemented by a thermal ground state and two invisible vector modes or excitations V$_\pm$ \cite{Hahn:2017yei}.
Thus called the SU(2)$_{\rm CMB}$. Furthermore, this SU(2) gauge theory can be coupled to an ultralight Planck scale axion field. Such an axion is a viable candidate for quintessence, i.e. dynamical dark energy, being associated with today's cosmological acceleration \cite{Wetterich:1987fm,Frieman:1995pm}.\nn

A direct fit of SU(2)$_{\rm CMB}$ to the CMB \cite{Hahn:2018dih} retrieves a Hubble constant of $H_0 = (74.24\pm1.46)$ km ${\rm s}^{-1}$ ${\rm Mpc}^{-1}$ which matches well with the locally measured Hubble constant of $H_0 = (73.24\pm1.74)$ km ${\rm s}^{-1}$ ${\rm Mpc}^{-1}$ \cite{Riess:2016jrr}. Furthermore, model-independent extraction of the co-moving sound horizon at baryon-velocity freeze-out from the relation $r_s\cdot H_0={\mbox const}$ obtains a similarly good match with the local Hubble constant \cite{Hahn:2017hjt}. The co-moving sound horizon is based on low-z observation \cite{Bernal:2016gxb,Riess:2016jrr}. Thereby, SU(2)$_{\rm CMB}$ seems to resolve the current discrepancy between the $\Lambda$CDM concordance model \cite{Planck2016} and the local measurements of $H_0$ \cite{Riess:2016jrr}.\nn
 
\newpage
 
Another open question of modern cosmology poses the \textit{missing baryon problem}: $\Lambda$CDM fits to the CMB fits and the primordial obtain a baryon density of $100\, \omega_{b,0} = 2.225 \pm 0.016$ \cite{Planck2016}, which is in a 1.9-$\sigma$ range of the weighted average from several Deuterium-to-Hydrogen ratios (D/H) in Big-Bang Nucleosynthesis (BBN) of $100\, \omega_{b,0} = 2.17 \pm 0.024$ \cite{Riemer_S_rensen_2015}. However, $^7$Li observations contradict predictions based on BBN; the 4–5-$\sigma$ mismatch is called the \textit{lithium problem} \cite{Fields:2011zzb}. In SU(2)$_{\rm CMB}$ fits to the CMB a low value of $100\,\omega_{b,0} = 1.73 \pm 0.02$ is obtained which could provide a theoretical solution the missing baryon problem. 
As of present, the low baryon density of SU(2)$_{\rm CMB}$ contradicts BBN. Note that the (anti)-screening function $G$ \cite{Falquez:2010ve} is not implemented in the CLASS code when fitting the CMB multipole spectrum. This could mitigate the offset at low multipoles.\nn

\begin{figure}[!h]
  \centering
  \includegraphics[width=14.8cm]{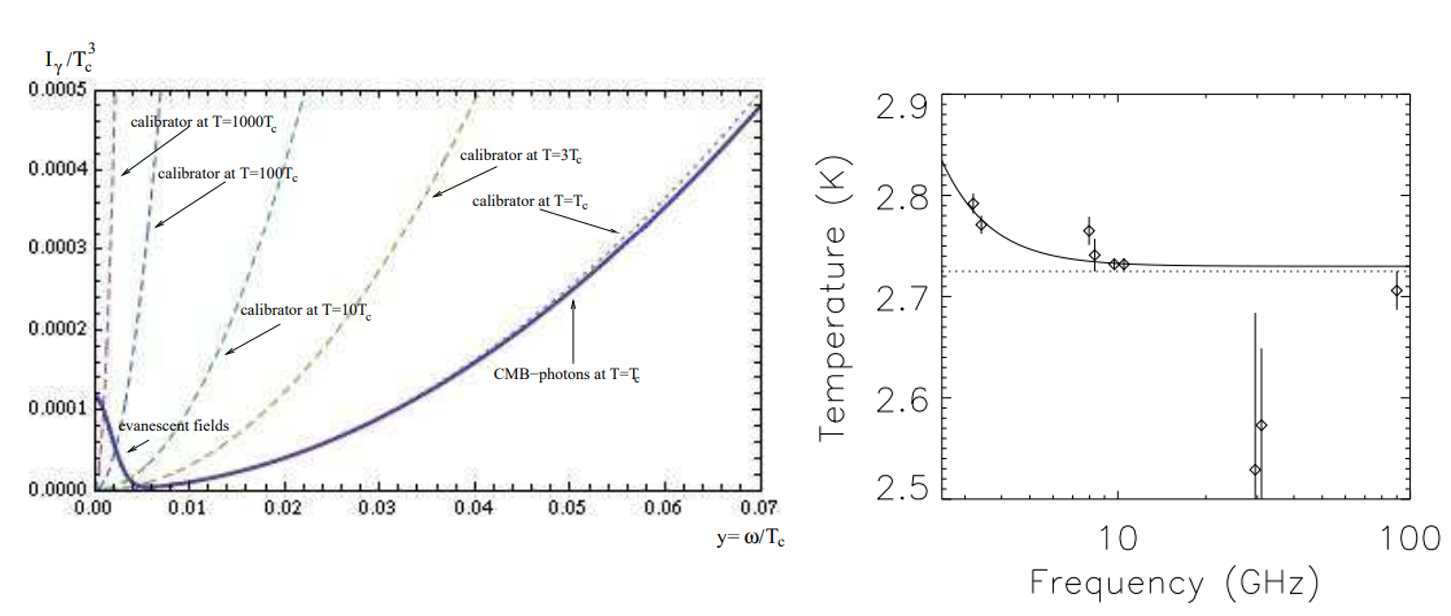}
  \caption{
  The normalized spectral intensities of CMB modes (thick line) at the critical temperature $T_c$ and of calibrator modes (dotted) at various temperatures on the left, compare \cite{Hofmann_2009}. The temperature offset in the microwave range as measure by ARCADE 2 on the right, \cite{Fixsen_2011}.
  }\label{fig:arcade}
\end{figure}

One of the original motivations for SU(2)$_{\rm CMB}$ was to explain the offset from the low-frequency CMB line temperature as measured by ARCADE 2’s \cite{Fixsen:2009xn}, right hand side of Fig.\,\ref{fig:arcade}, and earlier radio-surveys. Based on SU(2) Yang-Mills thermodynamics the excess at low frequencies could be explained by evanescent, non-thermal photon fields of the CMB. In this context, the excess is interpreted as a phase-boundary effect \cite{Hofmann_2009}.

The interpretation of Fig.\,\ref{fig:arcade} in the context of SU(2)$_{\rm CMB}$ cosmology is that the SU(2) theory which describes the photon is dynamically broken down to an U(1) symmetry. The idea is that at a critical temperature $T_c \sim 2.73$ electric monopoles emerge which dynamically break the U(1) symmetry and introduce a massive mode which leads to a photon mass. This is the same mechanism as in Superconductors Type II, Meissner massive photons. The only difference is that here instead of a material, e.g. the superconductor, the ground state of the vacuum changes adopts the new structure: A (anti)-monopole condensate. Furthermore, the integrated SU(2) anomaly generates a dipolar modulation which could explain large-angle anomalies in the CMB  \cite{Hofmann:2013rna}. This SU(2) anomaly can be falsified terrestrially by low frequency black body spectroscopy, as proposed in \cite{Falquez:2010ve}.

\section{SU(2) Yang-Mills Theory}

An SU(2) Yang-Mills theory can be described by a Lagrangian which is invariant under any local SU(2) transformation.\footnote{This is valid for any SU(N) Yang-Mills theory.} The pure Yang-Mills SU(2) Lagrangian $\mathcal{L}_{\rm YM}$ is given by

\vspace{1mm}
\eqb
\mathcal{L}_{\rm YM} = -\frac{1}{4}F^{a\mu\nu}{F}^a_{\mu\nu}= -\frac{1}{2} {\rm tr} F^{\mu\nu}{F}_{\mu\nu}\,.
\eqe
\vspace{1mm}

The pure Yang-Mills SU(2) Lagrangian is invariant under Lorentz transformations, gauge symmetry, is CP-invariant and renormalizeable \cite{tHooft:1972tcz,tHooft:1998yft}. Here, \textit{pure} refers to a Yang-Mills theory which contains only gauge fields without matter fields. Nonetheless, gauge bosons can still become massive by dynamical symmetry breaking via the Higgs mechanism\footnote{Note, that you need a transforming scalar field for that.}. 
As in Eq.\,(\ref{fieldstrength}), the field strength tensor $F_{\mu\nu}$ can be defined by the action of the commutator $[D_\mu,D_\nu]$ of the fundamental covariant derivative on a fundamental scalar $\phi$

\vspace{1mm}
\eqb
F_{\mu\nu} \phi=\frac{i}{g}[D_\mu,D_\nu]\phi\,.
\eqe
\vspace{1mm}

As previously in Maxwell electrodynamics, the covariant derivative $D_\mu$ for the SU(2) gauge is given in the fundamental representation by

\vspace{1mm}
\eqb
\label{covariant2}
D_{\mu} =  \mathbb{1}_{2\times2}\, \partial_\mu - i g A_\mu\,.
\eqe
\vspace{1mm}

In contrast to the previous U(1) theory, calculating the commutator in Eq.\,(\ref{fieldstrength2}) in order to obtain the fieldstrength with the covariant derivative Eq.\,(\ref{covariant2}) 
results in an extra term $- i g [A_\mu,A_\nu]$:

\vspace{1mm}
\eqb
\label{fieldstrength3}
F_{\mu\nu} = \partial_\mu A_\nu- \partial_\nu A_\mu - i g [A_\mu,A_\nu]\,,
\eqe
\vspace{1mm}

This term which is proportional to $g$ represents the self-interactions of the gauge field and is due to the non-abelian nature of the SU(2) theory. If the constraint of CP-invariance is relaxed, another term proportional to $F_{\mu\nu}^a \tilde{F}^{a \mu\nu}$ can be added to the Lagrangian. Here, $\tilde{F}^{a \mu\nu}$ is the dual fieldstrength tensor of the SU(2) theory which is defined by $\tilde{F}^{a \mu\nu}=\frac{1}{2}\varepsilon^{\mu\nu\rho\sigma}F_{\mu\nu}$. (Anti)self-dual solutions of the Yang-Mills equations are particularly interesting, because they correspond to minimae of the Yang–Mills action \cite{Rudolph:2017pug}. By integrating the SU(2) Yang-Mills fieldstrength  $\mathcal{L}_{\rm YM}$, we obtain the SU(2) Yang-Mills action $\mathcal{S}_{\rm YM}$ which is given by 

\vspace{1mm}
\eqb
\label{YMaction}
\mathcal{S}_{\rm YM} = \int {\rm d}^4 x\, \mathcal{L}_{\rm YM} = -\frac{1}{4} \int {\rm d}^4 x\,F^{a\mu\nu}{F}^a_{\mu\nu}\,.
\eqe
\vspace{1mm}

In the next section, SU(2) theory will be applied to derive a modified temperature-redshift relation.\nn

\newpage

\section{Modified temperature-redshift relation}\label{SU(2) Yang-Mills thermodynamics}

SU(2)\textsubscript{CMB} is a high-redshift modification which connects to the standard cosmological model $\Lambda$CDM at low redshifts.\nn

\par\noindent\colorbox{lightgray1}{
    \begin{minipage}{\textwidth}
    \vspace{1mm}
    \textbf{Redshift:} 
    Redshift $z$ is similarly to the Doppler-effect the deviation of emitted to received wavelength. It is defined as
    \eqb
\label{conservation}
\frac{\lambda_{obs}}{\lambda_{emit}}=1+\frac{\lambda_{obs}-\lambda_{emit}}{\lambda_{emit}}=1+z\,,
\eqe
where $\lambda_{emit}$ is the wavelength at the time of emission and $\lambda_{obs}$ is the wavelength at the time of observation, compare \cite[p. 9]{Bartelmann:2019}.
\vspace{1mm}
    \end{minipage}
    }\nn
\vspace{3mm}

A spatially flat Friedmann-Lemaître-Robertson-Walker (FLRW) universe is assumed for the background evolution. Let the energy density of the deconfining phase of SU(2) Yang-Mills thermodynamics be $\rho_{\rm YM}$ and pressure $P_{\rm YM}$. Then, if we assume energy conservation
the following equation has to be considered:\nn

\eqb
\frac{{\rm d}\rho_{\rm YM} }{{\rm d} a} = -\frac{3}{a}(\rho_{\rm YM}+P_{\rm YM})\,,
\eqe

where $a$ is the cosmological scale factor, normalised such that today $a(T_0) = 1$ and $T_0 = 2.725$\,K, the current temperature of the CMB, compare \cite{Hahn:2018dih}. If you solve for $a$, the solution of Eq.\,(\ref{conservation}) is

\eqb
\label{conservation2}
a=exp \left( -\frac{1}{3} \log \left(\frac{\mathpzc{s}_{\rm YM}(T)}{\mathpzc{s}_{\rm YM}(T_0)}  \right) \right)\,,
\eqe

\noindent where $\mathpzc{s}_{\rm YM}$ denotes the entropy density \cite{Hahn:2018dih}

\eqb
\mathpzc{s}_{\rm YM}\equiv \frac{\rho_{\rm YM}+P_{\rm YM}}{T}\,.
\eqe

\begin{figure}[H]\centering
\includegraphics[width=10cm]{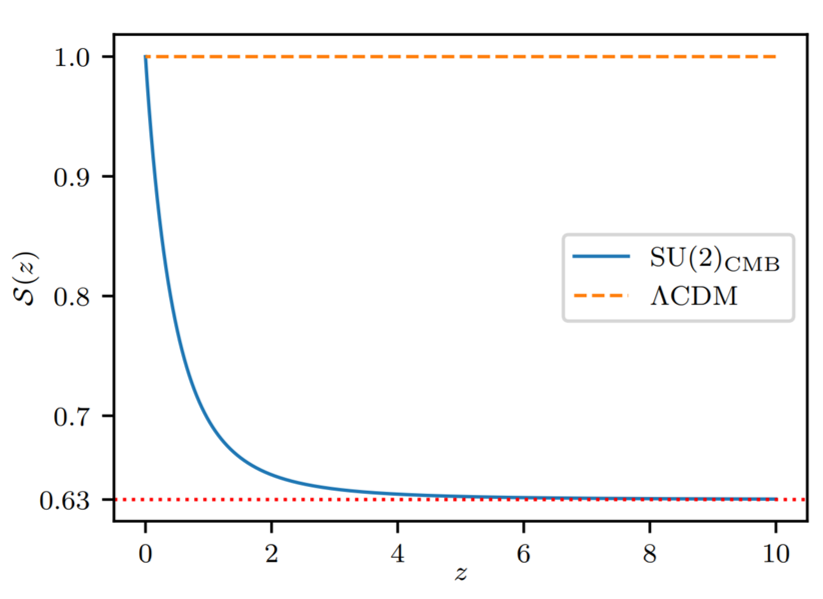}
\caption{The function $\mathpzc{S}(z)$ of Eq.\,(\ref{ScalingS}) which indicates the deviation from the linear $T-z$ relation. The curvature at low z is due to the breaking of scale invariance in the deconfining SU(2) Yang-Mills plasma for $T \sim T_0$. Notice the rapid approach towards $(\frac{1}{4})^{1/3} \approx 0.63$ with increasing $z$, compare \cite{Hahn:2018dih}.}
\label{Hahn}
\end{figure}

\noindent In figure \ref{Hahn} you can see the multiplicative deviation
$\mathpzc{S}(z)$ from linear scaling which is given as

\eqb
\label{ScalingS}
\mathpzc{S}(z)=\left(\frac{\rho_{\rm YM}(0)+P_{\rm YM}(0)}{\rho_{\rm YM}(z)+P_{\rm YM}(z)} \right)^{1/3}
\,.
\eqe

\noindent As you can see, $T \approx 0.63$ for $T\gg T_0$. With this scaling function $\mathpzc{S}(z)$, we are able to define the temperature as a function of $z$:

\eqb
T(z)= \left(\frac{1}{4}\right)^\frac{1}{3} T_0 \,(z + 1),
\eqe

and modify the temperature dependant radiation densities $\Omega_{\rm YM}(z)$ for the Yang-Mills plasma and $\Omega_{\nu}(z)$ for the neutrinos. This factor, $(1/4)^\frac{1}{3}$ is one of the main consequences of using an SU(2) group in order to describe the CMB photons instead of a U(1) group.\nn

\section{Dark sector of SU(2) Yang-Mills thermodynamics}

In \cite{Hahn:2018dih} a dark sector was introduced as a deformation to $\Lambda$CDM which describes a 
sudden transition from dark energy to dark matter at a redshift $z_{\rm p}$. Such a transition is required to 
reconcile high-$z$ cosmology, which is changed compared to $\Lambda$CDM 
as a result of an SU(2)-induced temperature-$z$ relation
\cite{Hofmann:2014lka} with $\Lambda$CDM at low $z$. 
For the entire dark sector we have 
\eqb
\label{edmdef}
\Omega_{\rm ds}(z)=\Omega_{\rm \Lambda}+\Omega_{\rm pdm,0}(z+1)^3+\Omega_{\rm edm,0} 
\left\{
\begin{array}{cc}
(z+1)^3,  & z<z_p \\
(z_p+1)^3, & z\geq z_p \\
\end{array}
\right.
\eqe


The density parameters are defined as: 

\eqb
\Omega_{r}(z)= \Omega_{\rm YM}(z)+\Omega_{\nu}(z)
\eqe

\eqb
\Omega_{\rm YM}(z)= 4 \frac{\pi^2}{30} \frac{T(z)^4}{\rho_{c}} 
\eqe
\eqb
\Omega_{\nu}(z)= \frac{21}{8} \left(\frac{16}{23}\right)^\frac{4}{3} \frac{\pi^2}{30} \frac{T(z)^4}{\rho_{c}}
\eqe

with $T_0=2.725\,$K$\cong 2.348 \times 10^{-4}\,$eV. The critical density $\rho_{c}$ is defined as 

\eqb
\rho_{c}=\frac{3 H_0^2}{8 \pi} M_p^2=1.76\times10^{-9}\, \text{eV}^4
\eqe

In the simplified model, we assume only one depercolation at $z_p$. This means that we consider primordial dark matter $\omega_{\rm pdm}$ always as a part of the matter density $\Omega_m(z)$:

\eqb
\Omega_m(z)= \frac{\omega_b + \omega_{\rm pdm}  }{\text{$\omega_{\rm b}$}+\text{$\omega_{\rm edm}$}+\omega_\Lambda +\text{$\omega_{\rm pdm}$}}
(z+1)^3
\eqe\nn

for the dark energy density we have

\eqb
\Omega_{\Lambda}(z)=\frac{\omega_\Lambda  }{\text{$\omega_{\rm b}$}+\text{$\omega_{\rm edm}$}+\omega_\Lambda +\text{$\omega_{\rm pdm}$}} (z_p+1)^3\,,
\eqe\nn

For emergent dark matter we assume

\eqb
\Omega_{edm}(z)=\left\{\begin{array}{cc}
(z+1)^3,  & z<z_p \\
(z_p+1)^3, & z\geq z_p \\
\end{array}\right.
\label{eq:depercolation}
\eqe\nn

Assuming that curvature $\Omega _{k}$ is zero, $H(z)$ is then calculated as following:\nn

\eqb
H(z) = H_0 \sqrt{
\Omega_{r}(z) +\Omega_{\Lambda}+\Omega_{m}(z)}
\eqe
\vspace{1mm}



\begin{figure}[h]\centering
\includegraphics[width=10cm]{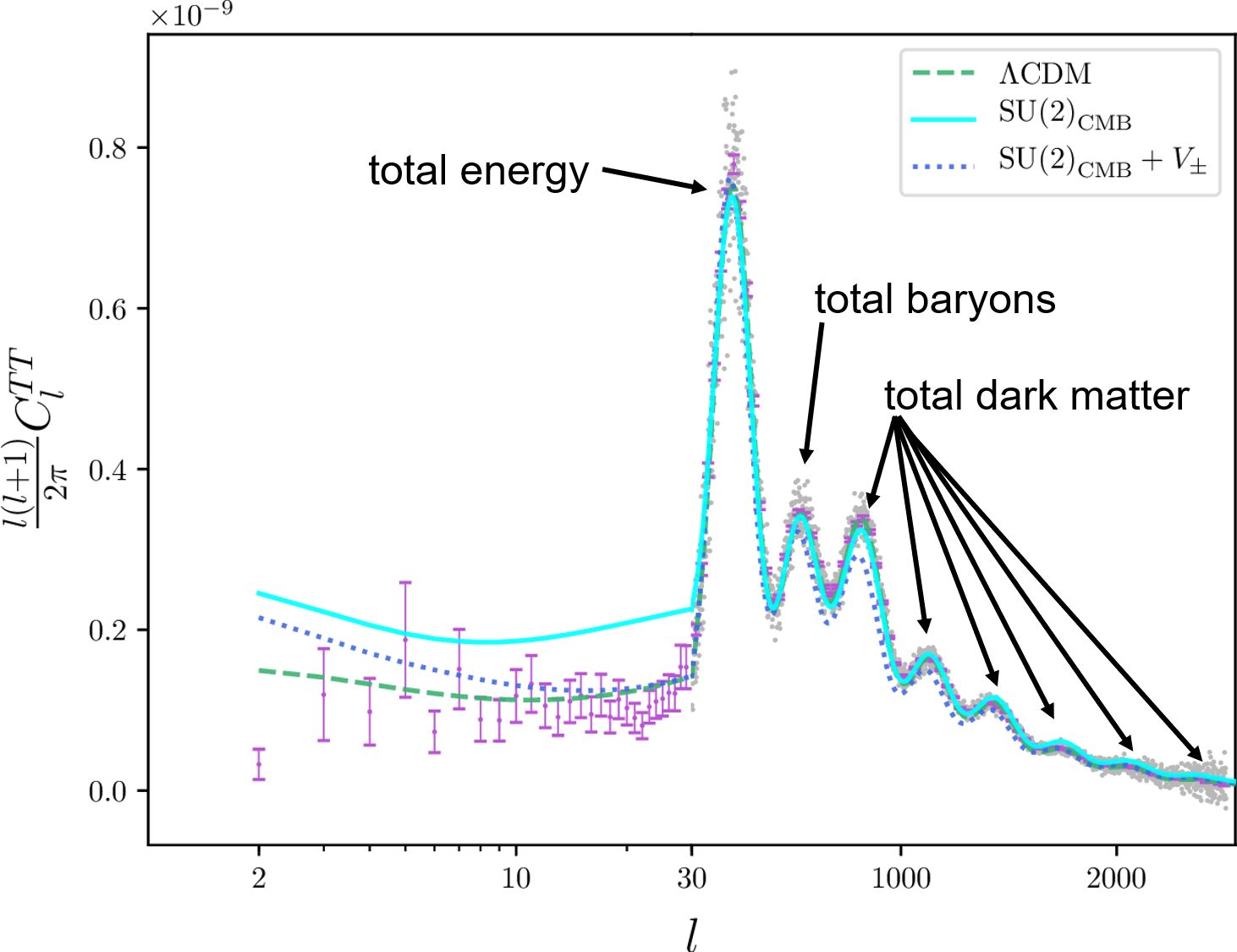}
\caption{ SU(2)\textsubscript{CMB} model fit (turquoise) in comparison to a $\Lambda$CDM fit (green). The 2015 Planck data points have been used for the fit \cite{Planck2016}.  
}
\label{fig:SU(2)fit}
\end{figure}

\begin{figure}[H]\centering
\includegraphics[width=10cm]{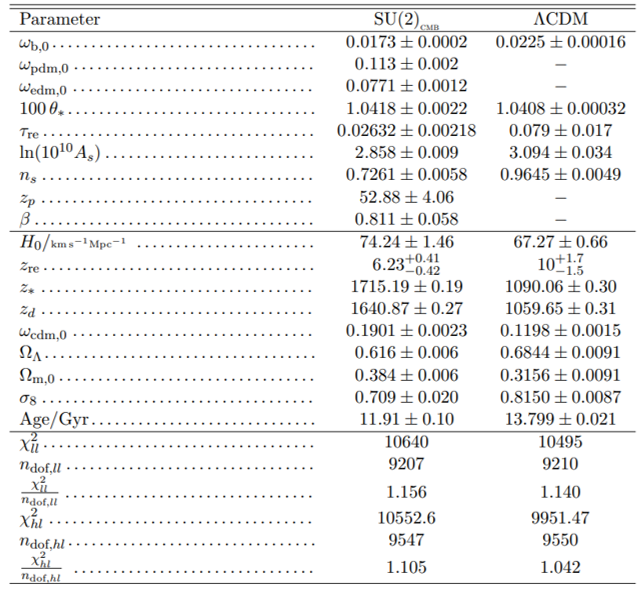}
\caption{Best-fit cosmological parameters of the SU(2)\textsubscript{CMB} model in comparison to $\Lambda$CDM, compare \cite{Hahn:2018dih}.}
\label{fig:tabelleDensities}
\end{figure}

Fitting the temperature fluctuations collected by the Planck collaboration (as seen in Fig.\,\ref{fig:SU(2)fit}), we get for the initial densities $\omega_\Lambda$, $\omega_{\rm b}$, $\omega_{\rm edm}$ and $\omega_{\rm pdm}$ the values in Fig.\,\ref{fig:tabelleDensities}. The SU(2)\textsubscript{CMB} fit as well as the $\Lambda$CDM model are compared in Fig.\,\ref{fig:SU(2)fit}.\nn

How can we deduce the energy content, and more over the specific contributions to the total energy of dark matter, dark energy, baryons from temperature fluctuations at the beginning of the universe, as seen in Fig.\,\ref{fig:SU(2)fit}? 
The temperature deviations from the background temperature, which is about 2.725 K today, arise from small density and radiation-pressure variations of matter in the early before the universe turned transparent. Fitting the CMB is a highly non-trivial and complicated task, many coupled Boltzmann equations have to be solved simultaneously.\nn

Programs such as the Cosmic Linear Anisotropy Solving System (CLASS) \cite{lesgourgues2011cosmic} are often employed when solving these equations. Roughly speaking, the first peak in Fig.\,\ref{fig:SU(2)fit} corresponds to the particle clouds which collapsed only once. It can give information about the total energy content of the universe. The second peak corresponds to the particle clouds which were able to collapse twice and the height of the peak gives information about the baryon content of the universe. All other peaks are linked to the dark matter content of the universe, compare \cite{YT2019}.\nn

\par\noindent\colorbox{lightgray1}{
    \begin{minipage}{\textwidth}
    \vspace{1mm}
    \textbf{Recombination} describes the point in time when the Universe turned transparent for photons\footnote{Note, that gravitational waves were able to propagate freely before as well as neutrinos.}. According to $\Lambda$CDM this happened at a redshift $z \sim 1090$ (see section \ref{SU(2) Yang-Mills thermodynamics} for a definition of redshift). The SU(2)\textsubscript{CMB} fit indicates a higher redshift, $z \sim 1715$. Before the recombination, the Universe was opaque because photons scattered of freely moving electrons. The Universe became transparent after those electrons \emph{combined} with protons to form hydrogen atoms.
\vspace{1mm}
    \end{minipage}
    }\nn
    
    \newpage

\begin{figure}[h]\centering
\includegraphics[width=10cm]{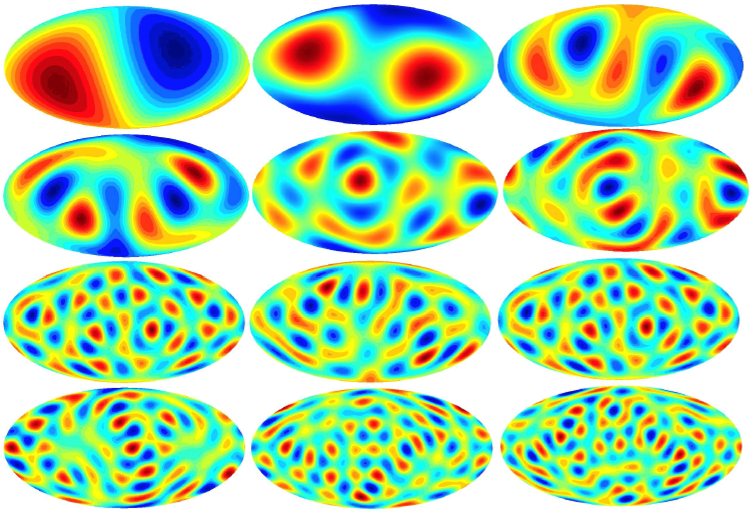}
\caption{Randomly generated multipoles from $l=1$ to $l=12$, top to bottom, left to right. The dipole ($l=1$) starts in the top corner left. Fitting the observed CMB temperature fluctuation enables one to fit those multipoles and conclude the total, energy density of the universe as well as the baryon and dark matter mass contributions, see Fig.\,\ref{fig:SU(2)fit}. The Fig.\,was made by Ville Heikkilä, compare https://astro.uni-bonn.de/$\sim$kbasu/ObsCosmo/Slides2019/CMB\_Part2.pdf.
}
\label{fig:multipoles}
\end{figure}

 The Cosmic Microwave background emerges from the first photons which are able to freely move long distances in the Universe after recombination. Sound waves in the photon-baryon fluid, so called baryonic acoustic oscillation (BAO) lead to small density fluctuations in the fluid which translates into small ($\sim \mu$K) temperature fluctuations in the CMB. Those osculations can be are linear-combination of multipoles as seen in Fig.\,\ref{fig:SU(2)fit}. An intuitive way of interpreting the multipoles is shown in Fig.\,\ref{fig:multipoles}, in the first panel (top,left) a randomly generated dipole is shown ($l=1$), followed by a quadrupole ($l=2$), an octupole ($l=3$), ($l=3\,\cdots\,12$).\nn

\par\noindent\colorbox{lightgray1}{
    \begin{minipage}{\textwidth}
    \vspace{1mm}
    \textbf{Baryonic Acoustic Oscillation (BAO)} are density waves in the photon-baryon fluid before recombination. Until the decoupling of matter and radiation, those sound waves were created by pressure and gravitational instability. Their maximum wavelength is the horizon size $r_H$ at the time of decoupling which can still be detected today by the anisotropies in the CMB spectrum and by the  distribution of galaxies in space.
\vspace{1mm}
    \end{minipage}
    }\nn
\vspace{3mm}

\begin{figure}[H]\centering
\includegraphics[width=12cm]{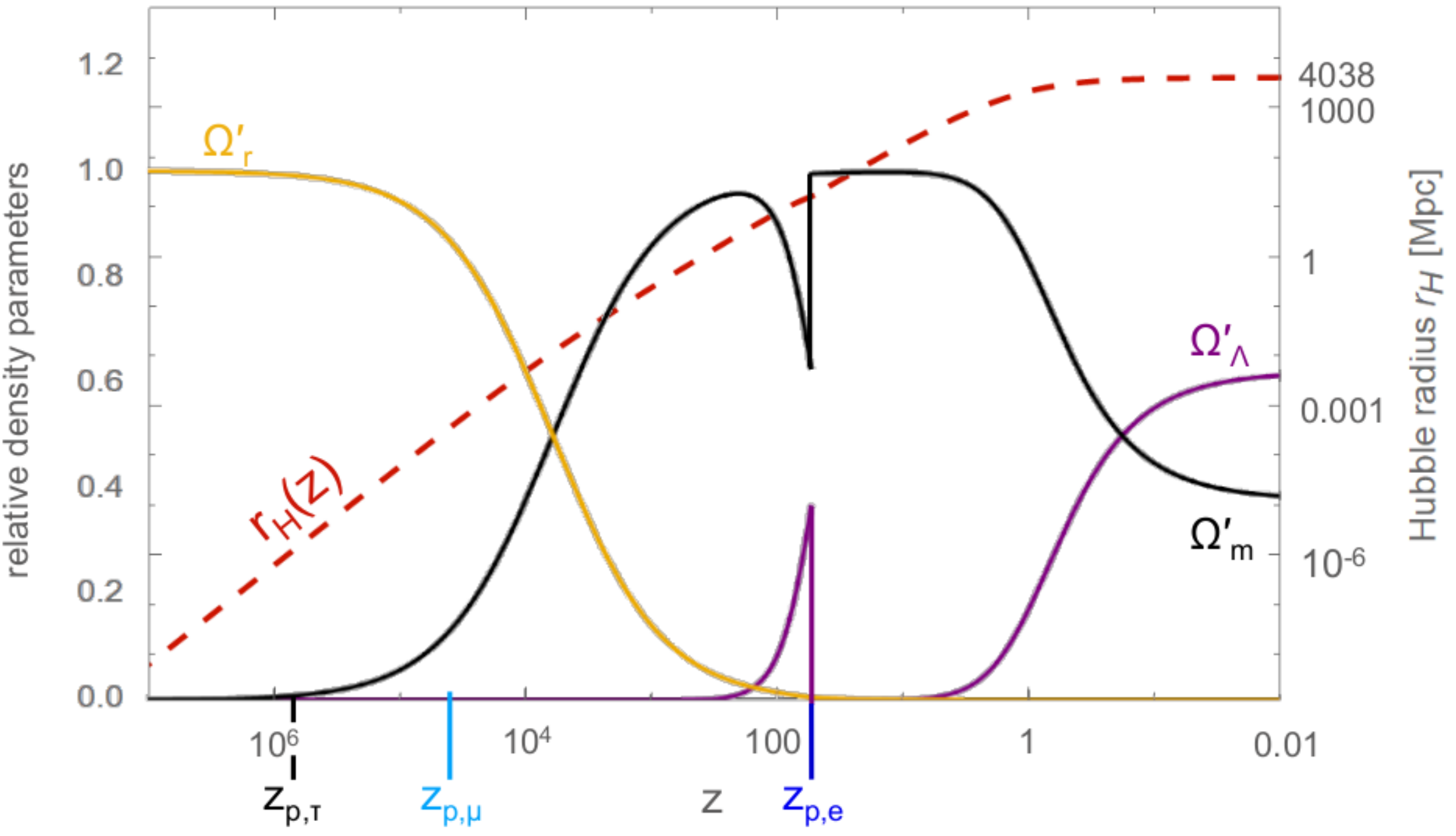}
\caption{
Cosmological model SU(2)$_{\rm CMB}$ of \cite{Hahn:2018dih} (with parameter values fitted to the TT, TE, and EE CMB Planck power spectra and taken from column 2 of Table 2 of that paper) in terms of relative density parameters as functions of redshift $z$. $\Omega^\prime_\Lambda$ stands for dark energy, $\Omega^\prime_{m}$ for total matter (baryonic and dark), and $\Omega^\prime_{r}$ for radiation (three flavours of massless neutrinos and eight relativistic polarisations in a CMB subject to SU(2)$_{\rm CMB}$). The dotted red line represents the Hubble radius of this model. The redshifts of $e$-lump, $\mu$-lump, and $\tau$-lump depercolations are indicated by vertical lines intersecting the $z$-axis. Only e-lump depercolation is taken into account explicitly within the cosmological model SU(2)$_{\CMB}$ since at $z_{p,\mu}=40,000$ and $z_{p,\tau}=685,000$ the Universe is radiation dominated.
}
\label{fig:rH}
\end{figure}

After this very brief overview of the CMB angular spectrum we can now visualize the density parameters of SU(2)$_{\rm CMB}$, see Fig.\,\ref{fig:rH}. The normalized energy composition at different redshifts $z$ is shown alongside the size of the universe $r_H=1/H(z)$ in Mpc. The most notable difference to $\Lambda$CDM model is the spike at $z=53$ which is induced by the Heaviside function of Eq.\,(\ref{eq:depercolation}). At this point dark energy spontaneously depercolates into dark matter. 
As one of the main results of this thesis we will argue in chapter \ref{chapter7} how to link the leptonic sector with three axions species. A fourth, the one corresponding to the photon SU(2) theory, has not yet depercolated. Consequently, not only one point of depercolation at $z=53$ but three are also indicated in the graphic. We get those high redshifts simply by comparing the size of the Universe $r_H$ with the gravitational Bohr radius of the axion species which we link to the electron scale $\Lambda_e$. This will be discussed in more detail in section \ref{Sec5}.
\nn

\chapter{ Dark Matter and Rotation Curves
}\label{chapter-DM}

\vspace{0.1cm}

\begin{minipage}{0.7\textwidth}
\begin{figure}[H]
\includegraphics[width=10cm]{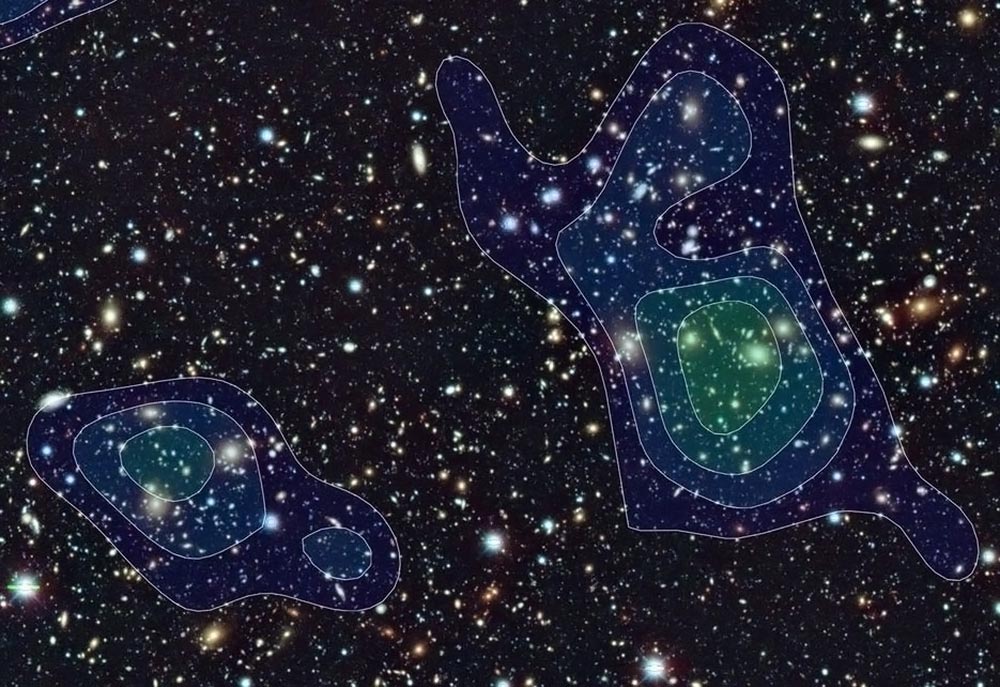}
\caption{A weak lensing image reaveals dark matter distribution which is shown in the coloured contours. Credit: Satoshi Miyazaki}
\end{figure}
\end{minipage} \hfill
\begin{minipage}{0.3\textwidth}
\begin{itemize}\sffamily

\item[\textcolor{gray!90}{\textbullet}] \textcolor{gray!90}{What is Dark matter?}

\item[\textcolor{gray!90}{\textbullet}] \textcolor{gray!90}{What motivates its existence?}

\item[\textcolor{gray!90}{\textbullet}] \textcolor{gray!90}{What could it be?}

\item[\textcolor{gray!90}{\textbullet}] \textcolor{gray!90}{What are axions?}
\end{itemize}
\end{minipage}\vspace{1.5cm}

\minitoc

\newpage

\section{Introduction}

The idea of dark matter has been around for well over one-hundred years \cite{Bertone_2018}. In 1933 the Swiss-American astronomer Fritz Zwicky found large discrepancies between estimates of dispersion velocities and measurements for the Coma galaxy cluster \cite{Zwicky:1933gu}. Therefore, he came to the conclusion that dark matter might be more abundant than ordinary matter.\nn

\begin{figure}[H]\centering
\includegraphics[width=10cm]{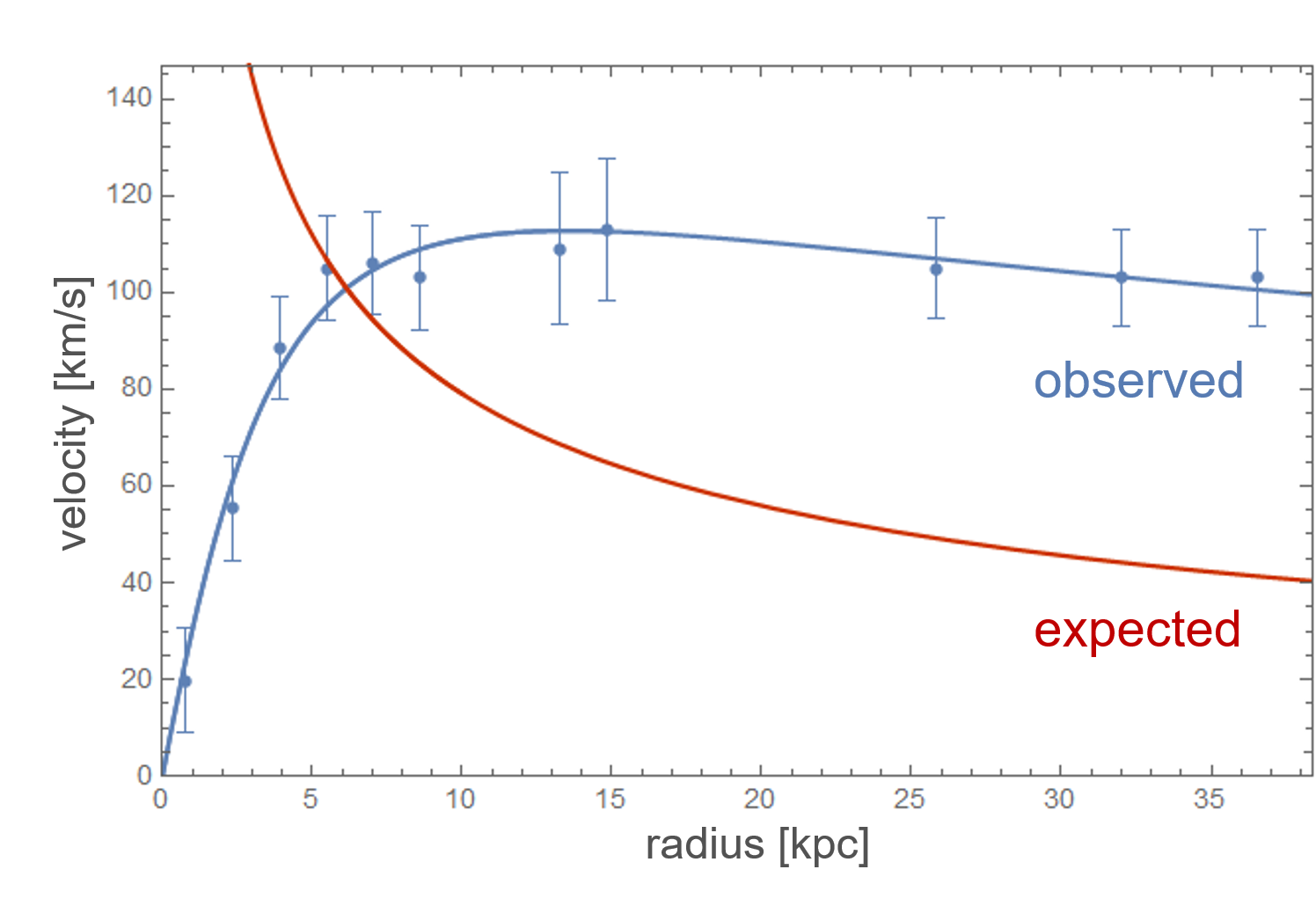}
\caption{
Further experimental evidence for dark matter came from measurements of rotation curves. Instead of a declining velocity, as expected from Kepler's laws  (red, $\propto 1/\sqrt{r}$), the rotation curve has a relatively constant slope (blue, plotted Burkert profile of galaxy UGC01230). This suggest that galaxies are heavier than expected based on baryonic matter. More matter is inside galaxies than we see: matter which does not interact with light and hence is dubbed dark matter.}
\label{RCRubin}
\end{figure}

The second key moment in the history of dark matter is the development of a spectroscope by Kent Ford in the 1960s which enabled him and Vera Rubin to measure rotation curves \cite{Bertone_2018}. Rotation curves of galaxies plot the circular velocity of visible stars or gas in a galaxy over their corresponding radii to the center of the galaxy. Kent and Rubin observed the rotation curve of the galaxy M31 in 1970 \cite{Rubin:1970zza}. To their surprise, the circular velocity did not show a reciprocal dependence on the radius $r$ as expected from Kepler's law, but was flat to the largest radius observed. This discrepancy between visible mass, based from hydrogen surface density profiles by e.g. Rogstad and Shostak in 1972 \cite{[262]}, and the total galaxy mass implied by rotation curves is nowadays one of the most convincing arguments for the existence of dark matter. This discrepancy is visualized for the galaxy UGC01230 in the figure \ref{RCRubin}, below.

Before focusing on axions in general and Planck-scale axions in particular, the main dark matter candidate of this thesis, we would like to briefly mention relatively new advances (within the last two decades) in search for dark matter. Namely, the cosmological standard model $\Lambda$CDM fit to the Planck satellite data, which as the name suggest accounts for cold dark matter, and the bullet cluster which is widely assumed to be a direct proof of the existence of dark matter.\nn

The Lambda Cold Dark Matter (LCDM or $\Lambda$CDM) model incorporates additionally to the Cold Dark Matter model -- as published in 1984 by George Blumenthal, Joel Primack, and Sandra Faber \cite{Blumenthal:1984bp} -- the cosmological constant $\Lambda$, which is associated with dark energy. The model, which is commonly also referred to as the standard cosmological model, has only the assumptions of spatial isotropy and homogeneity of the cosmos on the right scales for all observers and general relativity. Friedmann-Lemaître-Robertson-Walker Cosmology contains three major components for the present epoch, the aforementioned dark energy, cold dark matter and baryonic matter. A fit to the Cosmic Microwave Background as seen in Fig.\,\ref{fig:SU(2)fit} reveals their respective densities as seen in the table below. The $\Lambda$CDM fit densities are shown in Fig.\ref{fig:tabelleDensities}.\nn

\begin{figure}[H]\centering
\includegraphics[width=14.8cm]{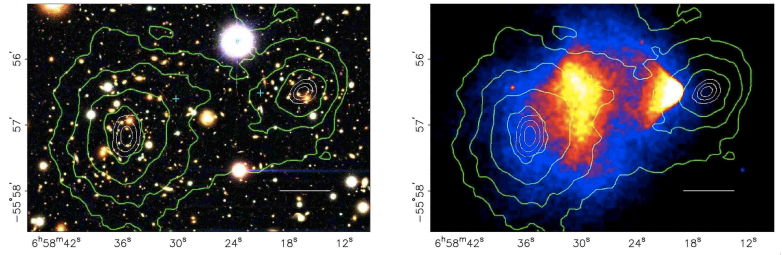}
\caption{
An image by the Chandra telescope of the bullet cluster, 1E 0657-558, overlayed with a weak gravitational image. The white bars in both pictures indicates a distance of 200 kpc, visible light is shown on the left picture, the hot gas detected by X-rays (red) and the dark matter distribution (blue) is shown on the right \cite{DelPopolo:2013qba}. 
The dark matter distribution was reconstructed using gravitational lensing which indicate the mass distribution determined by weak gravitational lensing  (green lines) \cite{Clowe_2006,DelPopolo:2013qba}.
}
\label{bullet}
\end{figure}

Arguably the strongest argument for the existence for dark matter comes from weak gravitational lensing. As shown on the right-hand side of Fig.\,\ref{bullet}, there are two former galaxy clusters which are now, after a collision, part of the galaxy cluster 1E 0657-56 \cite{Clowe_2006}. This galaxy cluster is also known as the “Bullet Cluster” because of the shape of the galaxy cluster on the right-hand side in Fig.\,\ref{bullet} resembles a bullet. The hot gas (red in Fig.\,\ref{bullet}) was detected by the Chandra telescope in the X-ray regime. The dark matter content in this galaxy cluster which is indicated in blue in Fig.\,\ref{bullet}, was reconstructed using weak gravitational lensing.\nn

\par\noindent\colorbox{lightgray1}{
    \begin{minipage}{\textwidth}
   \vspace{1mm}
     \textbf{Gravitational lensing} describes the distortion of light passing by a strong source of spacetime distortion. Such distortions can be caused by galaxy clusters, massive objects in galactic centers and black holes. They are predicted by General Relativity and can be used to estimate masses of the sources of those spacetime distortions. Special cases such as the multiplying of a light source is called Einstein's cross, as depicted below in Fig.\,\ref{fig:lensingcross}.\nn
     
    \begin{figure}[H]\centering
\includegraphics[width=8 cm]{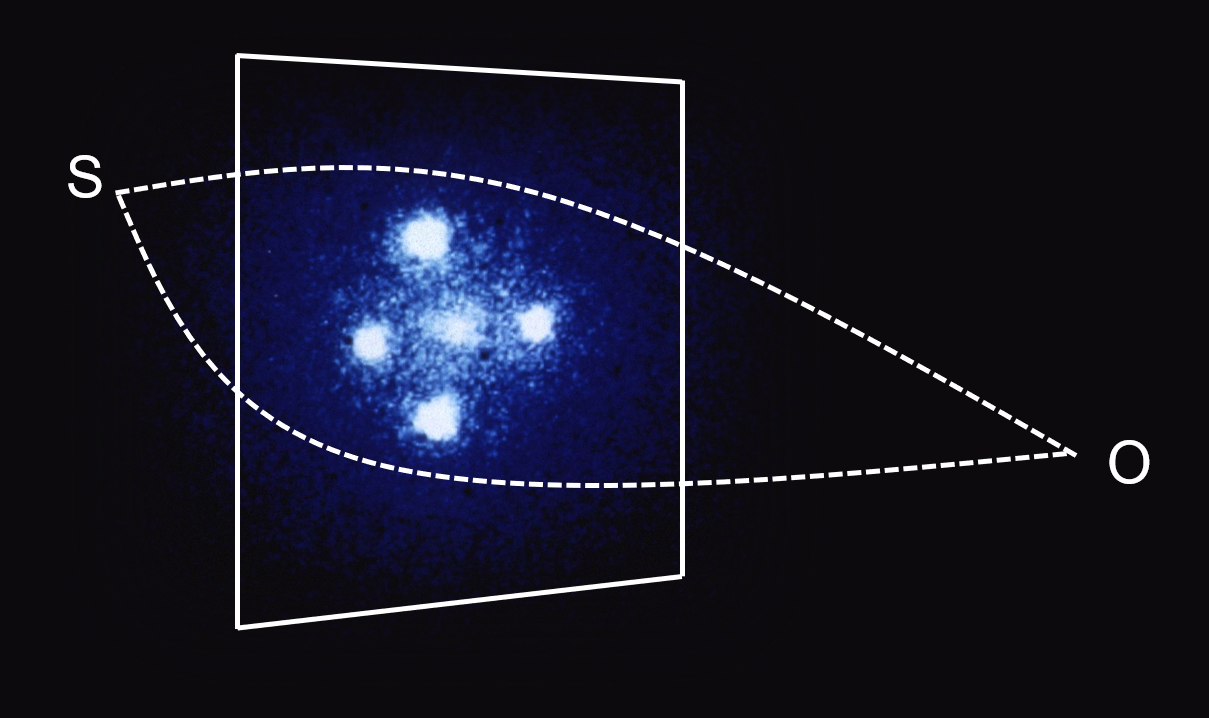}
\caption{ 
A gravitational lensing image (white rectangle) shows the quadrupling of the quasar QSO 2237+0305 (which started at a source S and is observed by an observer O) by a nearby galaxy, compare \cite{EinsteinCross}. Sometimes the multiplying is referred to as Einstein cross.
}
\label{fig:lensingcross}
\end{figure}

\vspace{1mm}
    \end{minipage}
    }
\vspace{3mm}

So far, the bullet cluster represents the strongest evidence for the existence of dark matter: The dominant mass portions of the merging galaxy clusters is ahead of the baryonic mass in their trajectories. This can  easily be explained by dark matter, which does not interact with the visible matter other than by gravitational interaction. They fly right through the hot gas lumps. This spatial offset of the baryonic mass distribution can not be explained by modified theories of gravity \cite{DelPopolo:2013qba}.\nn

\section{Axions}\label{axions}

\begin{figure}[H]\centering
\includegraphics[width=15cm]{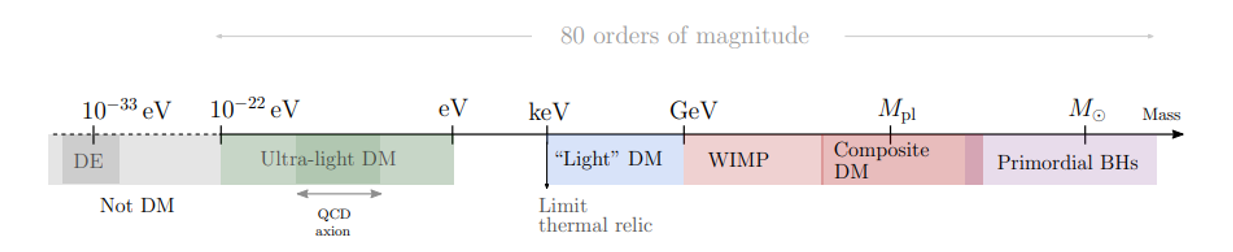}
\caption{
Sketch (not to scale) of the huge range of possible DM models that have been conceived. They span many orders of magnitude
in mass, with DM represented by very distinct phenomena, ranging from new elementary particles to black holes, compare figure 1 in \cite{ferreira2020ultralight}.
}
\label{fig:Orders80}
\end{figure}

Although dark matter in some form or another has been around for at least one-hundred years, there are plenty possible candidates covering 80 orders of magnitude (in eV), compare figure \ref{fig:Orders80}.
Unfortunately, it is not possible to give a comprehensive overview of all of them in 
this framework. We will focus on axions in general and on Planck-scale axions in particular (left green corner in figure \ref{fig:Orders80}).\nn

The idea of axions emerged as a byproduct of an attempt to solve the strong CP-problem in QCD. The strong CP-problem arises due to the non-trivial vacuum structure of non-Abelian gauge theories: Although the Standard model SU(3)$\times$SU(2)$\times$U(1) has no thermal phases, instantons and other topologically charged field configurations generate a non-trivial vacuum structure of the gauge theory \cite{Callan:1976je} which allows for (effective) CP-violating interactions of the strong and the weak force \cite{Peccei:1995ge}:\nn

\eqb
\label{maxxaxion}
\mathcal{L}_{\textrm CP viol.}= \theta_{\textrm strong}\, \frac{\alpha_s}{8 \pi} F^{\mu\nu}_a\widetilde{F}_{a\mu\nu}+
\theta_{\textrm weak}\, \frac{\alpha_2}{8 \pi} W^{\mu\nu}_a\widetilde{W}_{a\mu\nu}\,.
\eqe

Here $F^{\mu\nu}_a$ and $W^{\mu\nu}_a$ are the non Abelian gauge field strengths and $\widetilde{F}_{a\mu\nu}$, $\widetilde{W}_{a\mu\nu}$ their duals; $\alpha_s$ is the coupling constant of the strong force as $\alpha_2$ is the coupling constant of the weak force. The weak vacuum angle $\theta_{\textrm weak}$ can be rotated to zero since the electroweak theory is chiral \cite{Krasnikov_1979}. The CP-problem arises from the angle $\theta_{\textrm strong}$, which cannot be rotated away in the same fashion. More precisely, $\theta_{\textrm strong}$ gets modified due to the chiral rotation to minimize $\theta_{\textrm weak}$. The 'physical' angle is $\bar{\theta}= \theta_{\textrm strong} + \text{Arg det M}$ where M is the quark mass matrix \cite{Peccei:1995ge}. Therefore, the CP-violating term reads:\nn

\eqb
\mathcal{L}_{\textrm QCD}\supset \bar{\theta}\, \frac{\alpha_s}{8 \pi} F^{\mu\nu}_a\widetilde{F}_{a\mu\nu}\,.
\eqe

If $\bar{\theta}$ is of order one, the strong force would strongly violate CP-symmtery. However, measurements of the electric dipole moment of the neutron show no indications fur such a CP-violation \cite{Peccei:1995ge}: They imply a very small $\bar{\theta}$ of order $\sim 10^{-10}$. So the strong CP-problem boils down to the question of why $\bar{\theta}$ is so small?\nn

The idea by Helen Quinn and Roberto Peccei was to remove $\bar{\theta}$ dynamically by introducing a new, global chiral symmetry to the Standard Model, the so called, U(1)$_{\textrm PQ}$ symmetry. Similar to the electroweak theory with gauge group SU(2)$\otimes$U(1) where the chiral nature allows to rotate $\theta_{\textrm weak}$ away, the new chiral symmetry U(1)$_{\textrm PQ}$ allows to relax $\bar{\theta}$ to zero by dynamically and explicitly breaking it \cite{Peccei:1977hh,Peccei:1977ur,Peccei:1995ge}.
Frank Wilczek \cite{PhysRevLett.40.279} and Steven Weinberg \cite{Weinberg:1977ma} pointed out how the (pseudo) Nambu-Goldstone boson, the so axion, of dynamical U(1)$_{\textrm PQ}$ breaking couples to $F^{\mu\nu}_a\widetilde{F}_{a\mu\nu}$.\nn

\section{What are rotation curves?}

Rotation curves plot the averaged velocity of visible matter, i.e. stars which rotate around a galactic center, over their distances from the galactic center. The discrepancy between the classically expected rotation curve (blue) and the actually measured ones (green) in figure \ref{RCRubin} can be explained by postulating dark matter. Therefore, if one accepts the premise that dark matter exists, we can swap the argumentation and use rotation curves as a tool to look at the dark matter content of a given galaxy.\newline

As a warm-up we first look at a vacuum solution of Einstein's equations, the Schwarzschild solution \ref{sec:Rotation curves Sch.S.}. We compare an exact solution of the orbit equation in this background by utilizing elliptic functions with a post-Newtonian approximation, section \ref{sec:Rotation curves PN}. In the next chapter we will focuse on the nonrelativistic limit and fit rotation curves to data from the SPARC library \cite{Lelli:2016zqa}.
\newline

\section{Rotation curves for the Schwarzschild solution}
\label{sec:Rotation curves Sch.S.}

Based on an exact solution to the orbit equation for the bound trajectory of a test mass in the Schwarzschild background, obtained in \cite{Scharf_2011} in terms of a Weierstrass elliptic function, we numerically compute rotation curves (mean tangential velocity $\bar{v}_t$ as a function of proper distance $s_r$ to the horizon) in dependence of orbital eccentricity $\epsilon$. As it should, for $s_r\gg r_s$ the Newtonian behaviour $\bar{v}_t\propto 1/\sqrt{s_r}$ is reproduced.\vspace{12pt}


In his paper \cite{Scharf_2011}, G. Scharf obtained exact Schwarzschild geodesics in terms of elliptic functions. Building on his work, the aim of the present section is to derive rotation curves, that is, the dependence of the mean tangential velocity (averaged over one pseudo orbit) of a test mass on the proper distance to the Schwarzschild horizon, including the strong-field regime close to this horizon.\\

This could be of observational interest when the effects of mass distributions close to the gravitational center of a given spiral galaxy need to be disentangled from the strong-field effects \cite{Daod_2019}. Recall that, based on an estimate of luminous matter located in the central region, the observation of a flattening of rotation curves in spirals far away from the center serves as  an indication for the existence of a dark-matter halo \cite{Rubin:1980zd,Clowe_2006}.\\

The Schwarzschild metric in canonical coordinates is given as \cite{Carroll:1997ar}:
\begin{equation}
ds^2=\frac{r-r_s}{r} c^2 dt^2-\frac{r}{r-r_s} dr^2-r^2(d\theta ^2+sin^2\theta\, d\phi^2)
\label{SchSch}
\end{equation}\vspace*{1.5mm}

where the Schwarzschild radius $r_s$ is given as $r_s = {2 \,GM}/{c^2} $. With $G$ being the gravitational constant, $M$ the mass, and $c$ the speed of light. Usually, when discussing relativistic effects in the orbit of a test particle in the background of eq. \ref{SchSch}, one expands the metric coefficients in powers of $r/r_s$. Thanks to the knowledge on the exact solution $r(\phi)$ to the orbit equation (\cite{Scharf_2011}, eq. 2.11)

\begin{equation}
\frac{dr}{d\phi}= \sqrt{\frac{E^2-m^2}{L^2}\,r^4+\frac{m^2}{L^2}\,r_s r^3-r^2+r_s r}	\equiv \sqrt{f(r)}
\label{orbit}
\end{equation}\vspace*{1.5mm}

orbital parameters such as the perihelion shift could be obtained up to quadratic order in $r/r_s$ very economically by expanding this solution. Here $E$ and $L$ are constants of motion, energy and angular momentum and $m$ is the mass of the test-particle.
Here, we would like to exploit this knowledge to derive exact rotation curves in dependence of eccentricity $\epsilon$.\newline



The approach based on Weierstrass functions was used in \cite{Scharf_2011}. According to (\cite{Scharf_2011}, eq. 2.13) of this work, the radial coordinate $r(\phi)$ of a test particle in the background of the geometry governed by eq. \ref{SchSch} is 

\begin{equation}
r(\phi)= r_1 + \frac{f'(r_1)}{4\wp(\phi; g_2, g_3)-f''(r_1)/6}
\label{eq:refname1}
\end{equation}\vspace{1mm}

Where $f(r)$ is a quartic of the form (\cite{Scharf_2011}, eq. 2.16):

\begin{equation}
f(r) = a_0 \, r(r - r_1)(r - r_2)(r - r_3)
\end{equation}\vspace{1mm}

$a_0$ can related to the energy $E$, angular momentum $L$ and the mass of the test-particle $m$ as $a_0 = (E^2-m^2)/L^2$ (\cite{Scharf_2011}, eq. 2.18). The radial points $r_1$, $r_2$, and $r_3$ are zeros of this quartic where $r_1$ can be interpreted as perihelion, $r_2$ as aperihelion, and $r_3$ can be expressed as a combination of $r_1$, $r_2$ and $r_s$ (\cite{Scharf_2011}, eq. 2.21), as well as the invariants $g_2$ and $g_3$ (\cite{Scharf_2011}, eq. 2.22 f.)\newline

The first and second derivatives of the quartic are given as:

\begin{align}
f'(r_1) &= -\frac{r_1 (r_1 - r_2)\cdot(r_2 - r_3)}{r_2 r_1 + r_2 r_3 + r_3 r_1} \\ \vspace{1cm}
f''(r_1) &= \frac{r_1 (r_1-r_2) + r_1 (r_2-r_3) + (r_1-r_2)\cdot(r_2- r_3)}{r_2 r_1 + r_2 r_3 + r_3 r_1},
\label{eq:derivatives}
\end{align}\vspace*{1.5mm}

We get to the final form of $r(\phi)$ only depending on the perihelion $r_1$, $r_2$ and the angle $\phi$. However, it is convenient to define $r_2$ via the eccentricity $\epsilon$ as $r_2 = r_1 (1-\epsilon)/(1+\epsilon)$. In order to obtain a rotation curve we need to define the tangential velocity $\Bar{v_t}$ of the test-particle averaged over one pseudo orbit, as a function of the averaged distance $\Bar{s}$. The tangential velocity is:

\begin{align*}
v_t = r(\phi(\tau)) \, \frac{d\phi}{d\tau},
\end{align*}\vspace*{1.5mm}

where $r(\phi)$ is a solution of the orbit equation and $\phi(\tau)$ is a solution of the geodesic equations in dependence of the affine parameter $\tau$. For readability this dependence is notationally suppressed in the following. Then the averaged tangential velocity $\bar{v_t}$ is obtained by:
\begin{align}
\bar{v_t} &= \frac{1}{T} \int_{0}^{T} d\tau \, v_t\\
&=  \frac{r_s}{T} \int_{0}^{2\pi} d \phi \, r(\phi) 
\label{eq:vbar}
\end{align}\vspace*{1.5mm}

With $T$ the eigentime of the orbit. The averaged distance $\Bar{s}$ is:
\begin{align*}
\bar{s} &= \frac{1}{2 \pi} \int_{0}^{2 \pi} d\phi \, s_r(r(\phi)) 
\end{align*}
\vspace*{1.5mm}

$s_r$ is the average proper distance to the horizon, which is defined as:

\begin{align*}
 s_r(y) &= \int_{r_s}^{y} \frac{dy^\prime}{\sqrt{1-\frac{r_s}{y^\prime}} }
 = y \sqrt{1-\frac{r_s}{y}}+\frac{r_s}{2} \log \left(\frac{1+\sqrt{1-\frac{r_s}{y}}}{1-\sqrt{1-\frac{r_s}{y}}} \right) \\
\label{eq:refname4}
\end{align*}\vspace*{1.5mm}

 $\bar{v}(\bar{s})$ can now be calculated by inverting $\bar{s}(r(\phi))$ and plugging it in equation \ref{eq:vbar}. The only thing missing is $T$ which
 needs to be found by normalization, which here means comparing the general expression 
 
 \begin{equation}
T=  \frac{1}{L} \int_{0}^{2\pi} d\phi \, r(\phi)^2, \hspace{0,5cm} L=m \sqrt{ \frac{r_2 r_1 + r_2 r_3 + r_3 r_1}{r_2 + r_1 + r_3} \cdot r_s}
\ \end{equation}\vspace{1.5mm}
 
where $L$ is the angular momentum (\cite{Scharf_2011}, eq. 2.8 \& 2.19), with the physical orbital periods. In this case, normalization gets rid of the factor $2 \pi\, m$. Plugging $T$ in eq. (\ref{eq:vbar}) gives: 
 
\begin{align}
\bar{v_t} &=2 \pi\, m \sqrt{\frac{r_2 r_1 + r_2 r_3 + r_3 r_1}{r_2 + r_1 + r_3} \, \cdot r_s} \frac{\int_{0}^{2 \pi} d \phi \, r(\phi) }{\int_{0}^{2 \pi} d\phi \, r(\phi)^2}  
\end{align}\vspace*{1.5mm}


Because of the perihelion motion we have to integrate over $4 \omega$ instead of $2\pi$, where $\omega$ is the real half-period of the $\wp$-function which is given by (\cite{Scharf_2011}, eq. 3.12):
\begin{align*}
\omega = \pi  \left( 1+\frac{3 r_s}{4}\, \frac{r_1 + r_2}{r_1 r_2} + O(r_s^2)\right)
\end{align*}\vspace*{1.5mm}

Note, that here the post-Newtonian approximation has already been used. Together with the normalisation our tangential velocity finally is:

\begin{align}
\bar{v_t} &= \sqrt{\frac{r_2 r_1 + r_2 r_3 + r_3 r_1}{r_2 + r_1 + r_3}\cdot r_s} \frac{\int_{0}^{4\omega} d \phi \, r(\phi) }{\int_{0}^{4\omega} d\phi \, r(\phi)^2}  
\end{align}\vspace*{1.5mm}

\begin{figure}[H]
\centering
\includegraphics[width=11cm]{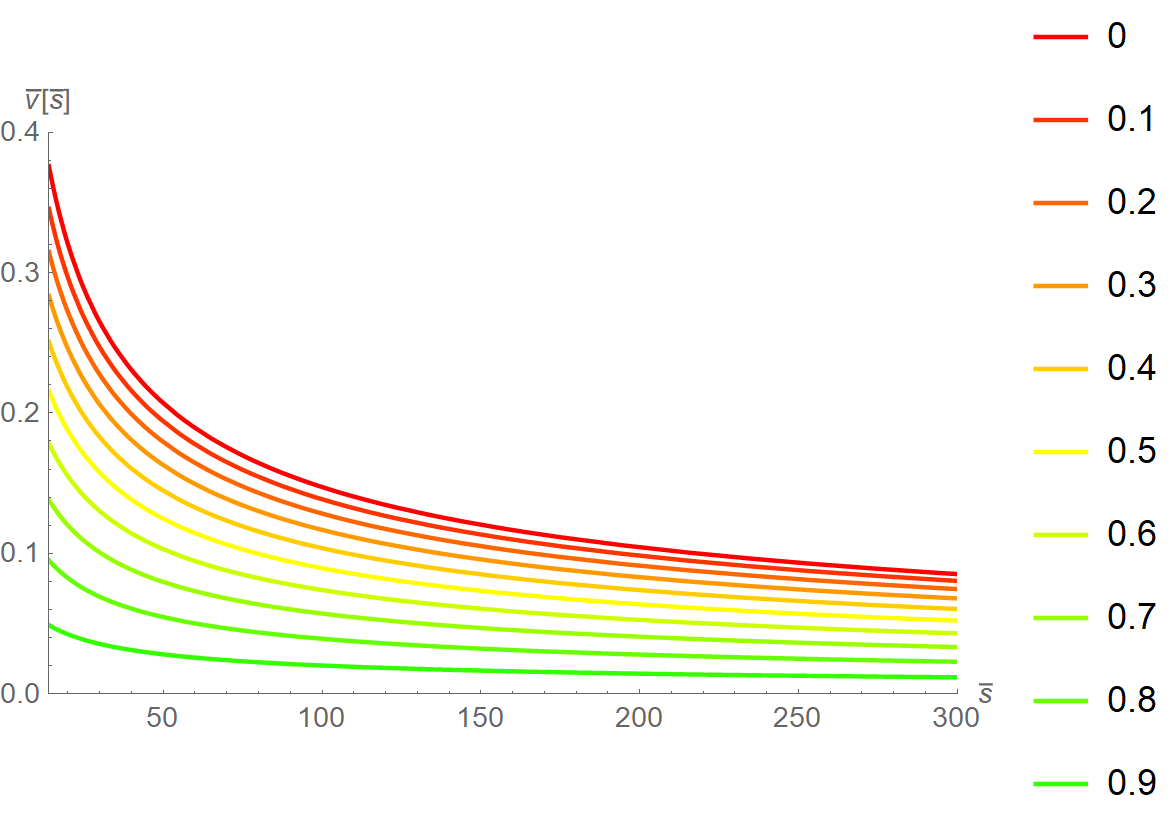}
\caption{Rotation curve for eccentricity from for different eccentricities $\epsilon \in \{0-0.9\}$ (red to green), $r_s=1$ for an exact solution of the Schwarzschild geodesics with elliptical functions. 
}
\label{fig:RotationCurves1}
\end{figure}

Calculating the averaged tangential velocity $\Bar{v_t}$ as a function of the averaged distance $\Bar{s}$ gives us the rotation curve of a test-particle. Figure \ref{fig:RotationCurves1} shows the rotation curve in natural units ($G=1, c=1$) for different eccentricities $\epsilon \in \{0-0.9\}$.

\section{Rotation curves for a post-Newtonian approximation}
\label{sec:Rotation curves PN}

Recall that the Weierstrass functions was used in the previous section and originally in \cite{Scharf_2011} in order to obtain a function for the orbit, compare \ref{eq:refname1}

\begin{equation}
r(\phi)= r_1 + \frac{f'(r_1)}{4\wp(\phi; g_2, g_3)-f''(r_1)/6}
\end{equation}\vspace{1mm}

The first and second derivative $f'(r_1),f''(r_1)$ can be expressed as \cite[p. 9, eq. 3.17]{Scharf_2011}

\begin{align}
f'(r_1) &= 2 r_1 \left(\frac{\epsilon}{1+\epsilon}-\frac{r_s}{r_1}\frac{3\epsilon+\epsilon^2}{(1+\epsilon)^2}\right) \\ \vspace{1cm}
f''(r_1) &= 2  \left(\frac{1-5\epsilon}{1+\epsilon}+3\frac{r_s}{r_1}\frac{1-4 \epsilon -\epsilon^2}{(1+\epsilon)^2}
\right)\,,
\end{align}\vspace*{1.5mm}

The Weierstrass function $\wp$ is given in terms of theta functions in order to account for the  relativistic corrections for $r(\phi)$, compare \cite[p. 650, eq. 18.10.5]{Handbook}

\begin{equation}
\wp(\phi)= e_2+\frac{\pi^2}{4 \omega^2}\left(
\frac{\theta_1'(0)}{\theta_3(0)}\frac{\theta_3(\phi)}{\theta_1(\phi)}
\right)^2\,,
\end{equation}\vspace{1mm}

where $\phi$ is given by $\phi=\frac{\pi}{2\omega}\varphi$ and the theta functions are given by \cite[p. 464]{Whittaker}

\begin{align}
\theta_1(z,q) & = 2 q^{1/4} (\text{sin}(z)-q^2 \text{sin}(3 z)+ q^6 \text{sin}(5 z) - \dots)\\
\theta_2(z,q) & = 2 q^{1/4} (\text{cos}(z)+q^2 \text{cos}(3 z)- q^6 \text{cos}(5 z) + \dots)\\
\theta_3(z,q) & = 1+2 q (\text{cos}(2 z)+q^3 \text{cos}(4 z)- q^8 \text{cos}(6 z) + \dots)\\
\theta_4(z,q) & = 1-2 q (\text{cos}(2 z)-q^3 \text{cos}(4 z)+ q^8 \text{cos}(6 z) - \dots)\,.
\end{align}\vspace{1mm}

In this context, $q$ is called a Nome \cite{Scharf_2011} and can be expressed as \cite[p. 591, eq. 17.3.21]{Handbook} 

\begin{equation}
q = k^2/16+ 8(k^2/16)^2+\dots\,. 
\end{equation}

Since $k$ is small the series is converging \cite[p. 9]{Scharf_2011}. This leads to the following Weierstrass function $\wp$ [ibid.]

\begin{equation}
\wp(\phi)=  \frac{f''(r_1) }{24} + \frac{1+\epsilon- 2 \epsilon \text{sin}^2(\phi)}{4 (1+\epsilon) \text{sin}^2(\phi)} \left(
1+\frac{r_s}{r_1}\frac{1}{1+\text{cos}(2 \phi)}\left(
-3-\frac{\epsilon}{2}(1-\text{cos}(\phi))+2\epsilon\frac{3+\epsilon}{1+\epsilon}\text{sin}^2(\phi)
\right)
\right)+\mathpzc{O}(\frac{r_s}{r_1})^2
\end{equation}\vspace{1mm}

Substituting this into \ref{eq:refname1}, one obtains the orbit given by

\begin{equation}
\text{r}(\theta) = \frac{r_1 (\epsilon +1)}{\epsilon  \cos \left(\frac{2 \theta }{\alpha}\right)+1}+\frac{ r_s \left(2 \epsilon  \sin ^2\left(\frac{\theta }{\alpha}\right)\right) }{\epsilon  \cos \left(\frac{2 \theta }{\alpha}\right)+1}\left(\frac{-\frac{2 \epsilon  (\epsilon +3) \sin ^2\left(\frac{\theta }{\alpha}\right)}{\epsilon +1}+\frac{1}{2} \epsilon  \left(1-\cos \left(\frac{\theta }{\alpha}\right)\right)+3}{\epsilon  \cos \left(\frac{2 \theta }{\alpha}\right)+1}-\frac{\epsilon +3}{\epsilon +1}\right)
\end{equation}

\begin{figure}[H]
\centering
\includegraphics[width=11cm]{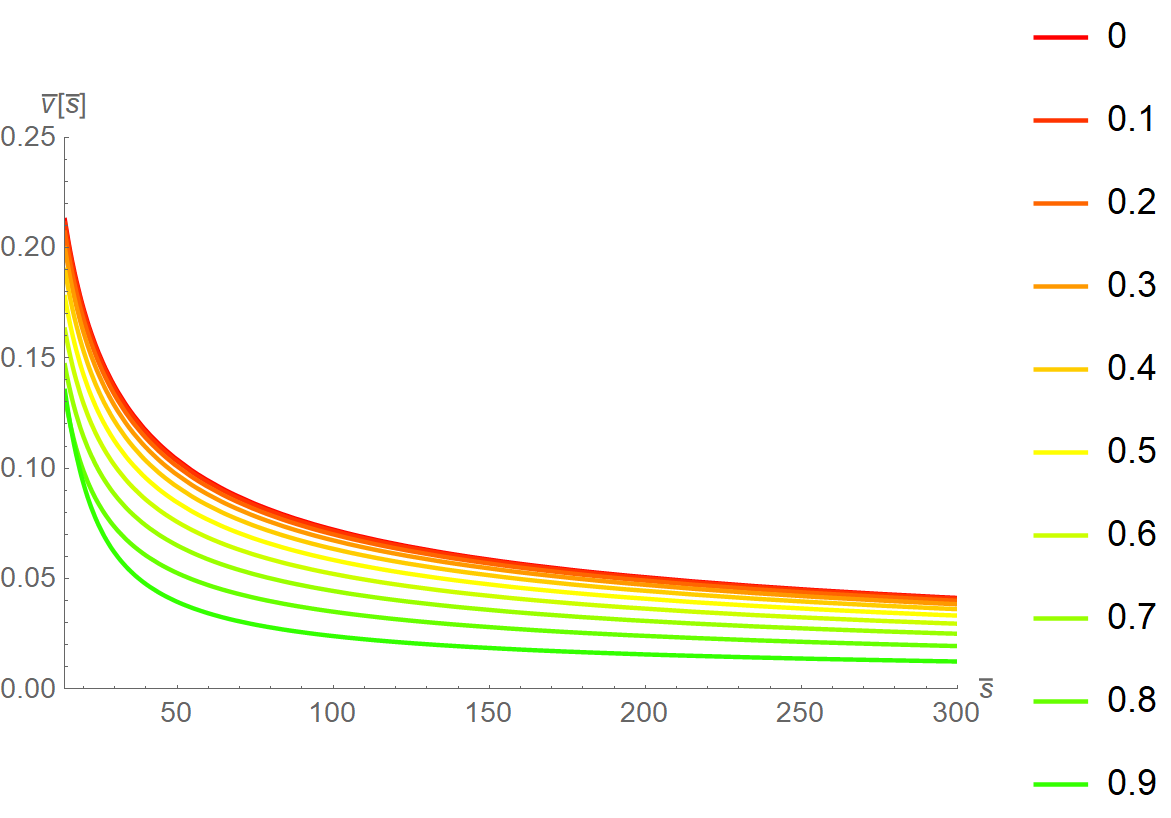}
\caption{Rotation curve for eccentricity for different eccentricities $\epsilon \in \{0-0.9\}$, $r_s=1$ in a post-Newtonian approximation.}
\label{fig:RotationCurves2}
\end{figure}

Where $\alpha$ is defined as:

\begin{align*}
\alpha = 2 \left(\frac{3 r_s (r_2+r_1)}{4 r_1 r_2}+1\right)
\end{align*}\vspace*{1.5mm}

Which results in the rotation curves as seen in figure \ref{fig:RotationCurves2}. Interestingly, for high eccentricities around $\epsilon = 0.9$ , the rotation curve seems to be steeper, i.e. crossing over RCs with less eccentricity at the center of the galaxy. However, this effect might be explained by insufficient numerical accuracy close to the center for high eccentricities.


\chapter{ Axion mass extractions}

\vspace{0.1cm}

\begin{minipage}{0.7\textwidth}
\begin{figure}[H]\centering
\includegraphics[width=9cm]{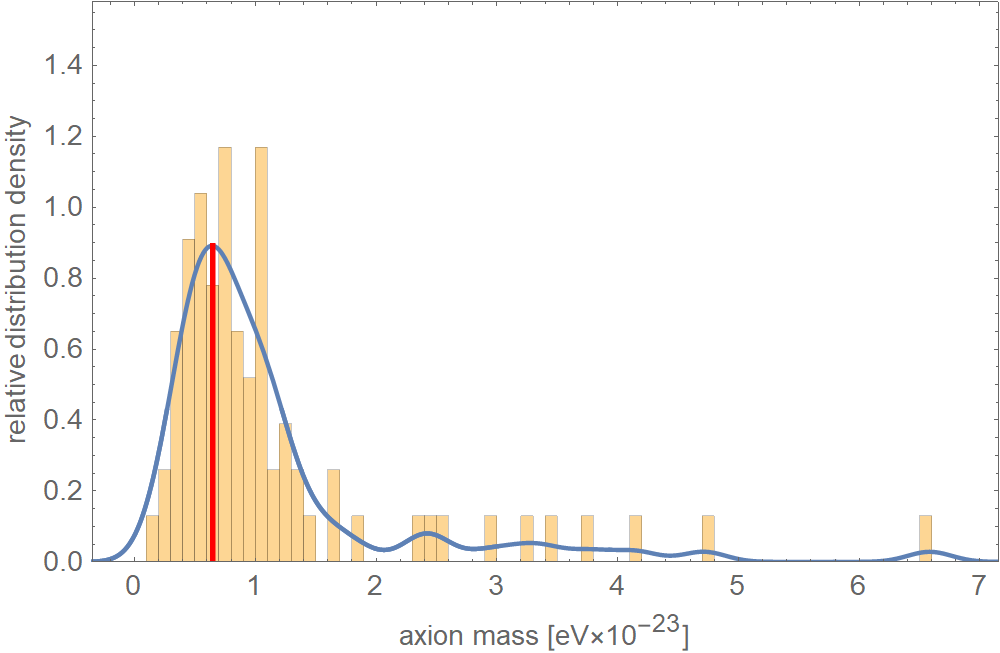}
\caption{Frequency distribution of 80 axion masses $m_{a,e}$, extracted from the Burkert-model fits of $M_{200}$ to the RCs of galaxies with a $\chi^2/\text{d.o.f.}<1$. The maximum of the smooth-kernel distribution (solid, blue line) is $m_{a,e}=(0.65\pm0.4)\times10^{-23}$\,eV (red, vertical line).}
\end{figure}
\end{minipage} \hfill
\begin{minipage}{0.3\textwidth}
\begin{itemize}\sffamily

\item[\textcolor{gray!90}{\textbullet}] \textcolor{gray!90}{Which axion mass do we extract?}

\item[\textcolor{gray!90}{\textbullet}] \textcolor{gray!90}{What are properties of Planck scale axions?}

\item[\textcolor{gray!90}{\textbullet}] \textcolor{gray!90}{Which model fits most galaxies?}

\item[\textcolor{gray!90}{\textbullet}] \textcolor{gray!90}{Do the models agree with each other?}
\end{itemize}
\end{minipage}\vspace{1.5cm}

\minitoc

\newpage

\section{Introduction}

Despite the outstanding success of the standard model of Cosmology which is based on cold dark matter (CDM) in explaining large-scale structures, cold dark matter has difficulties predicting small-scale structures \cite{Weinberg_2015}. Such small-scale problems are the prediction of more satellite dwarf galaxies around a host galaxy then currently observed, this is also known as the missing satellite problem\footnote{Similarly to the missing baryon problem and the missing intermediate massive black holes, frequently a detection is proposed and afterwards refuted. The author is currently not aware of a resolution in either one of those three problems.} \cite{M_ller_2020}. The so called cusp-core problem is another small-scale problem of cold dark matter \cite{Shi_2021}: The dark matter density is predicted to increase steeply towards the center of a galaxy with $\rho \sim r^\alpha$ and $\alpha \in \{-1,-1.5\}$ [ibid.]. This behaviour is called a \emph{cuspy} slope, however, a flatter profile is often observed \cite{Weinberg_2015}. Such flat behaviour could be explained by a core or possible also by baryonic physics [ibid.]. Nonetheless, the latter has not yet been observed in dwarf galaxies where the problem arises to begin with and seems to contradict the surprisingly high star formation efficiency in the center of galaxies (which is often referred to as the young star problem). The difference, between a cuspy and a core profile can be observed in Fig.\,\ref{fig:NFWvsSNFW}. The core-like behaviour is shown in the red line, a characteristic for the Soliton Navarro–Frenk–White profile (SNFW), while a cuspy slope can be observed in orange for the unmodified Navarro–Frenk–White profile (NFW).\nn

One framework which aims to resolve the aforementioned problems is so called fuzzy dark matter. The idea is that fuzzy dark matter consists out of non-thermal, non-relativistic ultralight bosons which form a Bose-Einstein condensate and behave like cold dark matter on large scales, i.e. larger than the de Broglie wavelength of usually 1 kpc \cite{Bar-Or:2018pxz}. However, on small scales their wavelike dynamics behaves different compared to cold dark matter, and one might argue that they fit better to the observed rotation curves. Typical values of axion masses in this framework lie between $m_a \sim 10^{-21}$ eV and $m_a \sim 10^{-22}$ eV \cite{Hui_2017}. In this chapter we will reproduce Matos et al. result of  $m_a \sim 5 \times 10^{-24}$ eV \cite{Bernal2017}. However, several studies doubt the feasibility of a dark matter model with a lower dark matter mass than $m_a \le 10^{-21}$ eV \cite{Bar-Or:2018pxz,Walther_2017}. Especially Ly$\alpha$ forest power spectra suggest that such a small mass would be indistinguishable from cold dark matter \cite{Walther_2017,Armengaud:2017nkf}. At first glance this would be unappealing since this would imply that fuzzy dark matter loses the desired feature of a core like behaviour. However, we argue that in the here presented three component dark matter model this constraint is not valid. The galactic core might form from two different axion types which correspond to the screened leptonic Yang-Mills scales in association to the muon and tau mass. Their masses of $m_{a,\mu} \sim 2.9 \times 10^{-19}$ eV and $m_{a,\tau} \sim 8.2 \times 10^{-17}$ eV are well above the constraint. Moreover, the scenario where a $\tau$-lump condensate gravitationally collapses, consumes the $\mu$-lump and becomes the central massive object of a given galaxy is still feasible. In this case the core of a galaxy is given by a complex succession of dark matter physics. This might also mitigate the cold dark matter behaviour on small scales by the e-lump with a mass of only $m_{a,e} \sim 6.7 \times 10^{-24}$ eV, since an explicit core already provides the desired core-like rotation curve behaviour.\nn

One way of extracting a mass estimate for dark matter due to fitting of rotation curves was discussed in the previous chapter. In this chapter we determine a Planck scale axion mass in the context of SU(2)$_{\rm CMB}$ cosmology, which was introduced in chapter \ref{chapter-Su(2)}. Therefore, we use low surface brightness galaxies of the SPARC catalogue \cite{Lelli:2016zqa}. Low surface brightness galaxies are especially interesting for us, because they can assumed to be dominated by dark matter. Rotation curves are fitted with four different models starting from section\,\ref{chapter:SNFW}:\nn

\begin{enumerate}[i)]
  \item The Soliton Navarro–Frenk–White profile (SNFW)
  \item The Navarro–Frenk–White profile (NFW) 
  \item The pseudo-isothermal model (PI) 
  \item and the Burkert model 
\end{enumerate}
\vspace{0.5cm}

Some results of this and the following chapter will be published soon. In particular, a preprint of the SNFW profile and a comparison to the Burkert model can be found under \cite{meinert2021axial}.


\section{Properties of a Planck-scale axion}\label{Sec2}

In this section, we would like to convey some basic facts about selfgravitating Planck-scale axion condensates: lumps. We start with the observation in \cite{Sin1994} that ultralight bosons necessarily need to occur in the form of selfgravitating condensates in the cores of galaxies. Halos of axion particles were formed around these cores, because they were separated in the course of nonthermal depercolation. Such a halo reaches out to a radius of $\sim r_{200}$ where its mass density starts to fall below 200 times the critical cosmological energy density of the spatially flat FLRW Universe. A key concept in describing such a fuzzy-dark-matter (FDM) system, e.g. a lump, is the gravitational Bohr radius $r_B$ defined as 
\eqb
\label{Bohr}
r_B\equiv \frac{M_{\rm P}^2}{M m_a^{2}}\,, 
\eqe
where $M$ is the mass of the lump which should coincide with the virial mass, say $M_{200}$. We use two FDM models of the galactic mass density $\rho(r)$ to describe low-surface-brightness galaxies and to extract the axion mass $m_a$: The Soliton-NFW model, see \cite{matos2016scalar} and references therein, and the Burkert model \cite{Burkert_1995,Salucci:2000ps}. Let 

\eqb
\label{Compton}
\lambdabar_C\equiv \frac{1}{m_a}
\eqe
denote the reduced Compton wavelength and  
\eqb
\label{intpartD}
d_a\equiv \left(\frac{m_a}{\bar{\rho}_{\rm }}\right)^{1/3} 
\eqe
the mean distance between axion particles within the spherically symmetric lump of mean dark-matter mass  
density $\bar{\rho}_{\rm }$. One has 
\eqb
\label{densmean}
\bar{\rho}_{\rm }\sim\frac{M}{\frac{4\pi}{3}r_B^3}\,. 
\eqe
A finite-extent lump self-consistently forms if the reduced Compton wavelength $\lambdabar_C$ - the correlation length in the condensate of free axion particles at zero temperature - is not very far from the gravitational Bohr radius $r_B$. More precisely, we can state that 
\eqb
\label{CvB}
\frac{r_B}{\lambdabar_C}=\kappa(\delta)\,,
\eqe
where $\kappa$ is a smooth dimensionless function of its dimensionless argument $\delta\equiv m_a/M_P$ with the property that $\lim_{\delta\to 0}\kappa(\delta)<\infty$. This is because the lump mass $M$, 
which enters $r_B$ via Eq.\,(\ref{Bohr}), is itself only a function of the two mass scales $m_a$ and $M_P$ for an isolated selfgravitating condensate of otherwise free particles of mass $m_a$. If $\delta$ turns out to be sufficiently smaller 
than unity, which will always be the case in what follows, we can treat the right-hand side of Eq.\,(\ref{CvB}) as a {\sl universal} constant. In practice, we will in Secs.\,\ref{Sec3} and \ref{Sec5} derive from the match of dark-matter halos of low surface-brightness galaxies with heuristic mass density models that $\kappa\sim 314$.\nn

Eq.\,(\ref{CvB}) together with Eqs.\,(\ref{Compton}), (\ref{Bohr}), and (\ref{maxxaxion}) imply for the mass $M$ of the lump 
\eqb
\label{MLambda}
M=\frac{1}{\kappa}\frac{M_P^3}{\Lambda^2}\,.
\eqe 
Eq.\,(\ref{MLambda}) is important because it predicts that the ratios of lump masses are solely determined by the squares of the ratios of the respective Yang-Mills scales or, equivalently \cite{Hofmann:2017lmu}, ratios of charged lepton masses $m_e$, $m_\mu$, and $m_\tau$. One has 
\eab
\label{lumpmassratios}
\frac{M_\tau}{M_\mu}&=&\left(\frac{m_\tau}{m_\mu}\right)^2\sim 283\,,\nonumber\\
\frac{M_\mu}{M_e}&=&\left(\frac{m_e}{m_\mu}\right)^2\sim 2.3\times 10^{-5}\,,\\ 
\frac{M_\tau}{M_e}&=&\left(\frac{m_e}{m_\tau}\right)^2\sim 8.3\times 10^{-8}\,.\nonumber
\eae
Moreover, Eqs.\,(\ref{maxxaxion}), (\ref{intpartD}), (\ref{Compton}), (\ref{densmean}), and (\ref{MLambda}) 
fix the ratio $\xi\equiv\frac{d_a}{\lambdabar_C}$ as 
\eqb
\label{xidef}
\xi=\left(\frac{4\pi}{3}\right)^{1/3}\left(\kappa\frac{\Lambda}{M_P}\right)^{4/3}\,.
\eqe
Since $\Lambda\ll M_P$ we have $\xi\ll 1$, and therefore a large number of axion 
particles are covered by one reduced Compton wave length. This assures that Planck-scale 
axions always occur in condensed form. A thermodynamical argument for the necessity of axion condensates throughout the Universe's expansion history is given in Sec.\,\ref{Sec5}. In \cite{Sin1994} the non-local and non-linear 
(integro-differential) Schr\"odinger-equation, obtained from a linear Schr\"odinger equation and a Poisson equation for the gravitational potential \cite{Bar_2018} and governing the lump, was analysed. An excitation of such a lump in terms of its wave-function $\psi$ containing radial zeros 
was envisaged in \cite{Sin1994}. Here instead, we assume the lump to be in its ground 
state parameterised by a phenomenological mass density 
$\rho(r)\propto |\psi|^2(r)>0$. Finally, Eq.\,(\ref{Bohr}) together with Eqs.\,(\ref{maxxaxion}) and (\ref{MLambda}) yield for the 
gravitational Bohr radius
\eqb
\label{Bohrradiusfund}
r_B=\kappa\frac{M_P}{\Lambda^2}\,.
\eqe


In this section, we extract the axion mass $m_{a,e}$ from observed RCs of low-surface-brightness galaxies which fix the lump mass $M_e$ and a characterising length scale -- the gravitational Bohr radius $r_{B,e}$. This, in turn, determines the Yang-Mills scale $\Lambda_e$ associated with the lump. We analyse RCs from the 
SPARC library \cite{Lelli:2016zqa}. 

\newpage

\section{Analysis of Rotation Curves (RCs)}\label{Sec3}

To investigate, for a given galaxy and RC, the underlying spherically symmetric 
mass density $\rho(r)$ it is useful to introduce the orbit-enclosed mass
\eqb
M(r) = 4\pi\int_{0}^{r}dr'\, r'^2\rho_{{\rm }}(r')\,.
\label{eq:Mtot}
\eqe
Assuming virialisation, spherical symmetry, and Newtonian gravity the orbital 
velocity $V(r)$ of a test mass (a star) is given as   
\begin{align}
V(r) = \sqrt{\frac{G M(r)}{r}}\,,
\label{eq:Vcirc}
\end{align}
where $M(r)$ is defined in Eq.\,(\ref{eq:Mtot}), and $G\equiv M_P^{-2}$ denotes Newton's constant. The lump mass 
$M$ is defined to be $M_{200}\equiv M(r_{200})$. Here $r_{200}$ denotes the virial radius defined such that 
\begin{align}
\label{NFWrange}
\rho_{\rm }(r_{200})=200\,\frac{3\,M_P^2}{8\pi}H_0^2\,,
\end{align}
where $H_0$ is the Hubble constant, and $\lambda_{\rm deB}=\lambda_{\rm deB}(r)$ indicates the de-Broglie wavelength of an axionic particle for $r_e<r<r_{200}$ where the NFW model applies. Note that within the core region $r<r_e$ the correlation length in the condensate is given by the reduced Compton wave length $\lambdabar_C=1/m_a$.\nn

We analyse the RCs from the 
SPARC library \cite{Lelli:2016zqa}. For a given galaxy, the value of $r$ in Eq.\,(\ref{eq:Mtot}) is assumed to coincide with the largest radius $r_B$ at which the orbital velocity $V$ (circular orbits) was determined observationally. This is only a weak constraint on generality because 
of the logarithmic dependence on $r$ of the integral in Eq.\,(\ref{eq:Mtot}).\\ 
 
\newpage

\section{Soliton Navarro–Frenk–White profiles}\label{chapter:SNFW}

For an extraction of $m_{a,e}$ and therefore the associated Yang-Mills scale governing the mass of a lump according to Eq.\,(\ref{MLambda}), we use the Soliton-Navarro-Frenk-White (SNFW) model. The consistency of the with Eq.\,(\ref{Bohr}) gained knowledge about the Bohr radius is tested with virial mass extractions of the NFW, PI and the Burkert model. The mass-density profile of the NFW-part of the SNFW-model is  
given as \cite{Navarro:1996gj}
\vspace*{1.5mm} 
\begin{align}
\rho_{{\rm NFW}}(r) = \frac{\rho_s}{\frac{r}{r_s}(1+\frac{r}{r_s})^2}\,,
\label{eq:RhoNFW}
\end{align}
\vspace*{1.5mm} 
where $\rho_s$ associates with the central mass density, and $r_s$ is a scale radius which represents the onset of the asymptotic cubic decay in distance $r$ to the galactic center. Note that profile $\rho_{\rm NFW}$ exhibits an infinite cusp as $r\to 0$ and that the orbit-enclosed mass $M(r)$ diverges logarithmically with the cutoff radius $r$ for the integral in Eq.\,(\ref{eq:Mtot}). In order to avoid the cuspy behavior for $r\to 0$, an axionic Bose-Einstein condensate (soliton density profile) is assumed to describe the soliton region $r\le r_e$. From the ground-state solution of the Schr\"odinger-Poisson system for a 
single axion species one obtains a good analytic description of the soliton density profile as \cite{Bernal2017}
\vspace*{1.5mm} 
\begin{align}
\rho_{{\rm sol}}(r) = \frac{\rho_c}{(1+0.091(r/r_c)^2)^8}\,,
\label{eq:SolitonNFW}
\end{align}
\vspace*{1.5mm} 
where $\rho_c$ is the core density \cite{Schive:2014dra}. On the whole, the fuzzy dark matter profile can than be approximated as 
\vspace*{1.5mm} 
\begin{align}
\rho_{{\rm SNFW}}(r) = \Theta(r_\epsilon-r)\rho_{{\rm sol}}+ \Theta(r-r_\epsilon)\rho_{{\rm NFW}}\,.
\label{eq:SolitonNFWwhole}
\end{align}
\vspace*{1.5mm} 
Using Eqs.\,(\ref{eq:Mtot}), (\ref{eq:Vcirc}), and (\ref{eq:SolitonNFWwhole}), 
we obtain the orbital velocity $V_{\rm SNFW}$ of the 
SNFW model \cite[Eq. (17)]{Robles:2012uy} which is fitted to observed RCs. 

\vspace*{1.5mm} 
\begin{equation} \label{eq:SolitonNFWvelocity}
\begin{split}
V_{{\rm SNFW}}(r) & =  \Theta(r_\epsilon-r)\,V_{{\rm sol}}(r) \\
& + \Theta(r-r_\epsilon)  
\left(
V_{{\rm sol}}(r_\epsilon)-V_{{\rm NFW}}(r_\epsilon)+V_{{\rm NFW}}(r)
\right)\,.
\end{split}
\end{equation}
\vspace*{1.5mm} 

Where $V_{{\rm sol}}(r)$ is given in Eq.\,(\ref{eq:Vsoliton})  

\vspace*{1.5mm} 
\begin{equation} \label{eq:Vsoliton}
\begin{split}
V_{{\rm sol}}(r) & =   \frac{r_c^3 \rho_c}{(r^2 + 10.9 r_c^2)^7} (1.9 r^{13} r_c + 1.4 \times 10^2 r^11 r_c^3 + 4.4 \times 10^3 r^9 r_c^5\\
& +   7.5 \times 10^4 r^7 r_c^7 + 7.5 \times 10^5 r^5 r_c^9 + 
   4.3\times 10^6 r^3 r_c^{11} - 3.4\times 10^6 r r_c^{13} \\
&   + (5.8\times 10^{-1} r^{12} +  4.5 \times 10 r^{12} r_c^2 + 1.4\times 10^3 r^{10} r_c^4 +  2.7\times 10^4 r^8 r_c^6 \\
& + 2.9\times 10^5 r^6 r_c^8 +  1.9\times 10^6 r^4 r_c^{10} + 
      7.2\times 10^6 r^2 r_c^{12} +  1.1\times 10^7 r_c^{14})\\ & \arctan{[(3\times 10^{-1} r)/ r_c]})\,,
      \end{split}
\end{equation}
\vspace*{1.5mm}

and $V_{{\rm NFW}}(r)$ is given by Eq.\,(\ref{eq:VNFW}) as

\vspace*{1.5mm} 
\begin{equation} \label{eq:VNFW}
\begin{split}
V_{{\rm NFW}}(r) & =  \frac{(0.99 r_c^{16} r_\epsilon (r_\epsilon +  r_s)^2 (r_s/(r + r_s) + \log{[r + r_s]}) \rho_c)}{(r_c^2 +  0.091 r_\epsilon^2)^8}\,.
      \end{split}
\end{equation}
\vspace*{1.5mm}

\begin{figure}[H]
    \centering
    \begin{minipage}{.5\textwidth}
        \centering
        \includegraphics[width=7cm]{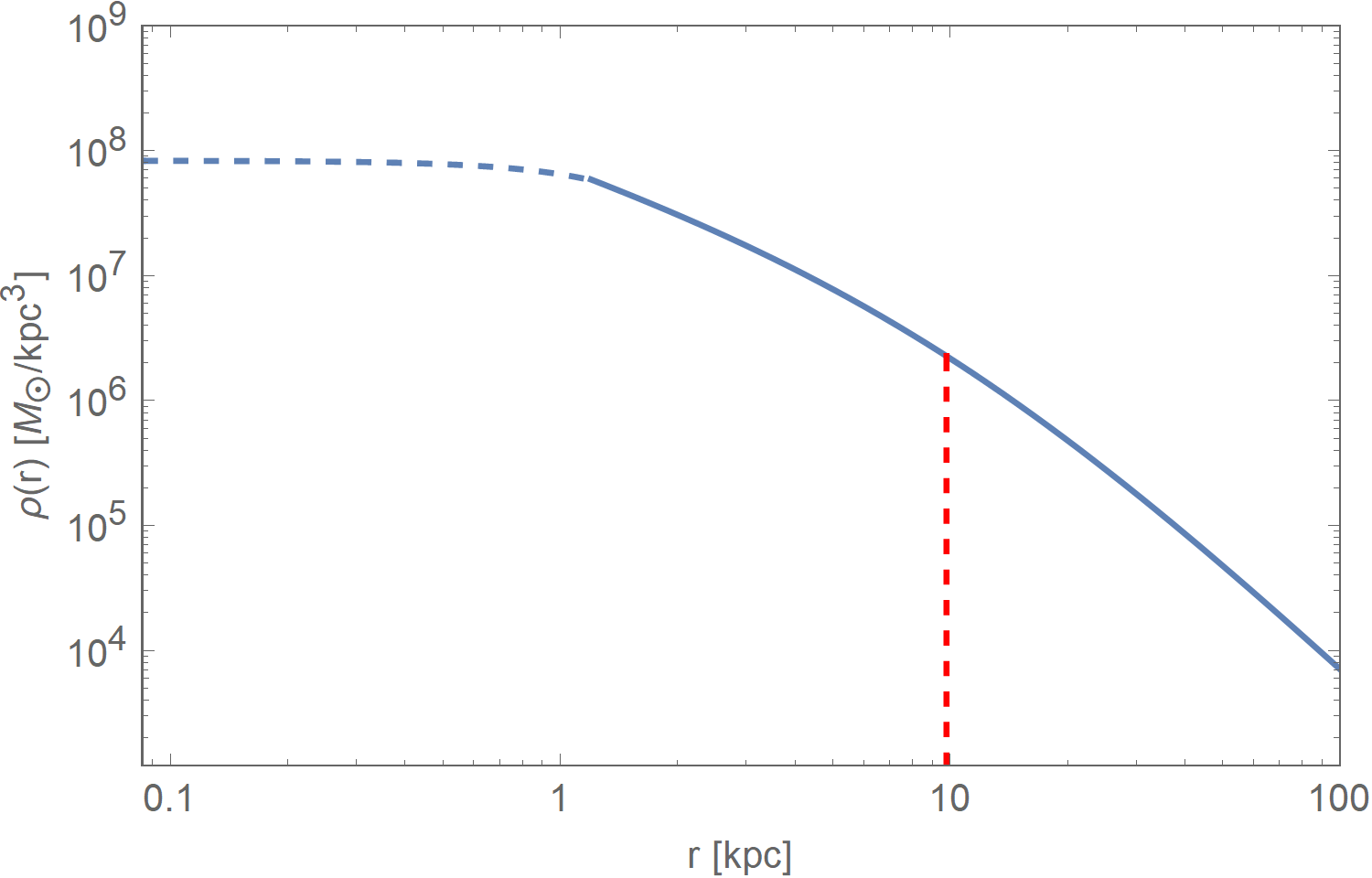}
    \end{minipage}%
    \begin{minipage}{.5\textwidth}
        \includegraphics[width=7cm]{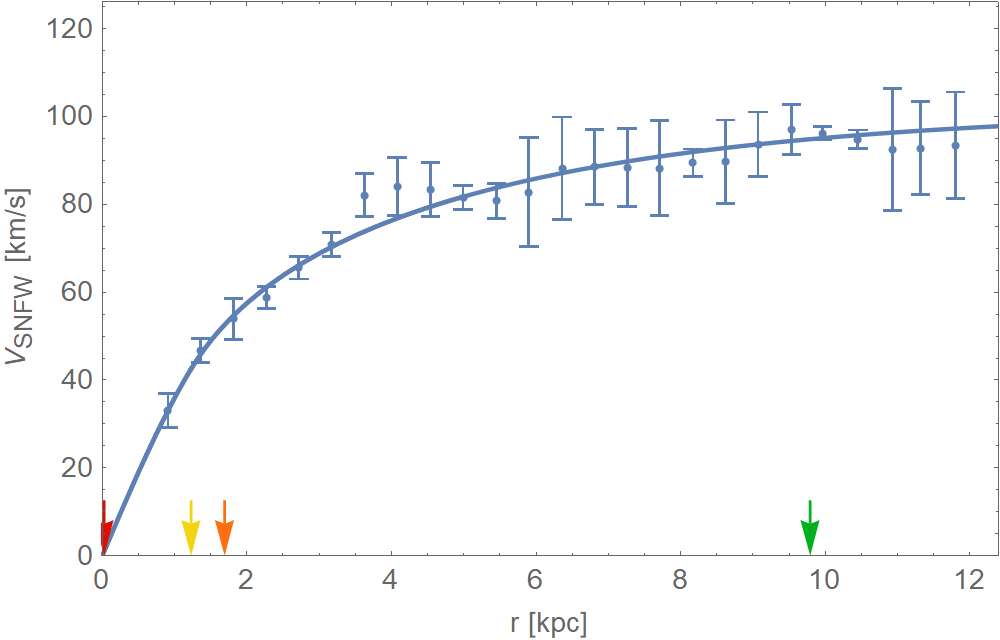}
    \end{minipage}
    \caption{
    An example for the characteristic density profile $\rho_{{\rm SNFW}}(r)$, Eq.\,(\ref{eq:SolitonNFWwhole}), of galaxy NGC0300 is given on the left-hand side. On the right-hand side the corresponding orbit velocity $V_{{\rm SNFW}}(r)$,
    Eq.\,(\ref{eq:SolitonNFWvelocity}), of the same galaxy is fitted to the data as obtained by the SPARC library \cite{Lelli:2016zqa}. The scale radius $r_s$ is left indicated by a vertical red, dotted line and right with a green arrow. 
    The transition radius from soliton to the NFW radius at $r_\epsilon$ is left indicted by the blue dotted region (left) and on the right hand side with a yellow arrow. The core radius $r_c$ is indicated by the orange arrow).}
     \label{fig:SNFWexample}
\end{figure}

Fig.\,\ref{fig:SNFWexample} shows an example for the characteristic density profile $\rho_{{\rm SNFW}}(r)$ of Eq.\, (\ref{eq:SolitonNFWwhole}), and the orbit velocity $V_{{\rm SNFW}}(r)$ of Eq.\,(\ref{eq:SolitonNFWvelocity}) for the galaxy NGC0300. As intended, the up to $r_\epsilon$ the density profile is relatively constant. The typical $r^{-3}$ behaviour is given by the NFW density profile.\nn

\begin{figure}[H]
    \centering
    \begin{minipage}{.5\textwidth}
        \centering
        \includegraphics[width=7cm]{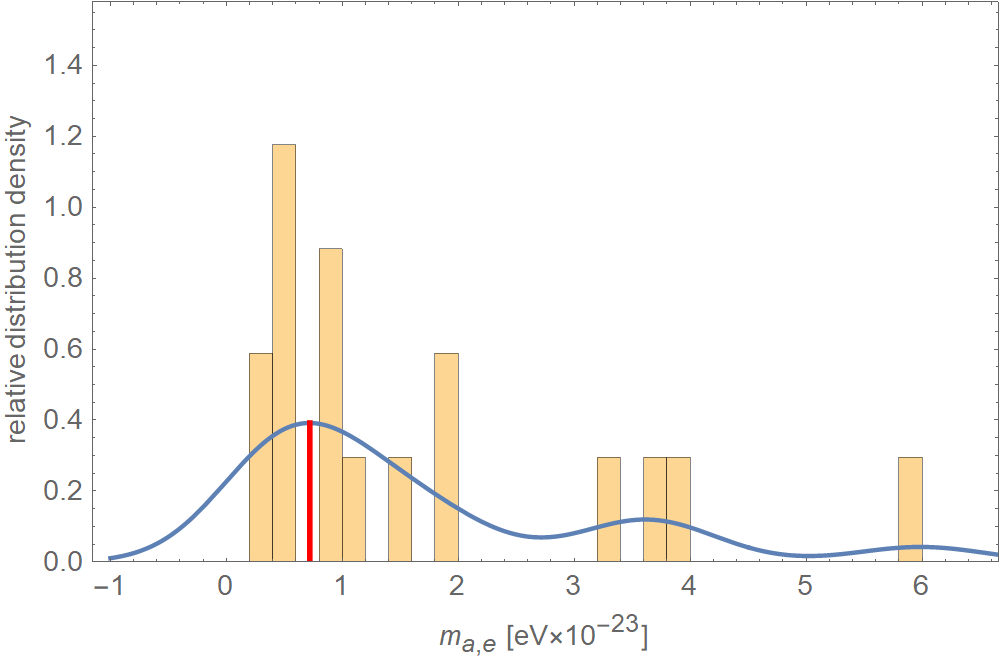}
    \end{minipage}%
    \begin{minipage}{.5\textwidth}
        \includegraphics[width=7.5cm]{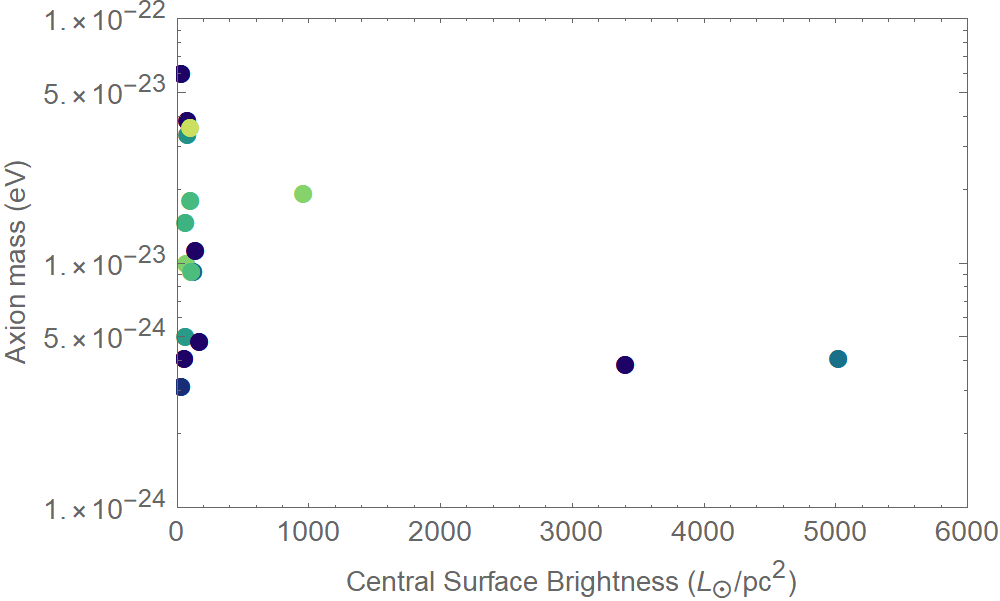}
    \end{minipage}
    \caption{
    left: Frequency distribution of axion mass $m_{a,e}$ as extracted from the SNFW model for 17 best fitting galaxies. The maximum of the smooth-kernel-distribution (solid, blue line) is at $m_{a,e}=(0.72\pm 0.5) \times10^{-23}$ eV  (red, vertical line).
   right: The same axion masses in a mass-luminosity plane.}
   \label{fig:AxionMassesSNFW}
\end{figure}

This determines the parameters $r_\epsilon$, $r_s$, and $\rho_c$. The density $\rho_s$ relates to these fit parameters by demanding continuity of the SNFW mass density at $r_\epsilon$ \cite{Bernal2017}. As a result, one has  
\vspace*{1.5mm} 
\eqb
\label{rhocrhos}
\rho_s(\rho_c,r_c,r_\epsilon,r_s)=\rho_c \frac{(r_\epsilon/r_s)(1+r_\epsilon/r_s)^2}{(1+0.091(r_\epsilon/r_c)^2)^8}\,.
\eqe
\vspace*{1.5mm} 
Examples of good fits with $\chi^2/\text{d.o.f.}<1$, see Tab.\,2, are shown in Tab.\,1. The derived quantity $m_{a,e}$ is extracted from the following equation  \cite{Schive:2014dra}
\vspace*{1.5mm} 
\eqb
\label{mafromrhoc}
\rho_c \equiv 1.9\times 10^9 (m_{a,e}/10^{-23} {\rm eV})^{-2} (r_c/{\rm kpc})^{-4} M_{\odot}{\rm kpc}^{-3}
\eqe
\vspace*{1.5mm} 
The other derived quantities $r_{200}$ and $M_{200}$ are obtained by employing Eqs.\,(\ref{NFWrange}) and (\ref{eq:Mtot}) with $M(r=r_{200})\equiv M_{200}$, respectively. In Fig.\,\ref{fig:AxionMassesSNFW} a frequency distribution of $m_{a,e}$ is shown, based on a sample of 17 best fitting galaxies, see Collage 1 for the fits to the RCs. The maximum of the smooth-kernel-distribution (solid line) is at 
\vspace*{1.5mm} 
\eqb
\label{maSNFW}
m_{a,e}=(0.72\pm 0.5)\times 10^{-23}\,{\rm eV}\,.
\eqe
\vspace*{1.5mm} 

\begin{figure*}
\begin{flushleft}
\textbf{Collage 1.} Best fits of SNFW to RCs of 17 SPARC galaxies. The arrows indicate the Bohr radius of the e-lump, $r_{B,e}$ (red), the core radius of the soliton $r_{c}$ (orange), the transition radius from the soliton model to the NFW model $r_{e}$ (yellow), and the scale radius of the NFW model $r_{s}$ (green). The dagger$^\dagger$ indicates galaxy NGC6195.
\end{flushleft}
\vspace{0.3cm}
\begin{minipage}{1\textwidth}
\includegraphics[width=15cm]{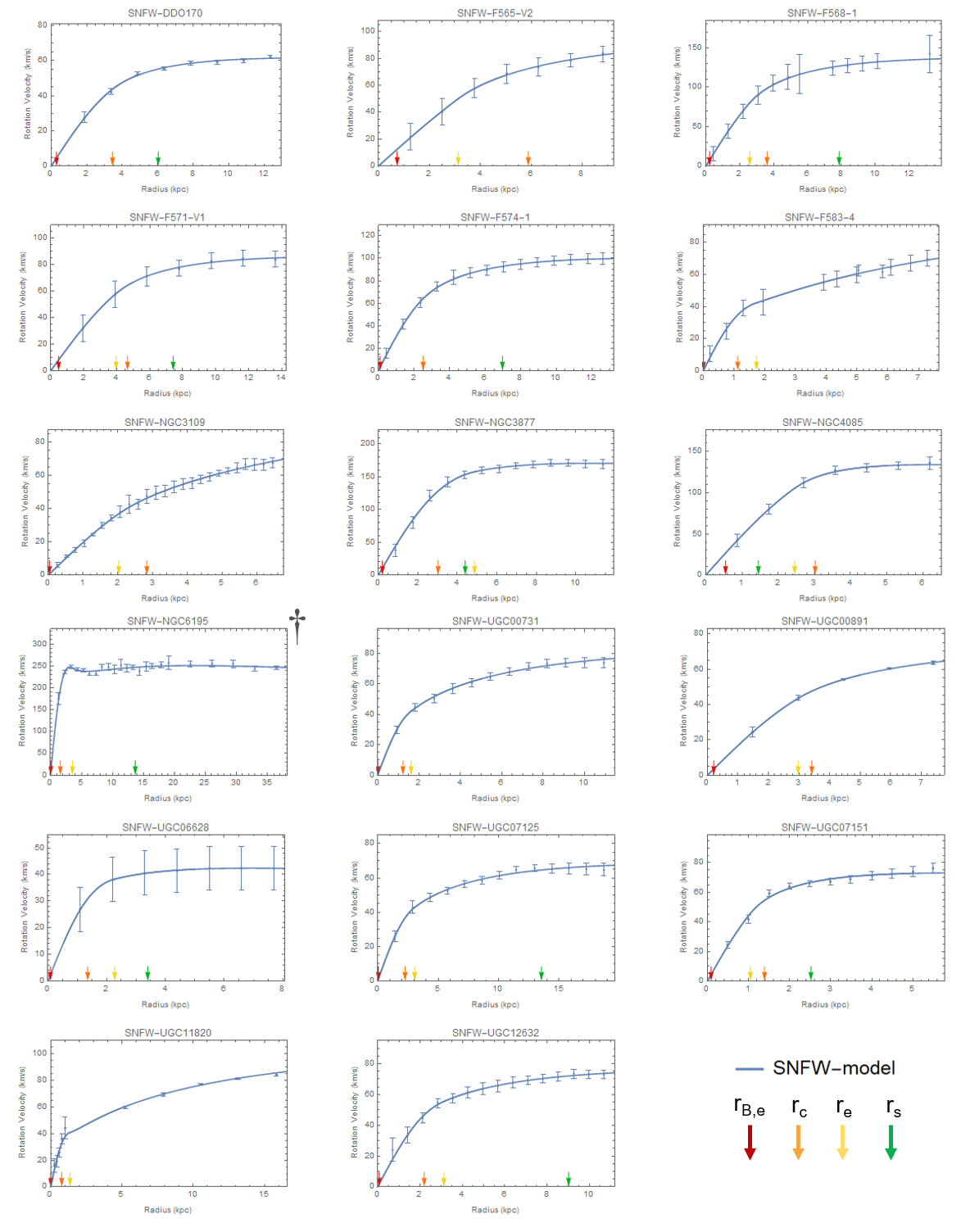}
\end{minipage}
\end{figure*}


\begin{figure}[H]
    \centering
    \begin{minipage}{.5\textwidth}
        \centering
          \label{fig:HistoSNFW}
        \includegraphics[width=7cm]{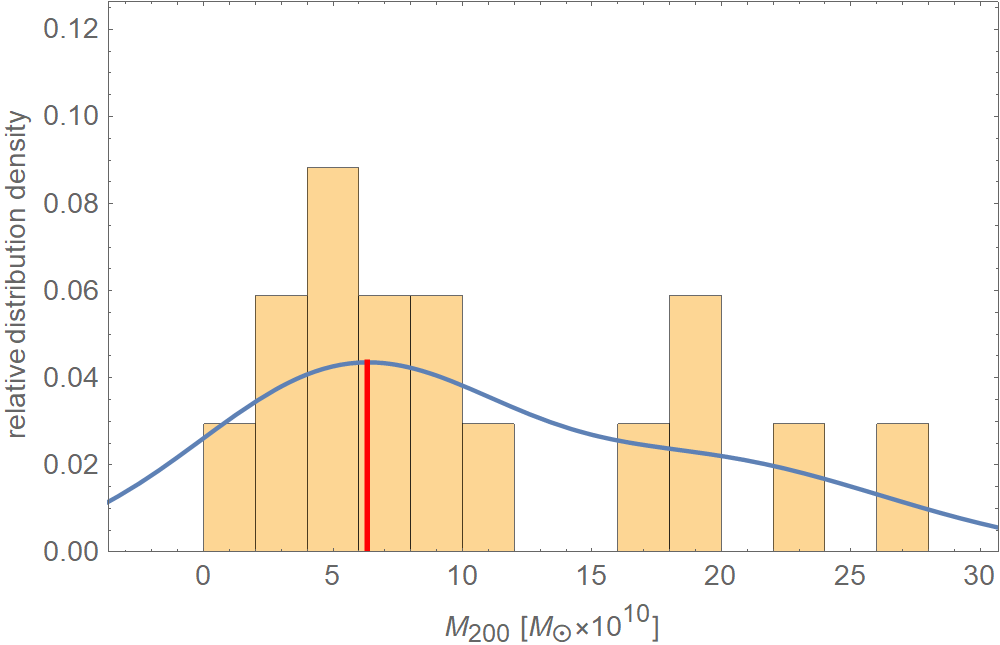}
    \end{minipage}%
    \begin{minipage}{.5\textwidth}
        \includegraphics[width=7.5cm]{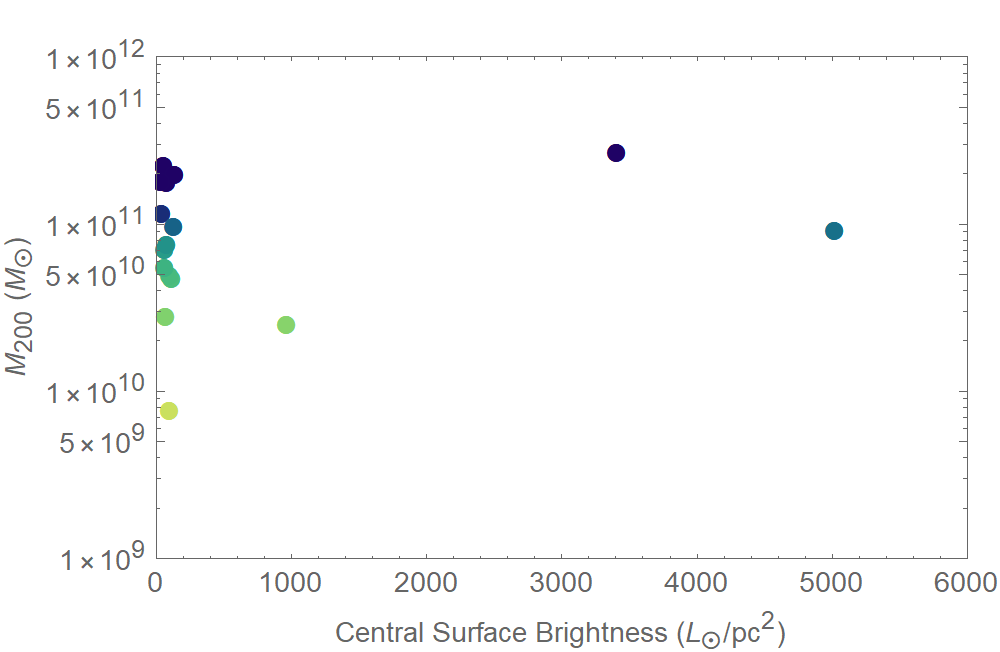}
    \end{minipage}
    \caption{left:Frequency distribution of the virial mass $M_{200}$ in units of solar masses $M_{\odot}$ from the 80 best-fitting galaxies in the Burkert  model. The maximum of the smooth-kernel-distribution (solid line) is at $M_{200}=(6.5\pm3)\times 10^{10}\,M_{\odot}$.
   right: Extracted virial masses $M_{200}$ in units of solar masses $M_{\odot}$ from sample of the 17 best-fitting galaxies in the NFW model.}
\end{figure}

The frequency distribution of axion mass $m_{a,e}$ as extracted from the SNFW model for 17 best fitting galaxies is shown in the left side of Fig.\,\ref{fig:AxionMassesSNFW}. On the right side those axion masses are shown in the surface-brightness plane. In Fig.\,\ref{fig:HistoSNFW} a frequency distribution of $M_{200}$ is shown for the 17 best-fitting galaxies on the left side, the right side depicts the distribution of these galaxies in the $M_{200}$ -- surface-brightness plane. The maximum of the smooth-kernel-distribution is at $M_{200}=(6.3\pm 3)\times 10^{10}\,M_{\odot}$. With Eq.\,(\ref{Bohr}) this implies a mean Bohr radius of  
\begin{align}
\label{maBurkert}
\nonumber r_{B,e}&=\frac{M_p^2}{(6.3\pm4) \times 10^{10}\,M_{\odot}\, ((0.72\pm 0.5) \times 10^{-23}\, {\rm eV})^2 }\\
&=(0.26\pm 0.1)\,{\rm kpc}\,.
\end{align} \nn
This value of $r_{B,e}$ is used in the following models in order to determine together with the model dependant virial masses an axion mass.\nn


On the right-hand side of Fig.\,\ref{fig:HistoSNFW}, the individual virial masses $M_{200}$ are depicted in the mass-luminosity plane. 
Table 1 shows the luminosity, axion mass $m_a$, $r_{200}$, virial mass $M_{200}$, central density $\rho_0^{\rm NFW}$, scale radius $r_s$, transition radius $r_e$, core radius $r_c$, and $r_e/r_c$ of those 17 best fitting galaxies in the SNFW model.\nn

The fitting constraints arise due to the many degrees of freedom. It is easy for two radii, e.g. $r_e$ and $r_c$ to take on the same value and "stick together", effectively decreasing the degrees of freedom of the underlying function again. Oftentimes, parameter as $r_{200}$ also took on unrealistic high values without any constraint. To avoid this undesired fitting behaviour, we have chosen those constraints. They are purely heuristic. Other fitting constraints we tried with less success were:

\begin{align*}
r_{200} < 200\,{\rm kpc}, r_e, r_c< 10\,{\rm kpc},r_e/r_c>0.1\\
200\,{\rm kpc} > r_{200} > r_e > r_c > 1\,{\rm kpc},r_e/r_c>0.1\\
\end{align*}

\begin{figure*}[h]
\begin{flushleft}
\textbf{Table 1.} Fits of RCs to NFW model: Galaxy name, Hubble Type, luminosity, $\chi^2/\text{d.o.f.}$, axion mass $m_a$, $r_{200}$, virial mass $M_{200}$, central density $\rho_0$, scale radius $r_s$, transition radius $r_e$, core radius $r_c$, and $r_e/r_c$. The fitting constraints are heuristic and motivated by the results of \cite{Bernal2017}, $r_{200} < 200$\,kpc, $r_e, r_c<$ 6\,kpc, and $r_e/r_c>0.1$.
\end{flushleft}
\begin{minipage}{1\textwidth}
\hspace{-2cm}\includegraphics[width=17cm,angle =0]{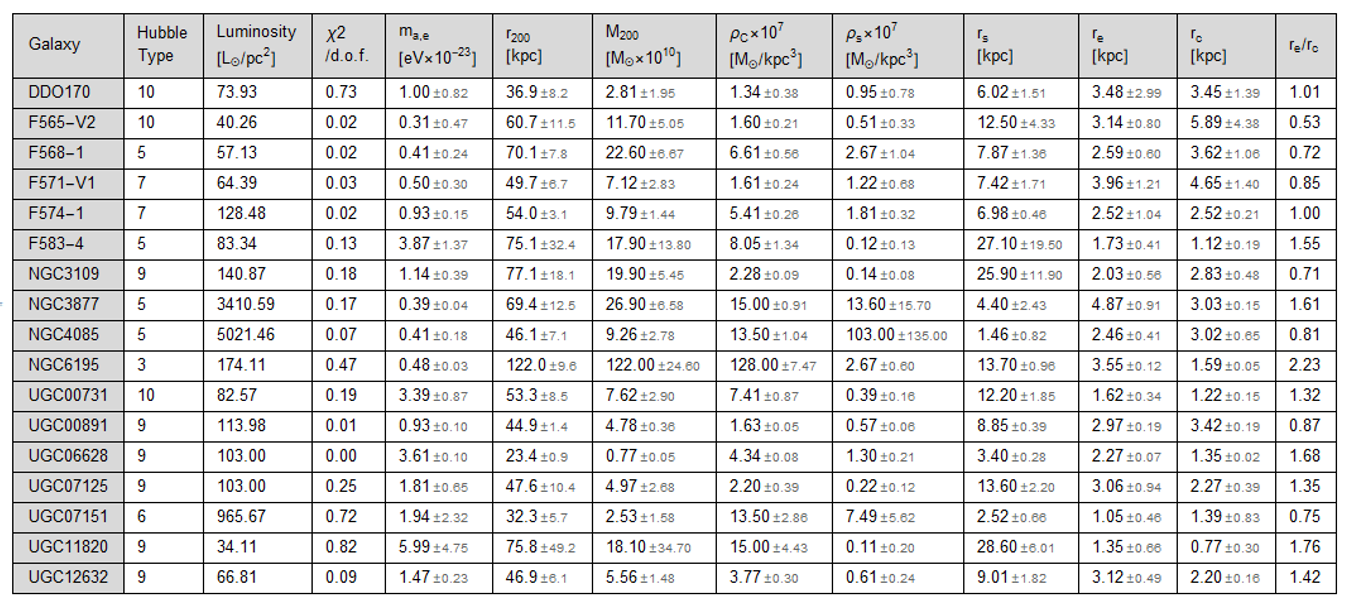}
\end{minipage}
\label{fig:GalaxyMassesSNFW}
\end{figure*}

\newpage

\section{Navarro–Frenk–White profiles}\label{chapter:NFW}

The Navarro–Frenk–White (NFW) profile is a spherical density profile of dark matter which was developed by Julio F. Navarro, Carlos S. Frenk and Simon D.M. White around 1995 \cite{Navarro:1995iw}. Their density profile leans on their findings of N-body simulation of cold dark matter. One of their findings is that dark matter halos are not well described by isothermal spheres. Therefore, they adopted the logarithmic slopes as proposed by Hernquist in 1990 for elliptical galaxies \cite{Hernquist:1990be},

\begin{align}
\rho_{{\rm Hernquist}}(r)  \propto  \frac{1}{r(1+\frac{r}{r_s})^3}\,,
\end{align}
\vspace*{1.5mm}

\begin{figure}[H]
    \centering
    \begin{minipage}{.5\textwidth}
        \centering
        \includegraphics[width=7cm]{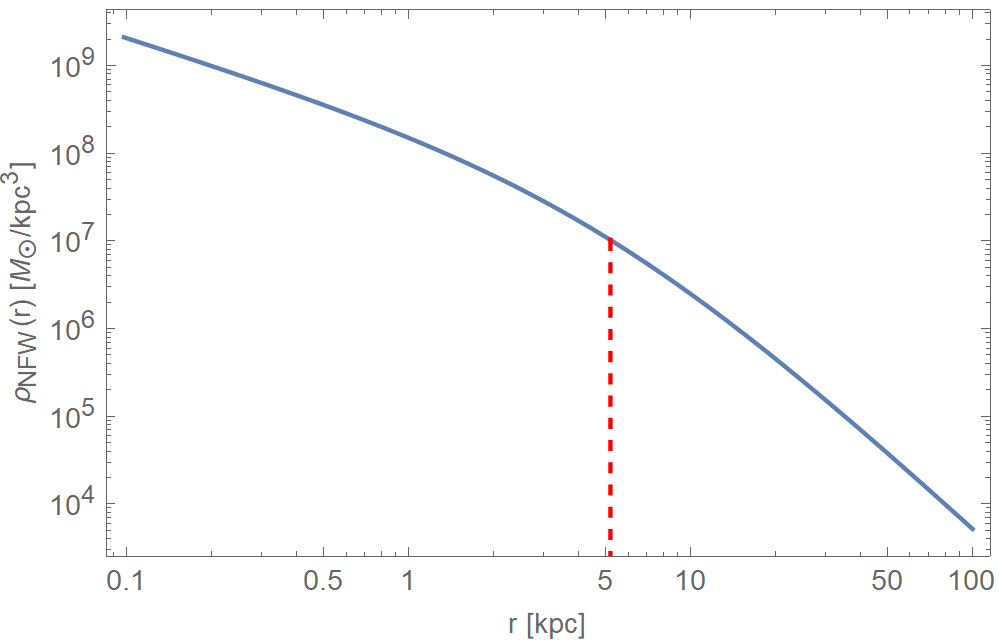}
    \end{minipage}%
    \begin{minipage}{.5\textwidth}
        \includegraphics[width=7cm]{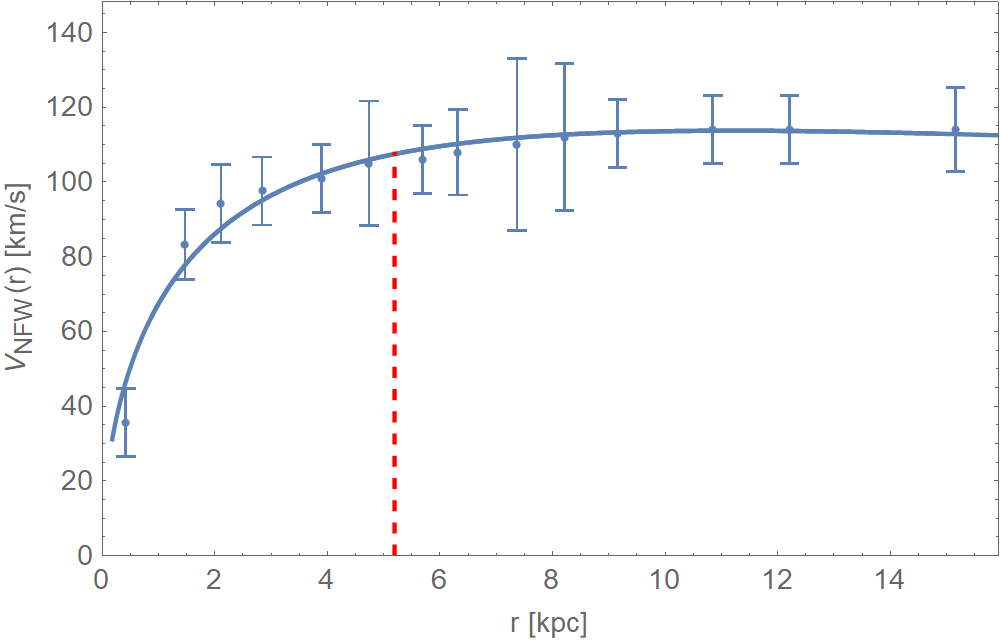}
    \end{minipage}
    \caption{
    An example for the characteristic density profile $\rho_{{\rm NFW}}(r)$, Eq.\,(\ref{eq:RhoNFW}), of galaxy F579-V1 is given on the left-hand side. On the right-hand side the corresponding orbit velocity $V_{{\rm NFW}}(r)$,
    Eq.\,(\ref{eq:NFW}), of the same galaxy is fitted to the data as obtained by the SPARC library \cite{Lelli:2016zqa}. The scale radius $r_s$ (red dotted line) indicates the transition from a $r^-1$ to a $r^-3$ behaviour of the density profile.}
     \label{fig:HistoNFWexample}
\end{figure}

Their mass-density profile of is only slightly modified. It is  
given as \cite{Navarro:1996gj}
\begin{align}
\rho_{{\rm NFW}}(r) = \frac{\rho_s}{\frac{r}{r_s}(1+\frac{r}{r_s})^2}\,,
\end{align}
\vspace*{1.5mm} 
where $\rho_s$ associates with the central mass density, and $r_s$ is a scale radius which represents the onset of the asymptotic cubic decay in distance $r$ to the galactic center. 
Note that profile $\rho_{\rm NFW}$ exhibits an infinite cusp as $r\to 0$ and that the orbit-enclosed mass $M(r)$ diverges logarithmically with the cutoff radius $r$ for the integral in Eq.\,(\ref{eq:Mtot}). Using Eqs.\,(\ref{eq:Mtot}), (\ref{eq:Vcirc}), and (\ref{eq:RhoNFW}), 
we obtain the orbital velocity of the 
NFW model \cite[Eq. (17)]{Robles:2012uy} as
\eqb
V_{\rm NFW}(r)=\sqrt{\frac{4 \pi  G}{r} \, \rho^{\rm NFW}_0 \, r_s^3 \left(\log \left(\frac{r}{r_s}+1\right)-\frac{r}{r_s \left(\frac{r}{r_s}+1\right)}\right)}\,.
\label{eq:NFW}
\eqe
The RC model of Eq.\,(\ref{eq:NFW}) is fitted to observed RCs to extract the values of the parameters $r_s$ and $\rho^{\rm NFW}_0$. Fig.\,\ref{fig:HistoNFWexample} shows an example for the characteristic density profile $\rho_{{\rm NFW}}(r)$ of Eq.\,(\ref{eq:RhoNFW}), and the orbit velocity $V_{{\rm NFW}}(r)$ of
Eq.\,(\ref{eq:NFW}) for the galaxy F579-V1. The scale radius $r_s$ (red dotted line) indicates the transition from a $r^-1$ to a $r^-3$ behaviour of the density profile.\nn

According to Eq.\,(\ref{eq:Mtot}), the total enclosed lump mass $M$ of this model is given as

\eqb
\label{MassGalNFW}
M_{{\rm NFW}}(r) =4 \pi  \rho_s r_s^3 \left(\frac{r_s}{r_s+r}+\log \left(\frac{r_s+r}{r_s}\right)-1\right)\,.
\eqe

\begin{figure}[h]\centering
\includegraphics[width=10cm]{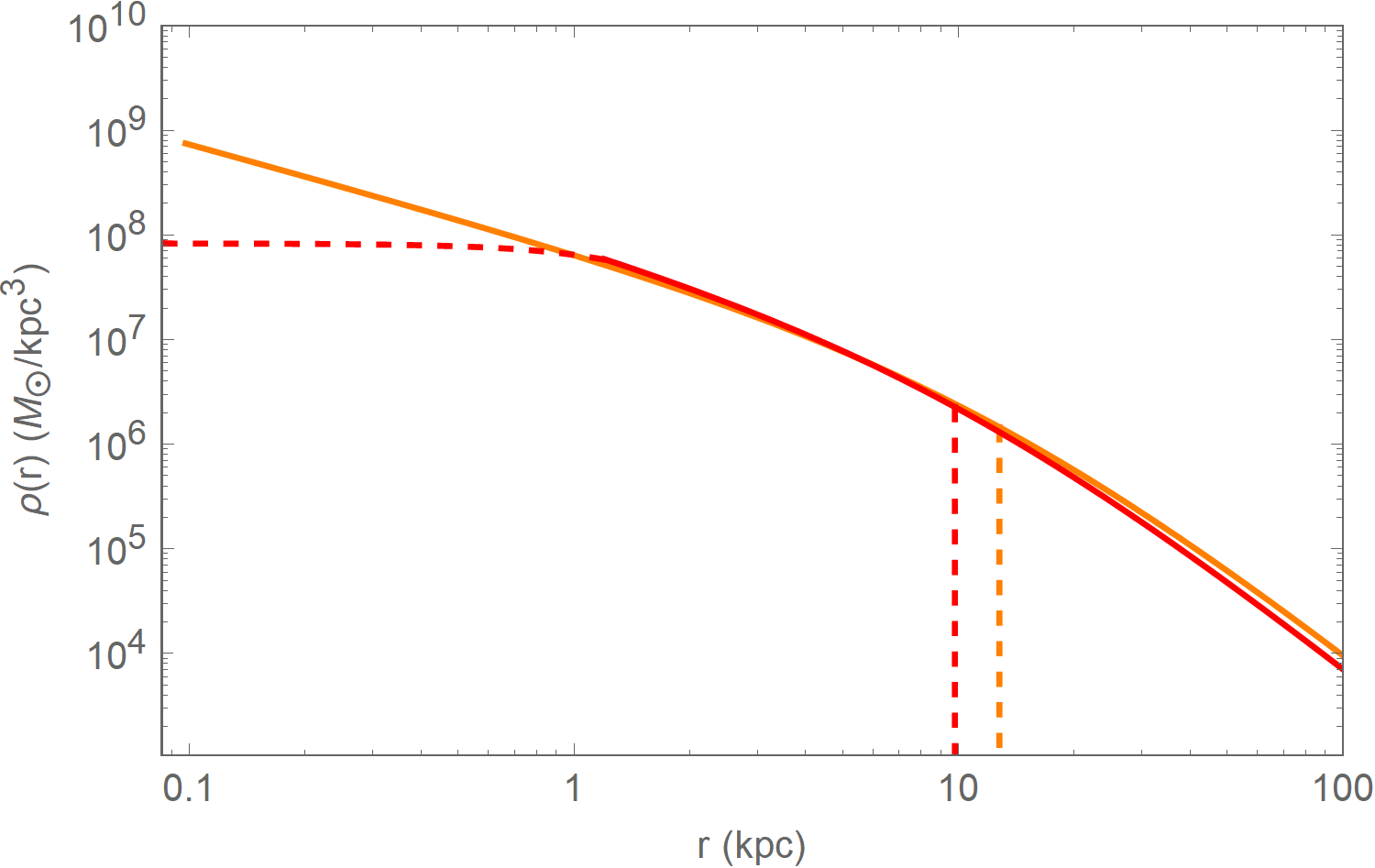}
    \caption{
    An example for the characteristic density profile $\rho_{{\rm SNFW}}(r)$, Eq.\,(\ref{eq:SolitonNFWwhole}), of galaxy NGC0300 is given in red. The red dotted, horizontal line indicates the description by the soliton density profile \ref{eq:SolitonNFW}, the red line represents the NFW part \ref{eq:RhoNFW}. For the same galaxy the NFW density profile was plotted in orange. The scale radii $r_s$ are indecated by vertical dotted lines in the corresponding colour. Notice the intended discrepancy in the galactic center.}
     \label{fig:NFWvsSNFW}
\end{figure}

A comparison between the NFW and the SNFW density profiles can be seen in Fig.\,\ref{fig:NFWvsSNFW} for galaxy NGC0300.\nn

\begin{figure}[H]
    \centering
    \begin{minipage}{.5\textwidth}
        \centering
        \includegraphics[width=7cm]{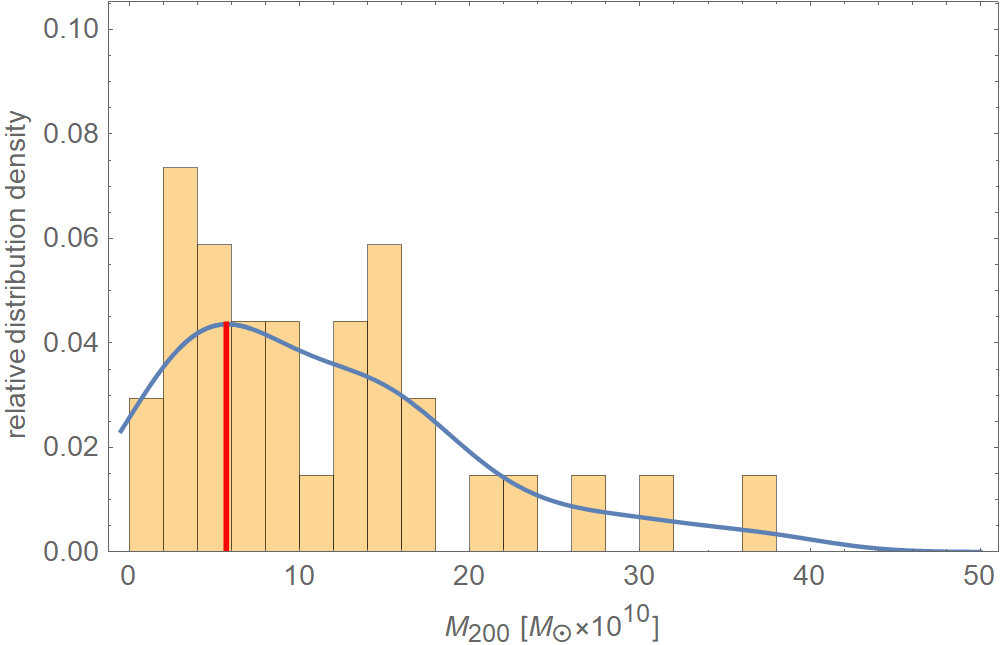}
    \end{minipage}%
    \begin{minipage}{.5\textwidth}
        \includegraphics[width=7.5cm]{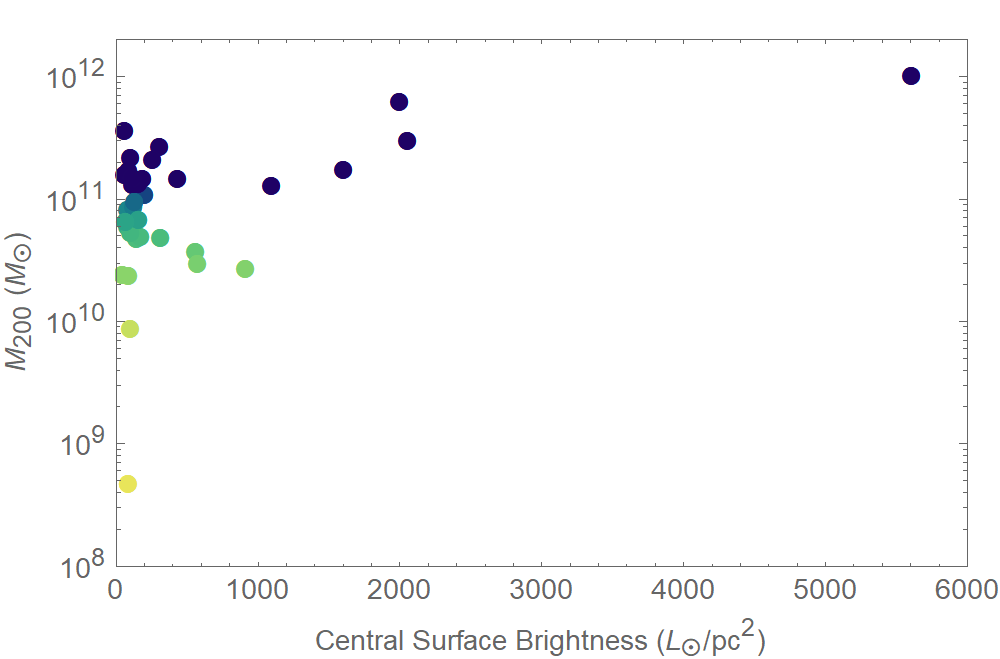}
    \end{minipage}
    \caption{
    left: Frequency distribution of the virial mass $M_{200}$ in units of solar masses $M_{\odot}$ from the 34 best-fitting galaxies in the NFW model. The maximum of the smooth-kernel-distribution (solid line) is at $M_{200}=(5.7\pm 3)\times 10^{10}\,M_{\odot}$.
   right: Extracted virial masses $M_{200}$ in units of solar masses $M_{\odot}$ from sample of the 34 best-fitting galaxies in the NFW model in the virial mass-luminosity plane.}\label{fig:HistoNFW}
\end{figure}

The left-hand side of Fig.\,\ref{fig:HistoNFW} shows the frequency distribution of the virial mass $M_{200}$ in units of solar masses $M_{\odot}$ from the 34 best-fitting galaxies in the NFW model. The maximum of the is at $M_{200}=(6.5\pm 3)\times 10^{10}\,M_{\odot}$. On the right-hand side of Fig.\,\ref{fig:HistoNFW}, the corresponding virial masses $M_{200}$ in units of solar masses $M_{\odot}$ are shown. Table 2 shows the luminosity, axion mass $m_a$, $r_{200}$, virial mass $M_{200}$, central density $\rho_0^{\rm NFW}$, scale radius $r_s$ of those 34 best fitting galaxies in the NFW model.

\begin{figure}[H]
    \centering
    \begin{minipage}{.5\textwidth}
        \centering
        \includegraphics[width=7cm]{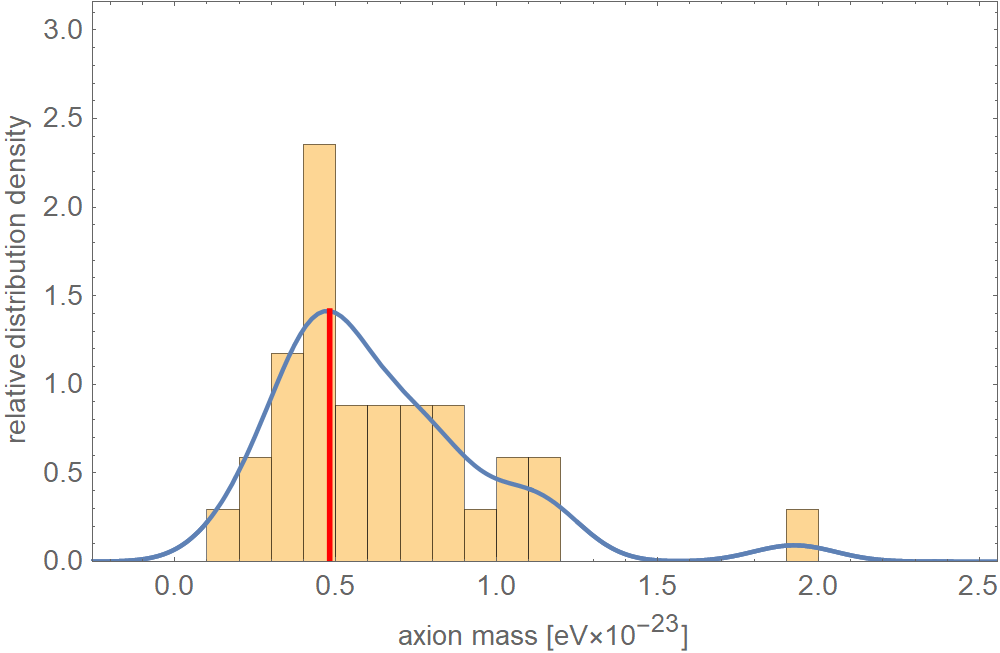}
    \end{minipage}%
    \begin{minipage}{.5\textwidth}
        \includegraphics[width=7.5cm]{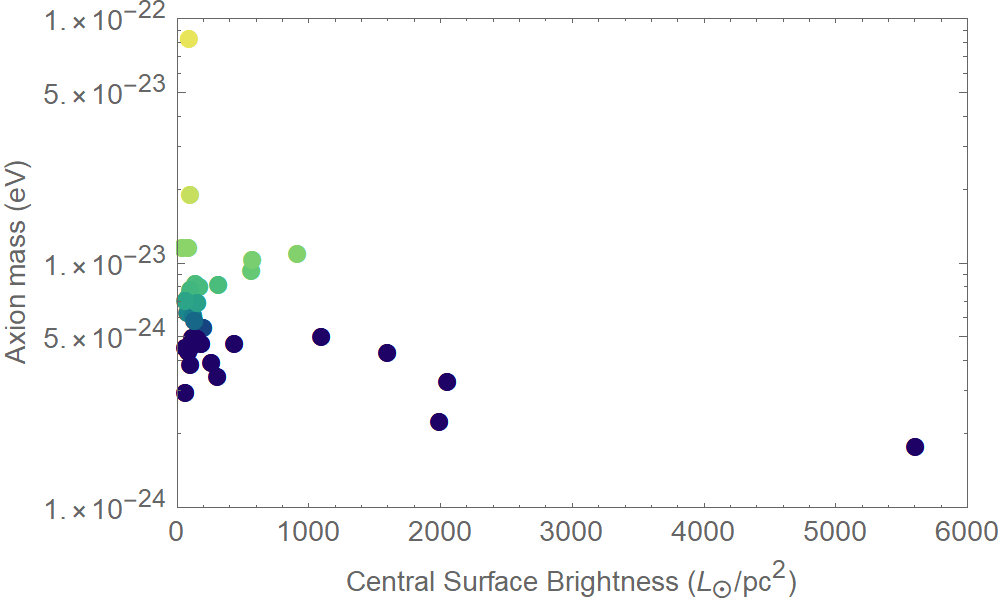}
    \end{minipage}
    \caption{
    left: Frequency distribution of axion mass $m_{a,e}$ as extracted from the PI model for 34 best fitting galaxies  and the Bohr radius $r_B=0.26$\,kpc. The maximum of the smooth-kernel-distribution (solid, blue line) is at $m_{a,e}=(0.48\pm 0.25) \times10^{-23}$ eV (red, vertical line). right: The same axion masses in a mass-luminosity plane.} \label{fig:HistoNFWAxions}
\end{figure}

In order to obtain an estimate for the axion mass $m_a$ from each fit, we use the relation between the Bohr radius $r_{B}$ and the axion mass $m_a$ of Eq.\,(\ref{Bohr}) and the Bohr radius of $r_{B}=0.26$\,kpc as determined in the SNFW fits together with the individually calculated virial mass $M_{200}$ as depicted in figure\,\ref{fig:HistoNFW}. Fig.\,\ref{fig:HistoNFWAxions} shows the frequency distribution of extracted axion masses $m_{a,e}$ from the NFW model for 34 best fitting galaxies. The maximum of the smooth-kernel-distribution is at $m_{a,e}=(0.44\pm 0.3) \times10^{-23}$ eV.

\begin{figure*}
\begin{flushleft}
\textbf{Table 2.} Fits of RCs to NFW model: Galaxy name, Hubble Type, $\chi^2/\text{d.o.f.}$, luminosity, axion mass $m_a$, $r_{200}$, virial mass $M_{200}$, central density $\rho_0^{\rm NFW}$, scale radius $r_s$.
\end{flushleft}
\begin{minipage}{1\textwidth}
\centering
\includegraphics[width=15cm]{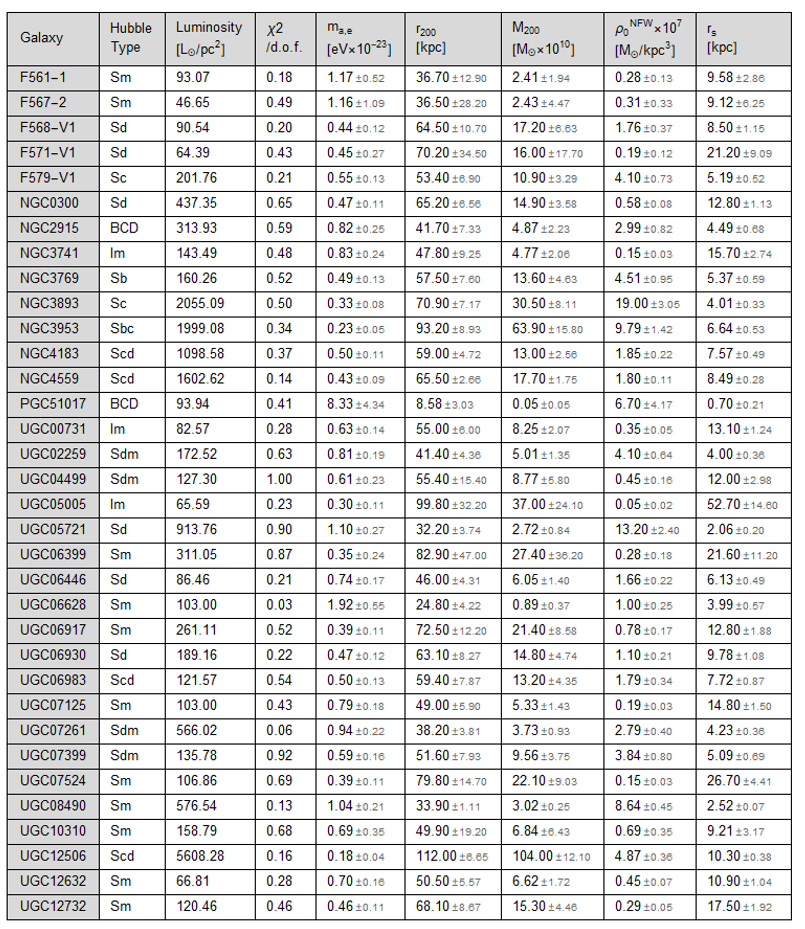}
\end{minipage}
\label{fig:GalaxyMassesNFW}
\end{figure*}

\newpage

\section{The pseudo-isothermal model}\label{chapter:PI}

In 2003, Jimenez et al. \cite{Jimenez_2003}, showed that a pseudo-isothermal (PI) profile fits well to many types of galaxies, not only dwarf galaxies as in the Burkert model.
For the PI model one assumes a mass-density profile of the form \cite{Begeman:1991iy}
\eqb
\label{eq:RhoPI}
\rho_{{\rm PI}}(r) = \frac{\rho_0^{{\rm PI}}}{1+(r/r_s)^2}\,
\eqe
where $\rho_0^{{\rm PI}}$ refers to the central mass density. Again, there is a scale 
radius $r_s$ which separates the core of the galaxy (ascending RC) from the region of quadratic asymptotic decay in $r$ where the RC is nearly constant. Note that the orbit-enclosed mass $M(r)$ now depends linearly on large $r$ such that our convention to set $r=r_B$ as a cutoff for the integral in Eq.\,(\ref{eq:Mtot}) is less controlled than in the NFW case. In analogy to the NFW model we obtain the rotational velocity of the PI model by using Eq.\,(\ref{eq:Mtot}), (\ref{eq:Vcirc}), and (\ref{eq:RhoPI})\,\cite[Eq.\,(15)]{Robles:2012uy}:\
\begin{align}
V_{\rm PI}(r)=\sqrt{4 \pi  G \, \rho_0^{\rm PI} \, r_s^2 \left( 1-\frac{r_s}{r} \arctan \left(\frac{r}{r_s}\right) \right)}\,.
\label{eq:PI}
\end{align}

\begin{figure}[H]
    \centering
    \begin{minipage}{.5\textwidth}
        \centering
        \includegraphics[width=7cm]{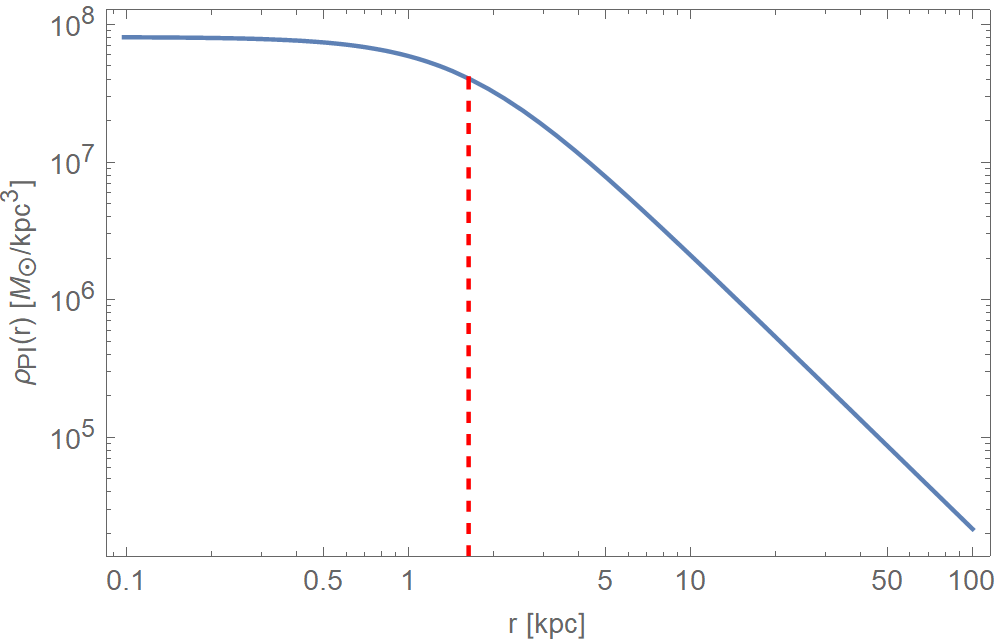}
    \end{minipage}%
    \begin{minipage}{.5\textwidth}
        \includegraphics[width=7cm]{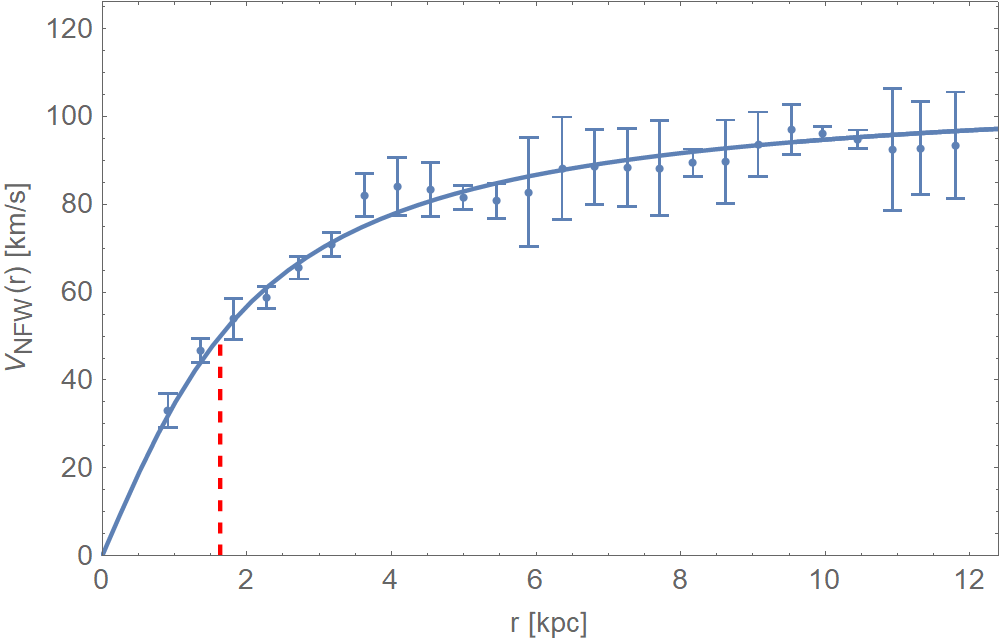}
    \end{minipage}
    \caption{
    An example for the characteristic density profile $\rho_{{\rm NFW}}(r)$, Eq.\,(\ref{eq:RhoNFW}), of galaxy NGC0300 is given on the left-hand side. On the right-hand side the corresponding orbit velocity $V_{{\rm NFW}}(r)$,
    Eq.\,(\ref{eq:NFW}), of the same galaxy is fitted to the data as obtained by the SPARC library \cite{Lelli:2016zqa}. The scale radius $r_s$ (red dotted line) indicates the transition from an almost constant to a $r^{-2}$ behaviour of the density profile.}
     \label{fig:HistoPIexample}
\end{figure}

Fig.\,\ref{fig:HistoPIexample} shows an example for the characteristic density profile $\rho_{{\rm PI}}(r)$ of Eq.\,(\ref{eq:RhoPI}), and the orbit velocity $V_{{\rm PI}}(r)$ of
Eq.\,(\ref{eq:PI}) for the galaxy NGC0300. The scale radius $r_s$ (red dotted line) indicates the transition from a $r^{-1}$ to a $r^{-3}$ behaviour of the density profile.

The orbit-enclosed mass $M_{\rm PI}$(r) of this model reads
\begin{align}
\label{amPI}
M_{\rm PI}(r) = 4 \pi  \rho_0 r_s^3 \left( \frac{r}{r_s} - \arctan \left(\frac{r}{r_s} \right) \right)
\end{align}

The axion mass $m_a$ is extracted by setting $r=r_B$ in Eq.\,(\ref{amPI}) and presuming that $r_B$ coincides with the largest radius of the observed RC. This is not a robust convention because of the linear dependence on large $r$ of $M_{\rm PI}(r)$ in Eq.\,(\ref{amPI}).\nn

\begin{figure}[H]
    \centering
    \begin{minipage}{.5\textwidth}
        \centering
        \includegraphics[width=7cm]{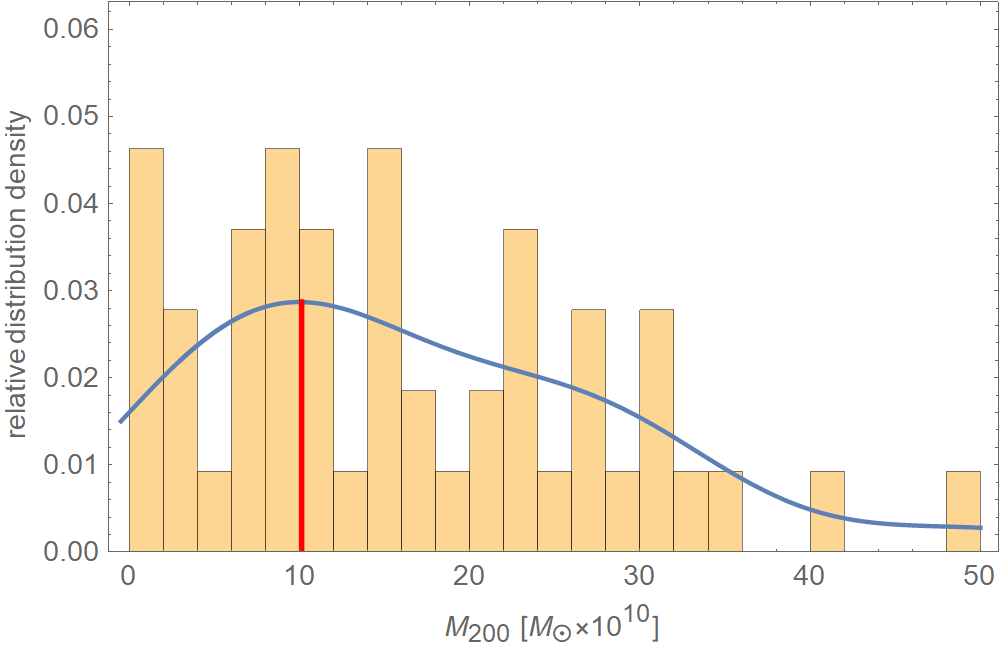}
    \end{minipage}%
    \begin{minipage}{.5\textwidth}
        \includegraphics[width=7.5cm]{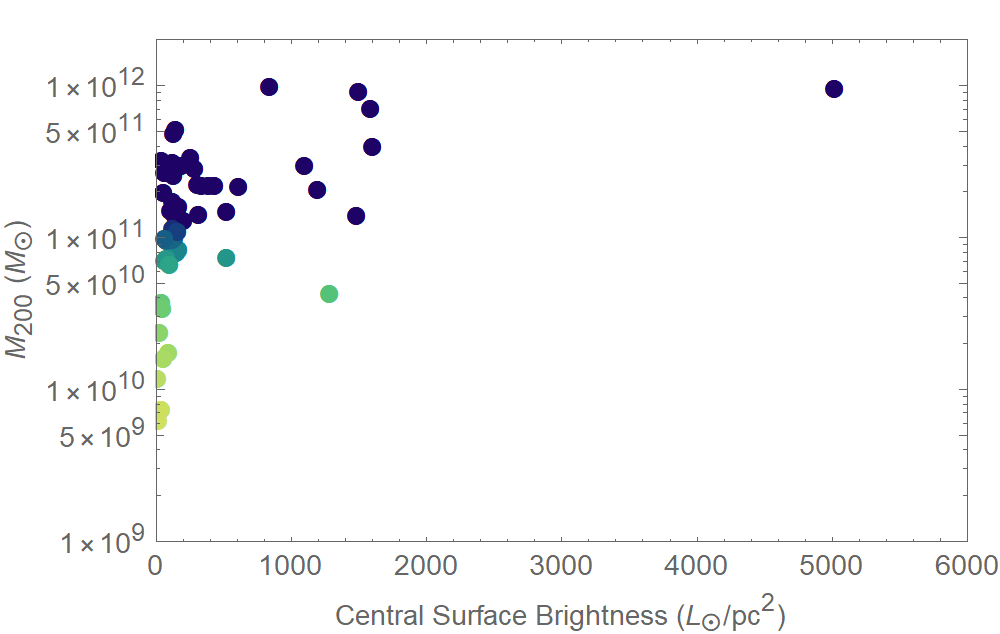}
    \end{minipage}
    \caption{
    left: Frequency distribution of the virial mass $M_{200}$ in units of solar masses $M_{\odot}$ from the 54 best-fitting galaxies in the NFW model. The maximum of the smooth-kernel-distribution (solid line) is at $M_{200}=(10.1\pm 5)\times 10^{10}\,M_{\odot}$.
   right: Extracted virial masses $M_{200}$ in units of solar masses $M_{\odot}$ from sample of the 34 best-fitting galaxies in the NFW model in the virial mass-luminosity plane.}\label{fig:HistoPI}
\end{figure}

\begin{figure}[H]
    \centering
    \begin{minipage}{.5\textwidth}
        \centering
        \includegraphics[width=7cm]{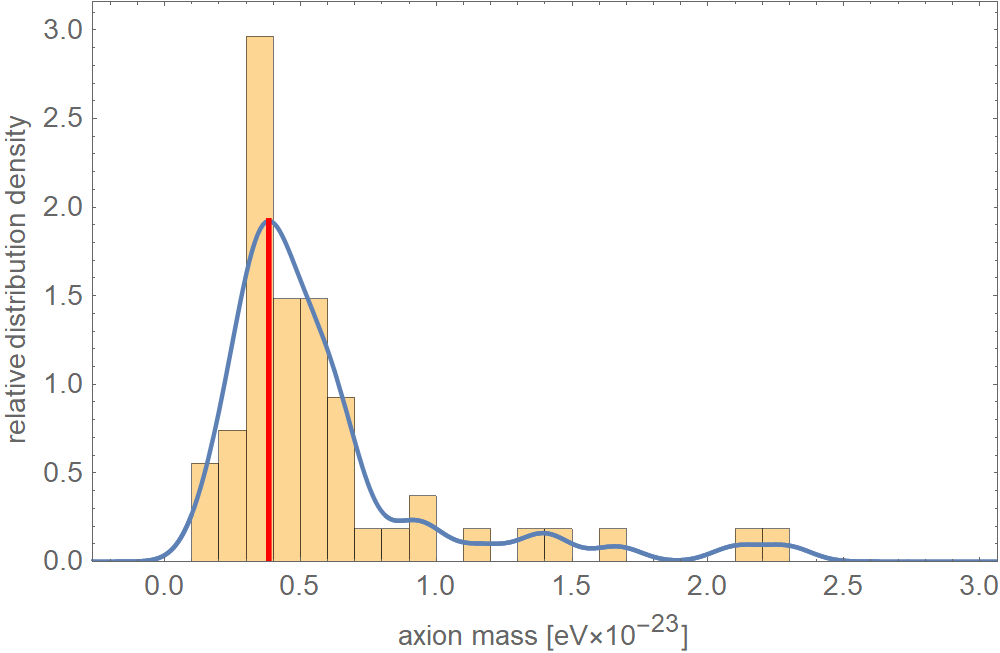}
    \end{minipage}%
    \begin{minipage}{.5\textwidth}
        \includegraphics[width=7.5cm]{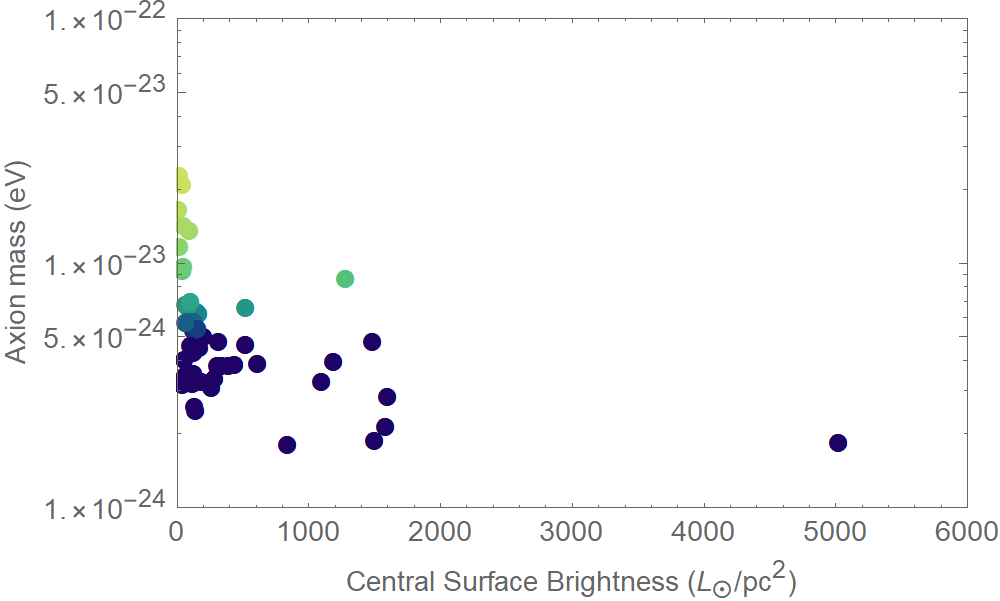}
    \end{minipage}
    \caption{
    left: Frequency distribution of axion mass $m_{a,e}$ as extracted from the PI model for 54 best fitting galaxies. The maximum of the smooth-kernel-distribution (solid, blue line) is at $m_{a,e}=(0.38\pm 0.2) \times10^{-23}$ eV  (red, vertical line). right: The same axion masses in a mass-luminosity plane.}
     \label{fig:HistoPIAxion}
\end{figure}

The left-hand side of Fig.\,\ref{fig:HistoPI} shows the frequency distribution of the virial mass $M_{200}$ in units of solar masses $M_{\odot}$ from the 54 best-fitting galaxies in the NFW model. The maximum of the is at $M_{200}=(10.1\pm 5)\times 10^{10}\,M_{\odot}$. On the right-hand side of Fig.\,\ref{fig:HistoPI}, the corresponding virial masses $M_{200}$ in units of solar masses $M_{\odot}$ are shown. Table 3 shows the luminosity, axion mass $m_a$, $r_{200}$, virial mass $M_{200}$, central density $\rho_0^{PI}$, scale radius $r_s$ of those 54 best fitting galaxies in the PI model. In order to obtain an estimate for the axion mass $m_a$ from each fit, we use the relation between the Bohr radius $r_{B}$ and the axion mass $m_a$ of Eq.\,(\ref{Bohr}) and the Bohr radius of $r_{B}=0.26$\,kpc as determined in the SNFW fits together with the individually calculated virial mass $M_{200}$ as depicted in Fig.\,\ref{fig:HistoPI}. Fig.\,\ref{fig:HistoPIAxion} shows the frequency distribution of extracted axion masses $m_{a,e}$ from the PI model for 54 best fitting galaxies. The maximum of the smooth-kernel-distribution is at $m_{a,e}=(0.38\pm 0.2) \times10^{-23}$ eV.\nn

\begin{figure*}\centering
\begin{flushleft}
\textbf{Table 3.} Fits of rotation curves to the PI model: Galaxy name, Hubble Type, $\chi^2/\text{d.o.f.}$, luminosity, axion mass $m_a$, $r_{200}$, virial mass $M_{200}$, central density $\rho_0^{\rm PI}$, scale radius $r_s$.
\end{flushleft}
\begin{minipage}{1\textwidth}
\centering
\includegraphics[width=14.5cm]{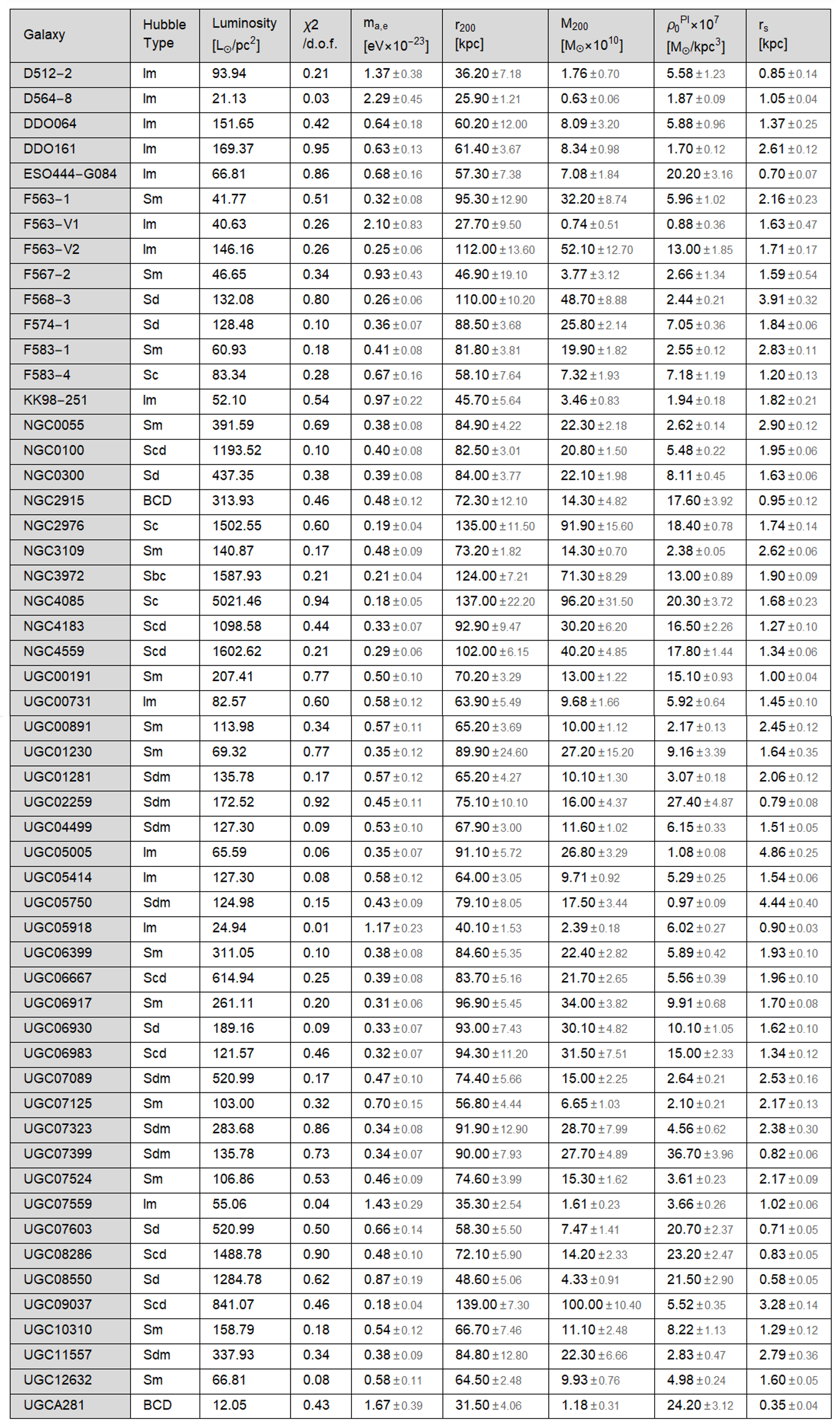}
\end{minipage}
\end{figure*}

\newpage

\section{Burkert model fits}\label{chapter:Burkert}

The Burkert profile was developed by Andreas Burkert and Paolo Salucci \cite{Burkert_1995,Salucci:2000ps}. It is an empirical model which is leaned on the pseudo-isothermal profile. The model is has a central core which is characterized by the core radius $r_0$ and by the central density $\rho_0$, the mass-density profile takes the form \cite{Burkert_1995,Salucci:2000ps}
\eqb
\label{eq:RhoB}
\rho_{{\rm Bu}}(r) = \frac{\rho_0\, r_0^3}{(r+r_0)(r^2+r_0^2)}\,.
\eqe

The rotational velocity of the Burkert model by using Eq.\,(\ref{eq:Mtot}), (\ref{eq:Vcirc}), and (\ref{eq:RhoB})\,:

\begin{align}
V_{\rm Bu}(r)=\sqrt{
\frac{ \pi\,  G\, \rho_0^{\rm Bu}\,  r_0^3}{r}\, 
\left(  2 \arctan{\frac{r}{r_0}} + 2 \log{\frac{r_0}{r+r_0}} + \log{\frac{r_0^2}{r^2+r_0^2}}\right)}\,.
\label{eq:B}
\end{align}

\begin{figure}[H]
    \centering
    \begin{minipage}{.5\textwidth}
        \centering
        \includegraphics[width=7cm]{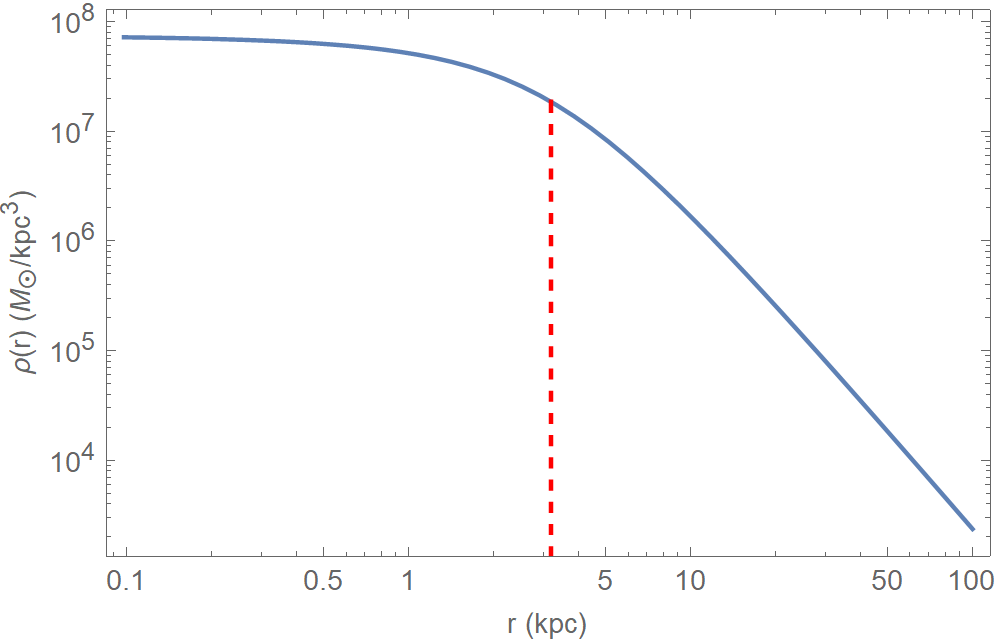}
    \end{minipage}%
    \begin{minipage}{.5\textwidth}
        \includegraphics[width=7cm]{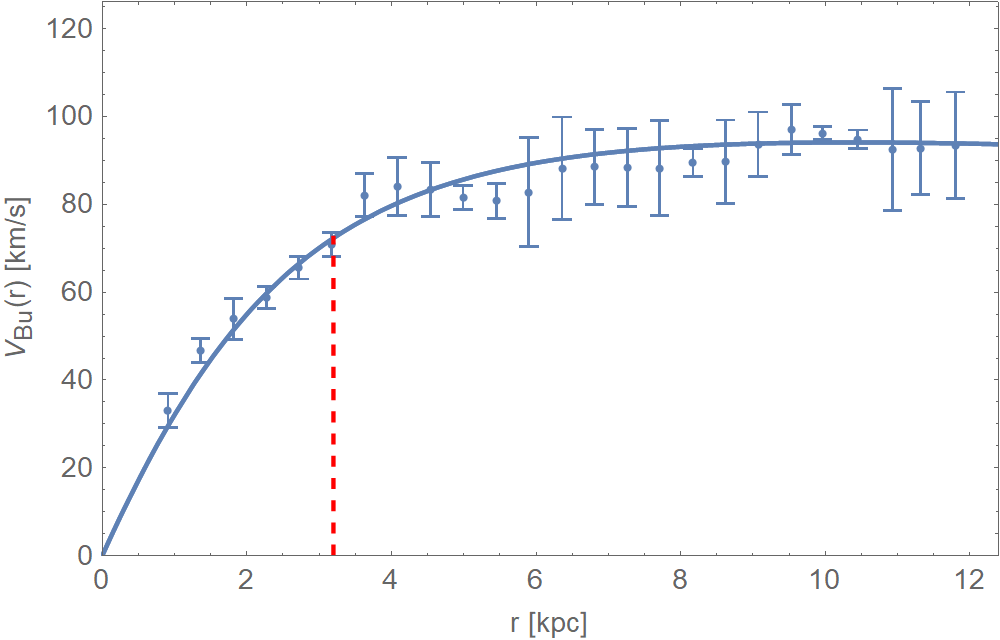}
    \end{minipage}
    \caption{
    An example for the characteristic density profile $\rho_{{\rm Bu}}(r)$, Eq.\,(\ref{eq:RhoB}), of galaxy NGC0300 is given on the left-hand side. On the right-hand side the corresponding orbit velocity $V_{{\rm Bu}}(r)$,
    Eq.\,(\ref{eq:B}), of the same galaxy is fitted to the data as obtained by the SPARC library \cite{Lelli:2016zqa}. The scale radius $r_s$ (red dotted line) indicates the transition from a $r^{-1}$ to a $r^{-3}$ behaviour of the density profile.}
     \label{fig:HistoBexample}
\end{figure}

Fig.\,\ref{fig:HistoBexample} shows an example for the characteristic density profile $\rho_{{\rm Bu}}(r)$ of Eq.\,(\ref{eq:RhoB}), and the orbit velocity $V_{{\rm Bu}}(r)$ of
Eq.\,(\ref{eq:B}) for the galaxy NGC0300. Starting relatively constant, the scale radius $r_s$ (red dotted line) indicates the transition from a $r^{-1}$ to a $r^{-3}$ behaviour of the density profile similarly to the NFW profile.

Collage 2 depict the fits of the Burkert model to the 17 RCs used in the SNFW fits. Table 4 a) indicates that three out of these 17 RCs are fitted with a $\chi^2/{\rm d.o.f.}>1$. Therefore, we resort to a sample of 80 galaxies which fit with  $\chi^2/{\rm d.o.f.}<1$, their results are shown in Table 4 a) and b).\nn

Recall that our strategy is to demonstrate that the extracted axion mass $m_{a,e}$ from the SNFW is consistent with the other three empirical models. Therefore, we calculate a Bohr radius Eq.\,(\ref{Bohr}) from that axion mass and together with the virial masses $M_{200}$ extracted from RC fits in the corresponding model we retrieve a model (and SNFW)-dependant axion mass. Overall, an ensemble of 80 SPARC galaxies fits well to the Burkert model.\nn 

\begin{figure}[H]
    \centering
    \begin{minipage}{.5\textwidth}
        \centering
        \includegraphics[width=7cm]{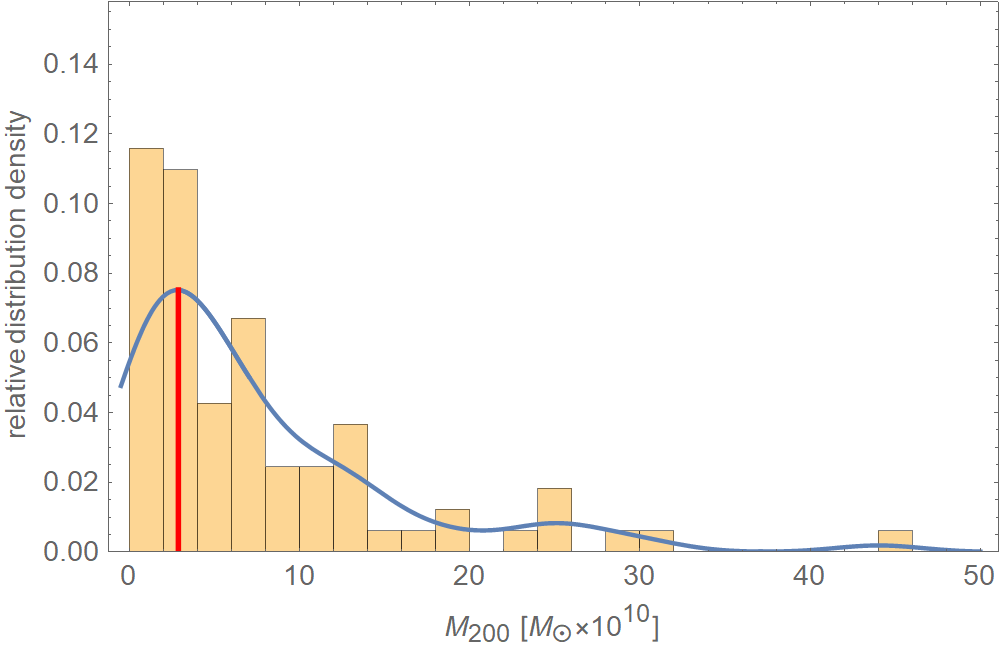}
    \end{minipage}%
    \begin{minipage}{.5\textwidth}
        \includegraphics[width=7.5cm]{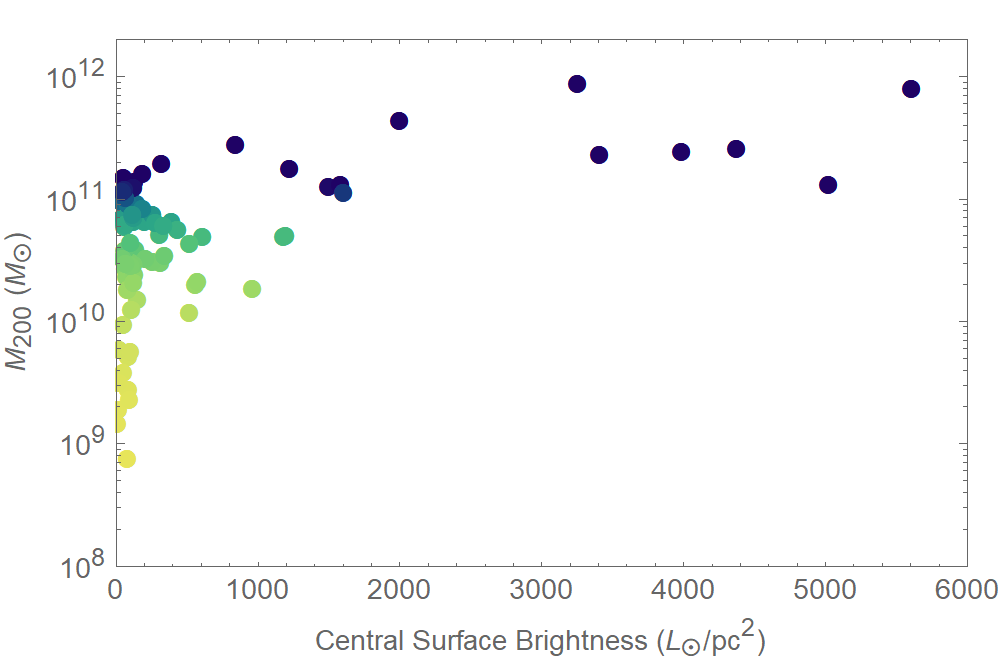}
    \end{minipage}
    \caption{
    left: Frequency distribution of the virial mass $M_{200}$ in units of solar masses $M_{\odot}$ from the 80 best-fitting galaxies in the Burkert  model. The maximum of the smooth-kernel-distribution (solid line) is at $M_{200}=(2.9\pm4)\times 10^{10}\,M_{\odot}$.
   right: Extracted virial masses $M_{200}$ in units of solar masses from Burkert-model fits to 80 RCs with $\chi^2/\text{d.o.f.}<1$ vs. the respective galaxy's central surface brightness in units of $L_{\odot}/\text{pc}^2$.}
   \label{fig:M200Burkert}
\end{figure}

\begin{figure}[h!]
    \centering
    \begin{minipage}{.5\textwidth}
        \centering
        \includegraphics[width=7cm]{HistoB8.png}
    \end{minipage}%
    \begin{minipage}{.5\textwidth}
        \includegraphics[width=7.5cm]{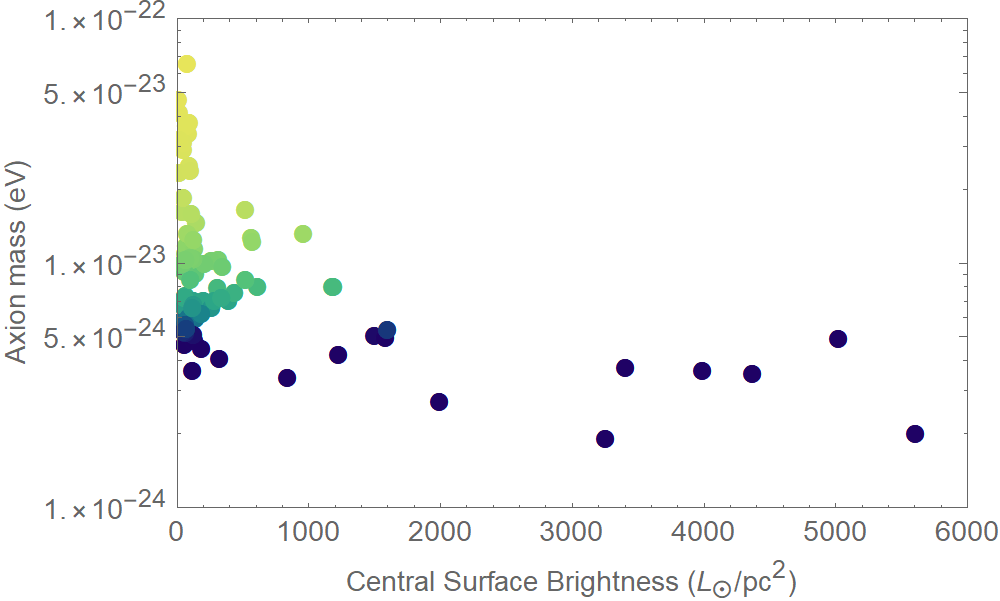}
    \end{minipage}
    \caption{
    left: Frequency distribution of 80 axion masses $m_{a,e}$, extracted from the Burkert-model fits of $M_{200}$ to the RCs of galaxies with a $\chi^2/\text{d.o.f.}<1$. The maximum of the smooth-kernel distribution (solid, blue line) is $m_{a,e}=(0.65\pm0.4)\times10^{-23}$\,eV (red, vertical line).
   right: The same axion masses in a mass-luminosity plane.}
   \label{fig:AxionMassesBurkert}
\end{figure}

With a typical virial mass of $M_{200}=(2.9\pm4)\times 10^{10}\,M_{\odot}$, Fig.\,\ref{fig:M200Burkert}, the Burkert model yields a frequency distribution of $m_{a,e}=(0.65\pm 0.4)\times 10^{-23}\,$eV as shown in Fig.\,\ref{fig:AxionMassesBurkert}. Thus, the maximum of the smooth-kernel distribution is compatible with that in the SNFW model $m_{a,e}=(0.72\pm 0.5)\times 10^{-23}\,{\rm eV}$. Notice how $M_{200}$ clusters around the value $M_{200}\sim 5\times 10^{10}\,M_{\odot}$.\nn

In our treatment Sec.\,\ref{Sec5} of cosmological and astrophysical implications we appeal to the mean value of $m_{a,e}$-extractions in the SNFW and the Burkert model as 
\eqb
\label{maBurkertSNFW}
m_{a,e}=0.675\times 10^{-23}\,{\rm} {\rm eV}\,.
\eqe 

\begin{figure*}[h!]
\begin{flushleft}
\textbf{Collage 2.} Burkert-model fits to the 17 best fitting SNFW-model galaxies. The purple arrow indicates the value of $r_{0}$.\nn
\end{flushleft}
\vspace{0.3cm}
\begin{minipage}{1\textwidth}
\includegraphics[width=15cm]{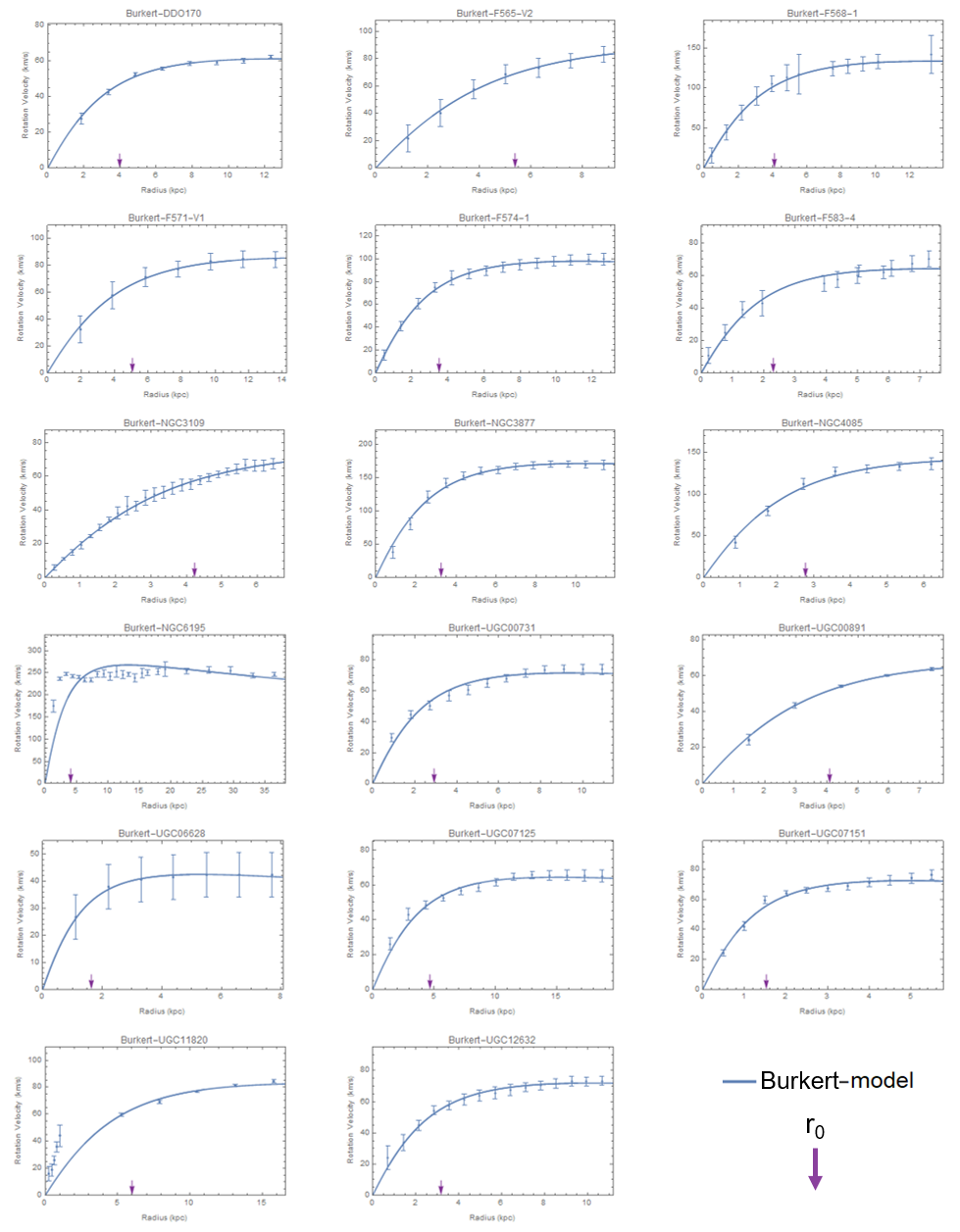}
\end{minipage}
\label{GalaxyMassesBurkert}
\end{figure*}

\begin{figure*}[h!]
\begin{flushleft}
\textbf{Table 4 a)} Burkert model: Galaxy name, Hubble Type, $\chi^2/\text{d.o.f.}$, luminosity, axion mass $m_a$, $r_{200}$, virial mass $M_{200}$, core density $\rho_0$, and core radius $r_0$. In the blue frame the 17 galaxies used for the SNFW fit are highlighted.
\end{flushleft}
\begin{minipage}{1\textwidth}
\centering
\includegraphics[width=15cm]{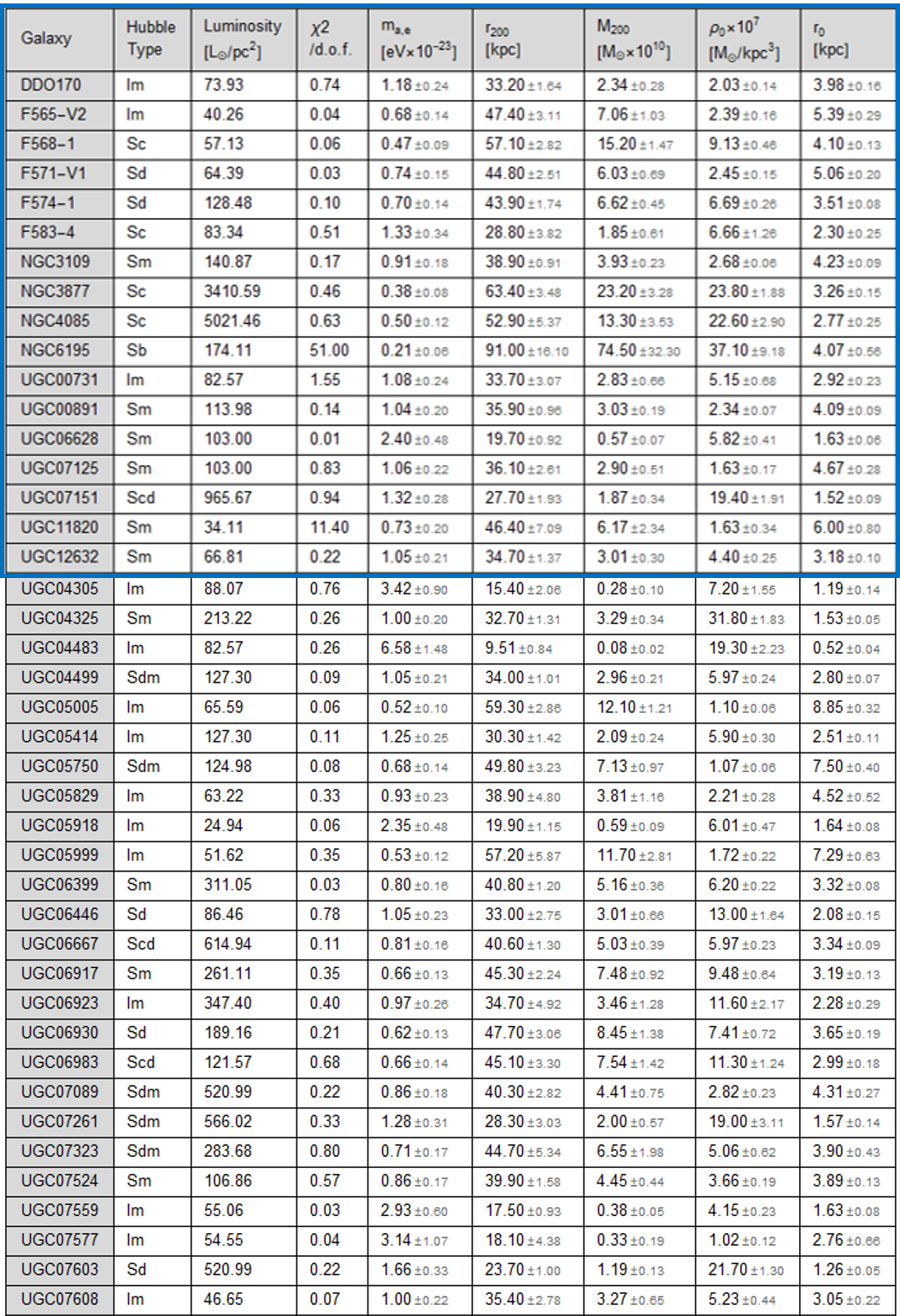}
\end{minipage}
\end{figure*}

\begin{figure*}[h!]
\begin{flushleft}
\textbf{Table 4 b)} Burkert model with $\chi^2/{\rm d.o.f.}<1$: Galaxy name, Hubble Type, $\chi^2/\text{d.o.f.}$, luminosity, axion mass $m_a$, $r_{200}$, virial mass $M_{200}$, core density $\rho_0$, and core radius $r_0$ (ordered as in the SPARC library).
\end{flushleft}
\begin{minipage}{1\textwidth}
\includegraphics[width=15cm]{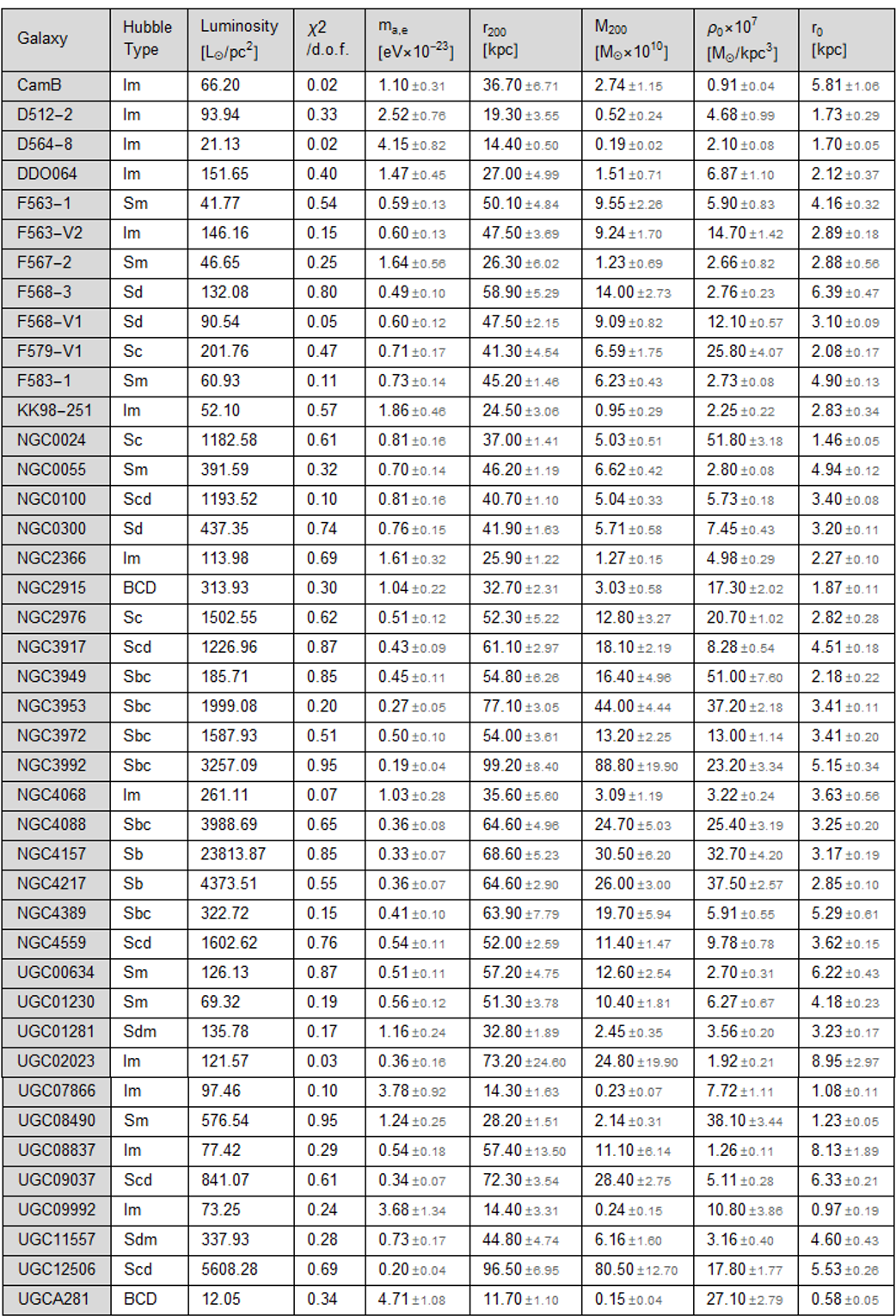}
\end{minipage}
\end{figure*}

\newpage


\chapter{ Interpretation of the results}\label{chapter7}

\vspace{0.1cm}

\begin{minipage}{0.7\textwidth}
\begin{figure}[H]\centering
\includegraphics[width=8cm]{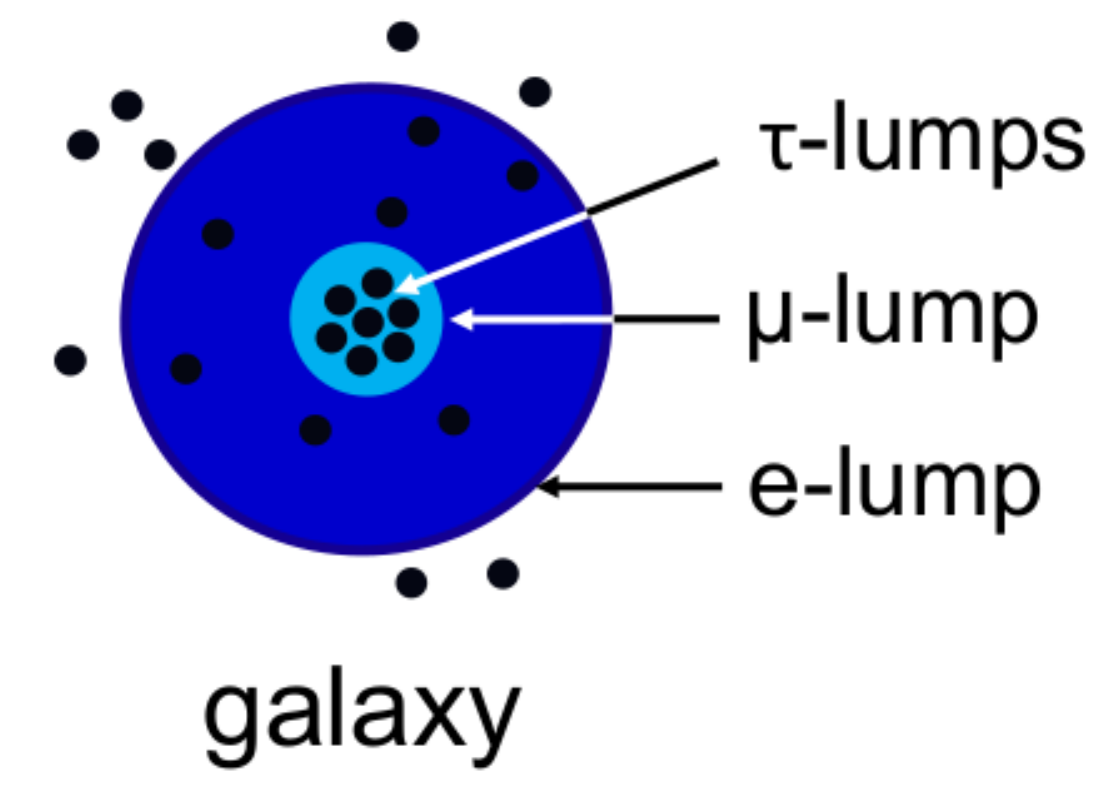}
\caption{Three axion species in condensate (lump) form suggests that in the halo of the Milky Way can be associated with an e-lump, the central compact massive object could originate from a $\mu$-lump, and that the $\tau$-lumps provide the dark sources for the accretion of stars in globular clusters.}
\end{figure}
\end{minipage} \hfill
\begin{minipage}{0.3\textwidth}
\begin{itemize}\sffamily

\item[\textcolor{gray!90}{\textbullet}] \textcolor{gray!90}{What is the dynamical symmetry braking scale?}

\item[\textcolor{gray!90}{\textbullet}] \textcolor{gray!90}{What is the connection to particle physics?}

\item[\textcolor{gray!90}{\textbullet}] \textcolor{gray!90}{Are there any indications about typical lump sizes and masses?}

\item[\textcolor{gray!90}{\textbullet}] \textcolor{gray!90}{What does this mean?}
\end{itemize}
\end{minipage}\vspace{1.5cm}

\minitoc

\newpage

\section{Model comparison}

\begin{figure}[h]\centering
\includegraphics[width=9cm]{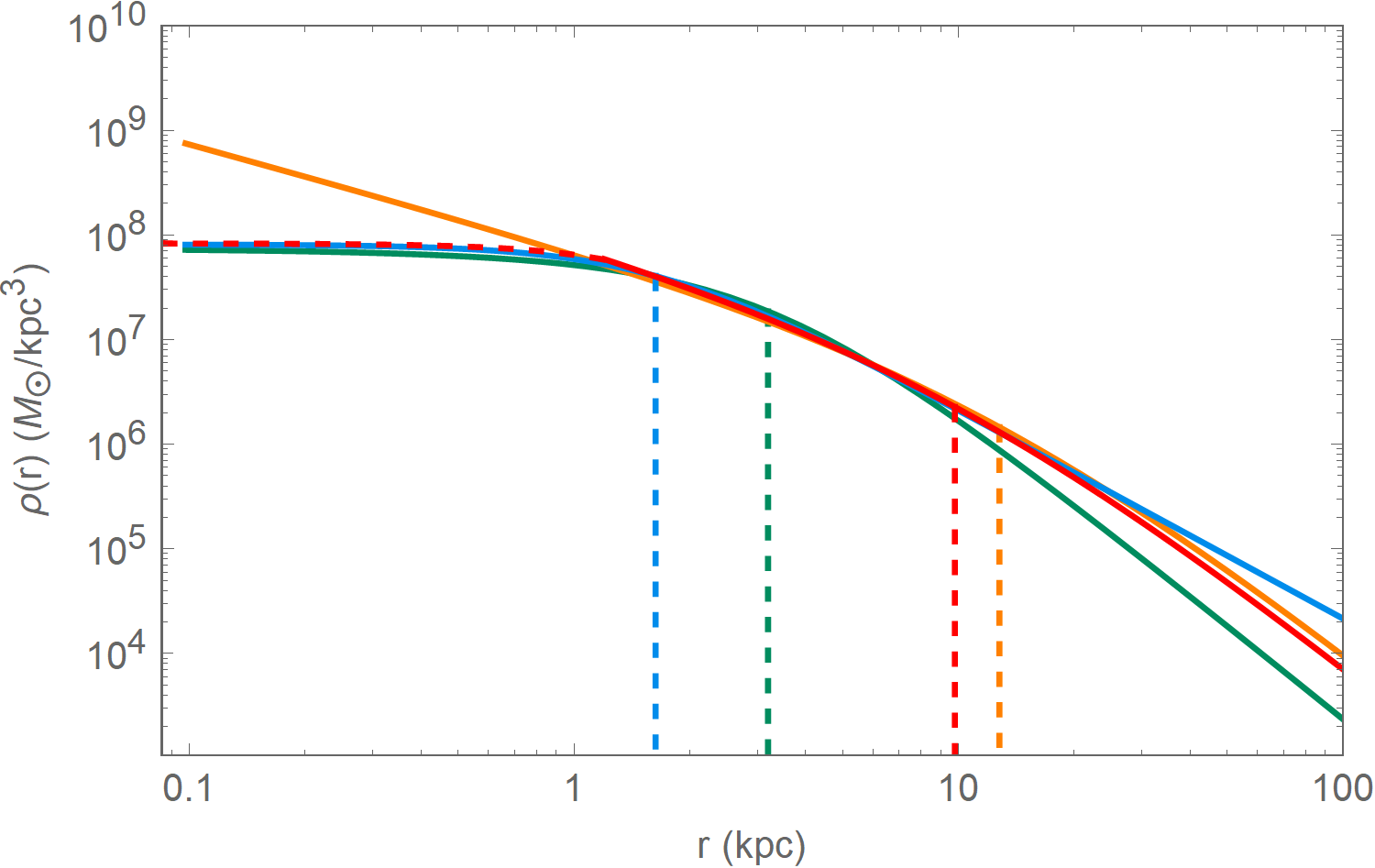}
    \caption{
    An example for the characteristic density profile $\rho_{{\rm SNFW}}(r)$, Eq.\,(\ref{eq:SolitonNFWwhole}), of galaxy NGC0300 is given in red. The red dotted line indicates the description by the soliton density profile \ref{eq:SolitonNFW}, the red line represents the NFW part \ref{eq:RhoNFW}. For the same galaxy the NFW density profile was plotted in orange. The scale radii $r_s$ are indecated by vertical dotted lines in the corresponding colour. 
    }
     \label{fig:Allfit}
\end{figure}

A comparison between the SNFW, NFW, PI and Burkert model density profiles for an explicit galaxy, NGC0300 is shown in Fig.\,\ref{fig:Allfit}.\vspace{-1mm}

\begin{figure}[H]
    \centering
    \begin{minipage}{.5\textwidth}
        \centering
        \includegraphics[width=7cm]{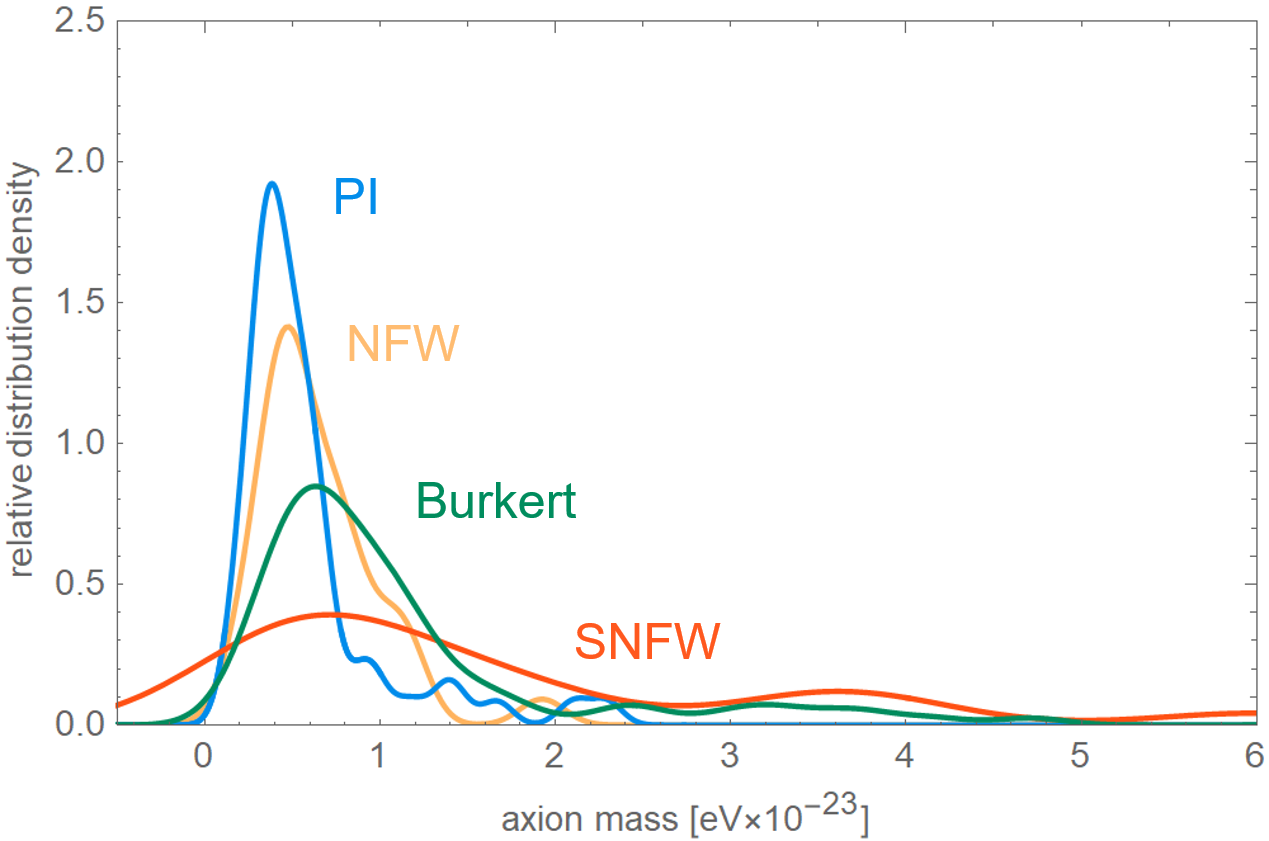}
    \end{minipage}%
    \begin{minipage}{.5\textwidth}
        \includegraphics[width=7cm]{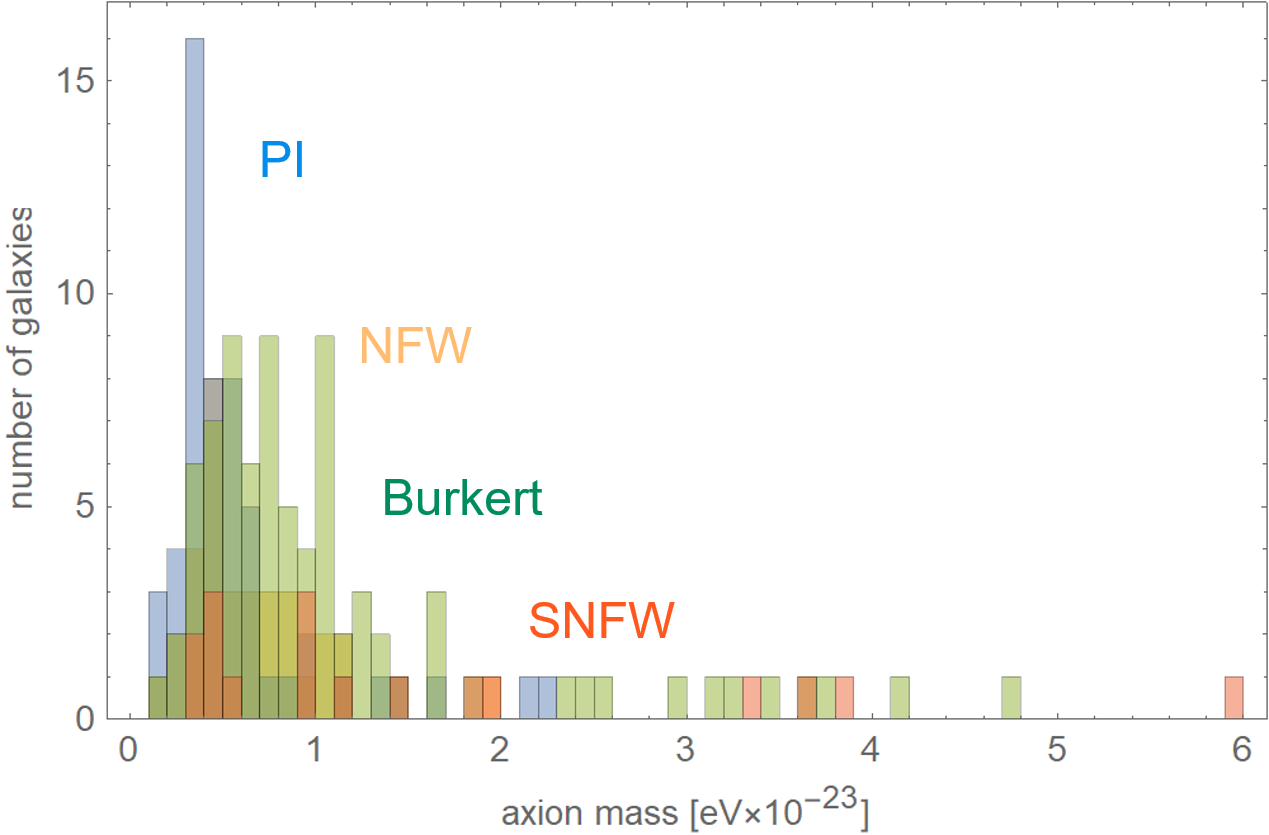}
    \end{minipage}
    \caption{
    Axion mass probability density fits for the PI (blue), NFW (orange), Burkert (green) and SNFW (red) model for a Bohr radius of $R_{B}=0.26$\,kpc (left) next to the histogram of in the same colour coding (right)}
    \label{fig:Comparison}
\end{figure}

\begin{figure}[H]
    \centering
    \begin{minipage}{.5\textwidth}
        \centering
        \includegraphics[width=7cm]{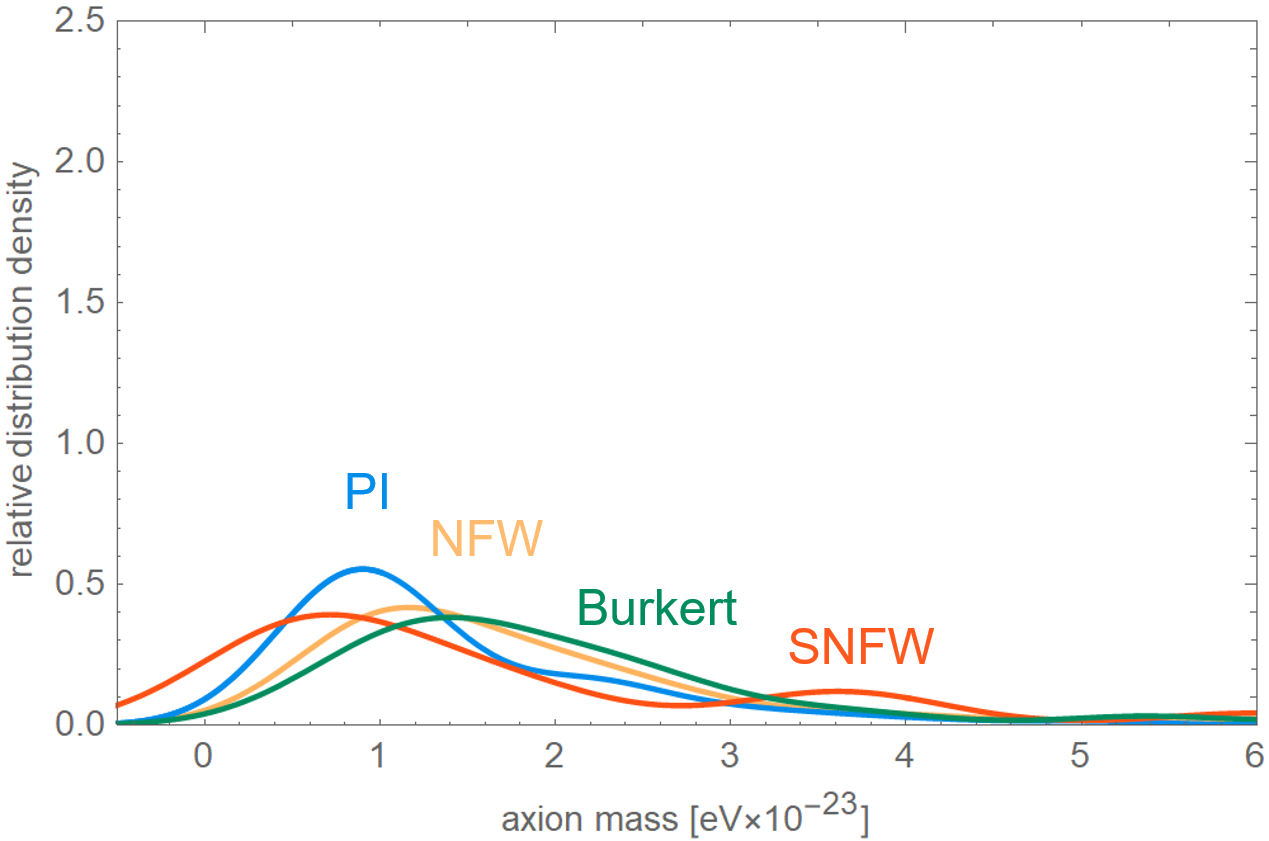}
    \end{minipage}%
    \begin{minipage}{.5\textwidth}
        \includegraphics[width=7cm]{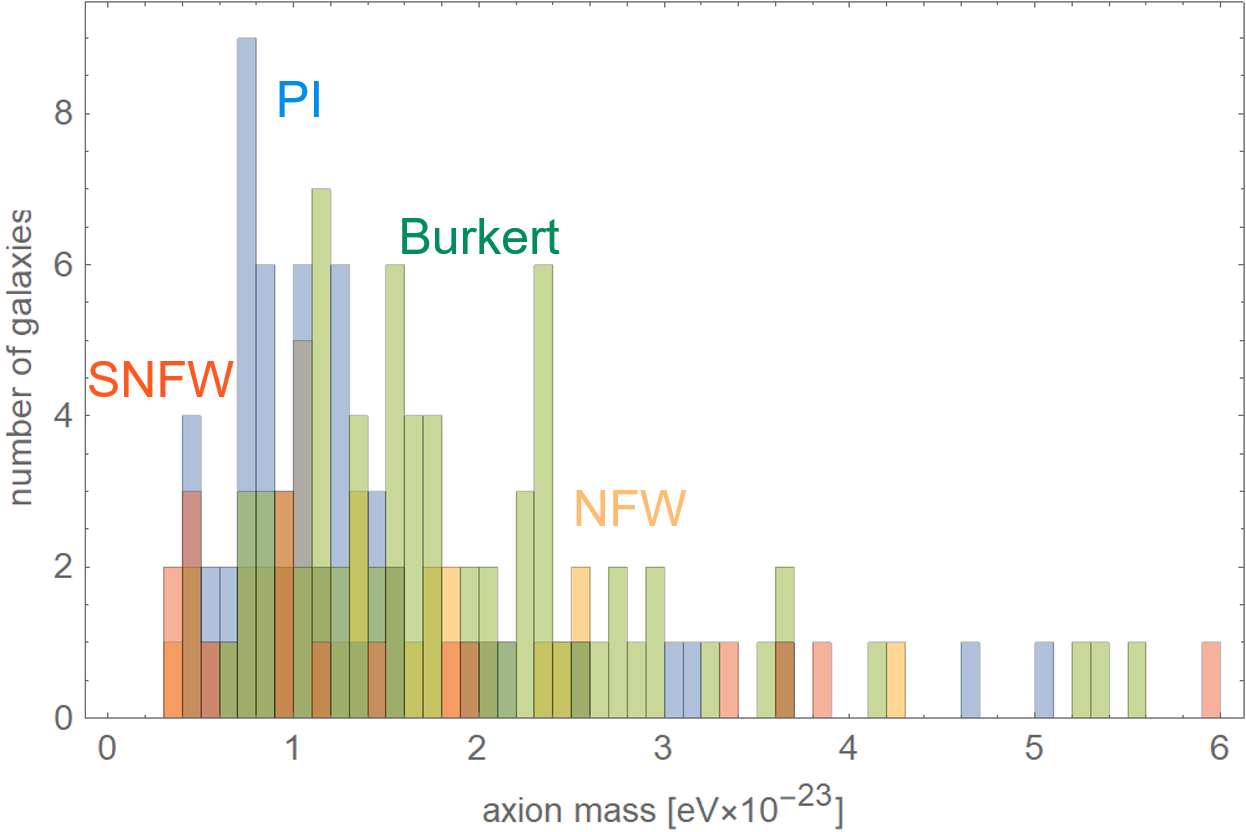}
    \end{minipage}
    \caption{
     Axion mass probability density fits for the PI (blue), NFW (orange), Burkert (green) and SNFW (red) model for a Bohr radius of $R_{B}=0.05$\,kpc (left) next to the histogram of in the same colour coding (right)}
     \label{fig:AxionMassesAllComparison2}
\end{figure}

The fitted probability density functions for the axion masses in the SNFW, NFW, PI and Burkert model are shown and compared in Fig.\,\ref{fig:Comparison} and Fig.\,\ref{fig:AxionMassesAllComparison2}, alongside their histograms. Note, that there are two ways of extracting a Bohr radius $r_B$ from the SNFW fits for the other three models: 
(i) fitting a probability density functions to the axion masses $m_a$ and the virial masses $M_{200}$, determine their maxima and using equation \ref{Bohr} to calculate the gravitational Bohr radius. This results in a gravitational Bohr radius of $R_{B}=0.26$ kpc, or
(ii) calculating for each galaxy the gravitational Bohr radius with the axion mass $m_a$ and the virial masses $M_{200}$ of that particular galaxy and using the maxima of the probability density fit to all Bohr radii; This results in a gravitational Bohr radius of $R_{B}=0.05$ kpc.\vspace{-1mm}

Figure \ref{fig:Comparison} compares the axion mass distribution for the four different profiles, while a gravitational Bohr radius of $R_{B}=0.26$ kpc was assumed for the NFW, PI and Burkert model, while Fig.\,\ref{fig:AxionMassesAllComparison2} shows compares the axion mass distribution under the assumption of $R_{B}=0.05$\,kpc. Interestingly, the standard deviation of the PI, NFW and the Burkert fit are smaller for $R_{B}=0.26$ kpc, as well as the spread of the maxima with regard to maximum of the SNFW fit compared to  $R_{B}=0.05$. Furthermore, this seems not to be a binning effect, as the shape and the individual bin counts do not vary with a re-scaling.

\begin{figure}[H]
\floatbox[{\capbeside\thisfloatsetup{capbesideposition={right,top},capbesidewidth=6cm}}]{figure}[\FBwidth]
{\caption{Probability density fit for all Bohr radii in the SNFW model. The smoth kernel distribution (blue) has a maxima of  $R_{B}=0.05$\,kpc (red).}\label{fig:BohrHistoSNFW}}
{\includegraphics[width=7cm]{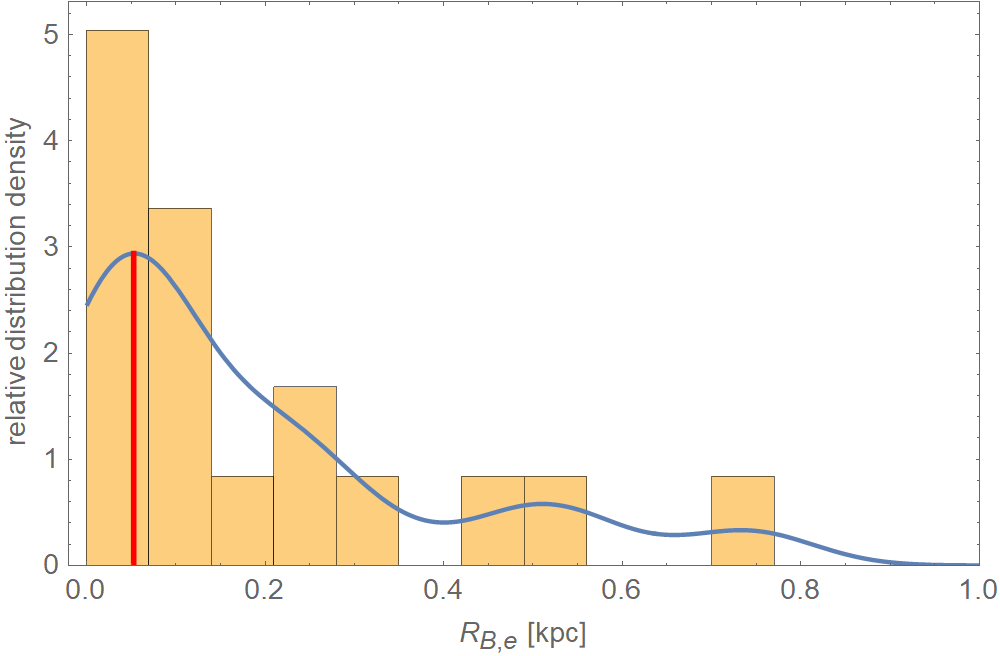}}
\end{figure}

Figure \ref{fig:BohrHistoSNFW} illustrates the second approach of retrieving a Bohr radius, based on a direct Histogram fit of all 17 Bohr radii as obtained in the SNFW model. This plot might immediately explain the qualitatively worse prediction, as it clearly takes a non-Gaussian form and the smallest bin is overpopulated. Unfortunately, re-binning does not resolve this problem. Moreover, it is interesting to see that although the maxima of the three heuristic models smaller than the SNFW maxima in Fig.\,\ref{fig:Comparison} and bigger in Fig.\,\ref{fig:AxionMassesAllComparison2}, the order does not change: The PI model obtains the smallest axion value, followed by NFW and Burkert.\nn

Consequently, a gravitational Bohr radius of $R_{B}=0.26$\,kpc will be used in the following. The overlap of the SNFW axion mass with the extracted axion masses in the NFW, PI and Burkert model were a consistency check. In the next chapter we look roughly at some implications of a three component dark matter model with lump types of dark matter.\nn

\begin{figure}[H]\centering
\includegraphics[width=14cm]{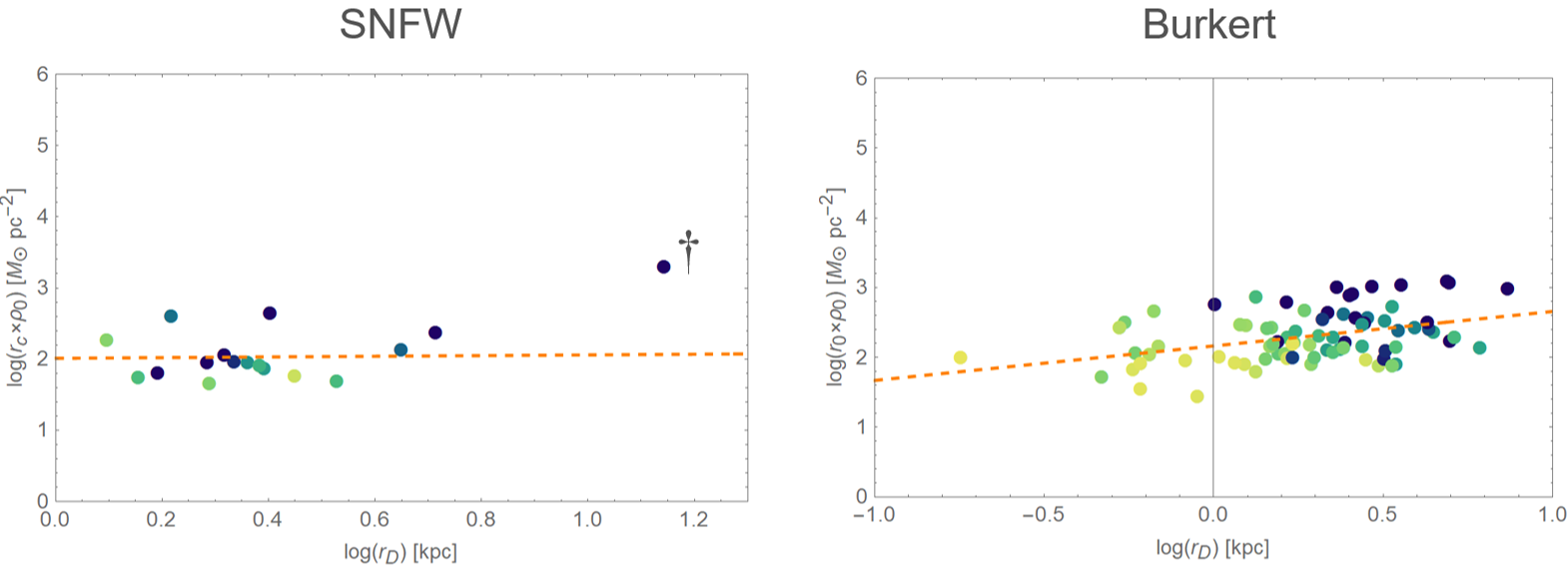}
    \caption{Constant relationship of the central halo surface density $r_0 \times \rho_0$ in the SNFW model left, and the Burkert model right. The dagger$^\dagger$ indicates galaxy NGC6195, its high baryonic content may cause the offset.}
     \label{fig:Relation}
\end{figure}

Comparing different dark matter halo profiles is a good way of testing consistency. However, it is prone to systematic errors. Another test of the validity of the fits is given by an empirically found constant relationship of the central halo surface density $r_0 \times \rho_0$, \cite{Donato_2009}. Interestingly this constant relationship holds throughout many galaxy types and dark matter halo profiles \cite{Donato_2009}. Apparently, it is as of right now unclear whether fuzzy dark matter can reproduce this as well. Nonetheless, we are able to reproduce this empirical found relation in the Burkert model as well as in the SNFW model, see Fig.\,\ref{fig:Relation}. Note that the one galaxy in the SNFW sample marked with a dagger$^\dagger$ is galaxy NGC6195. As it is shown previously in Table 1, this galaxy has a steep increase in orbital velocity which implies a high baryonic content of this galaxy. This might be an explanation for the relatively large offset as we assume dark matter dominated galaxies. A counter argument for this explanation is the fact that the constant behaviour of $r_0 \times \rho_0$ is valid for a broad range of galaxy types \cite{Donato_2009} regardless the baryon density.
\vspace{2mm}

\section[screened Yang-Mills scales]{A link to screened Yang-Mills scales in association to lepton masses}

In chapter \ref{chapter-Su(2)} a brief summary about SU(2)\textsubscript{CMB} Cosmology was given. Recall that the SU(2)\textsubscript{CMB} fit, Fig.\,\ref{fig:SU(2)fit}, has a smaller dark matter density than $\Lambda$CDM. This requires a sudden increase of dark matter which can be accomplished by a transition from dark energy into dark matter. The sudden transition\footnote{Although the model description of the transition in Eq.\,(\ref{eq:depercolation}) is in fact assumed to be instantaneous, which is also represented by the spike in Fig.\,\ref{fig:rH}, it is reasonable to assume a process which took a few million years. Therefore, \textit{sudden} should be interpreted in the context of cosmological time scales.}
which is called \textit{depercolation} suggests that dark energy is an ultra-light pseudo-Nambu-Goldstone boson field as well, see e.g. \cite{Frieman:1995pm,Kim:2002tq,Wilczek:2004cr,Hall:2005xb,Kaloper:2005aj,Barbieri:2005gj}.\nn

The idea is that an axion field $\phi$, which is generated by Planckian physics, develops a small mass due to topological defects of a Yang-Mills theory. More precisely, the mechanism to produce axions is by means of the chiral anomaly which is already shown to be implemented by nature in the $\pi_0 \rightarrow \gamma \gamma$ decay \cite{Adler:1969er,Bell:1969ts}. The chiral anomaly occurs on top of a dynamical breakdown of the global U$_A$(N) symmetry. The U$_A$(N) symmetry is carried by fermions $\psi$. By integrating out the groundstate portion of the gauge field $F^i_{\mu\nu}$, the field $\phi$ acquires a potential of the form \cite{Wilcek1983,Peccei:1977ur}

\begin{equation}\centering
\left( 1-\cos\frac{\phi}{f}\right) \Lambda^4_{\rm }\,,
\end{equation}

where $\Lambda_{}$ refers to the explicit symmetry breaking scale and $f$ the spontaneous symmetry breaking scale.
If the chiral pseudo-Goldstone field emerges as a consequence of gravitationally induced, chirally-invariant fermion interactions, it is plausible to associate the spontaneous symmetry breaking scale $f$ in the axion potential with the Planck scale $M_P$ \cite{Frieman:1995pm, Giacosa:2008rw}. In the latter, \cite{Giacosa:2008rw}, it has already been shown that at the Planck scale gravity and in a de Sitter Universe is able to break chiral symmetry dynamically. With the relation

\eqb
\Lambda^2=m_a\, f\,=m_a\, M_{P}\,,
\eqe


it is possible to extract $\Lambda_{}$ for a given axion mass $m_a$. As shown in the previous sections, we extracted axion masses in the range of $m_a \in (0.38-0.72)\times 10^{-23}\,$eV. With an averaged mass between Burkert and SNFW of $m_a \sim 0.675\times 10^{-23}\,$eV, we retrieve an explicit symmetry breaking scale of  

\eqb
\label{YMeextr}
\Lambda_e=287\,\mbox{eV}\,.
\eqe

This is by only a factor 15 smaller than the scale $\Lambda_e=m_e/118.6$ ($m_e=511\,$keV the mass of the electron) of an SU(2) Yang-Mills theory proposed in \cite{Hofmann:2017lmu} to originate the electron's mass in terms of a fuzzy ball of deconfining phase. This deconfining region is immersed into the confining phase and formed by the selfintersection of a center-vortex loop. Considering an undistorted Yang-Mills theory for simplicity\footnote{The chiral dynamics at the Planck scale, which produces the axion field, to some extent resolves the ground states of Yang-Mills theories: axions become massive by virtue of the anomaly because of this very resolution of topological charge density.}, the factor of 15 could be explained by a stronger screening of topological charge density -- the origin of the axial anomaly -- in the confining ground state, composed of round, pointlike center-vortex loops, versus the deconfining thermal ground state, made of densely packed, spatially extended (anti)caloron centers subject to overlapping peripheries \cite{bookHofmann}.\nn

\begin{figure}[h]
  \centering
  \includegraphics[width=8cm]{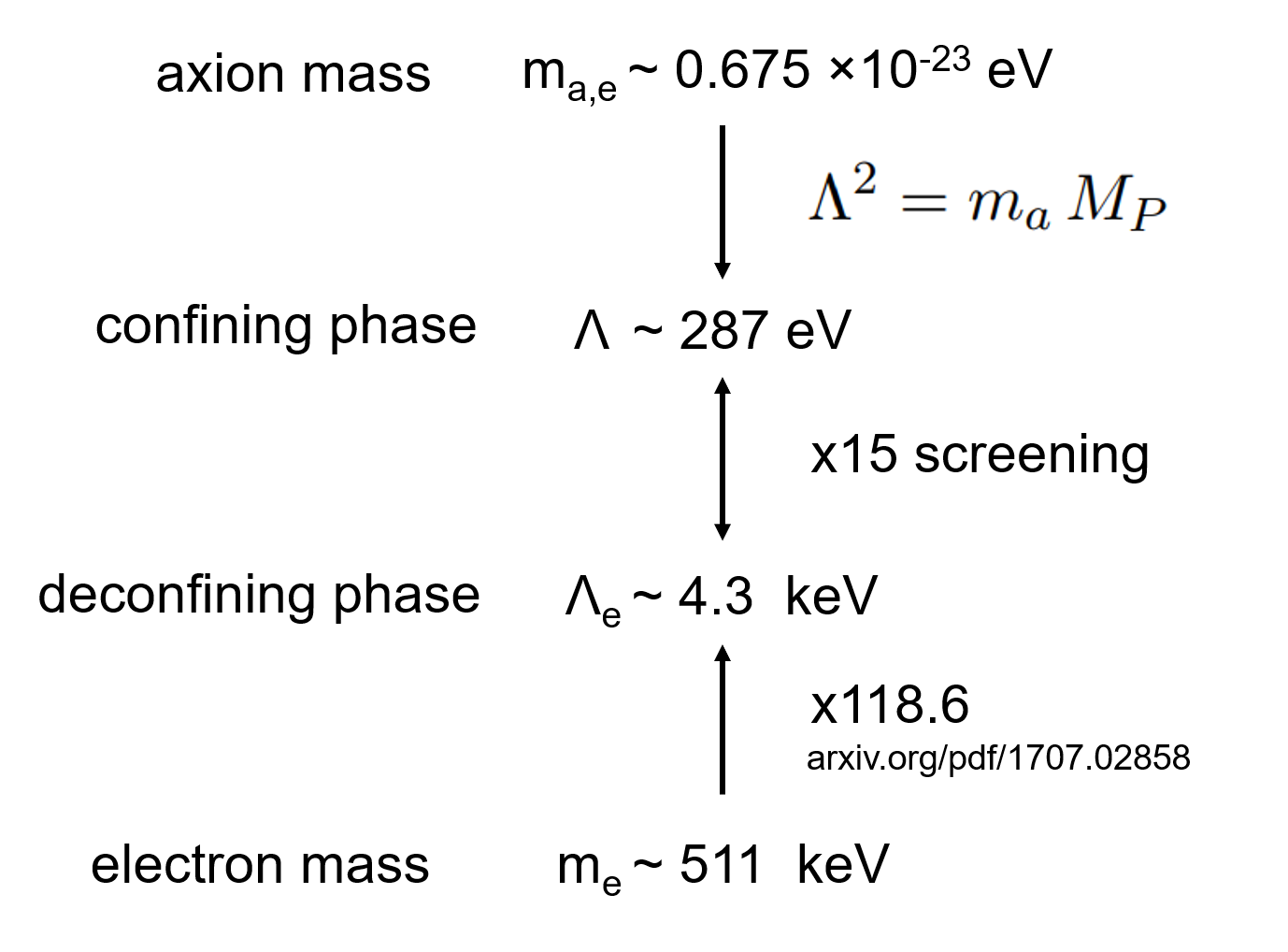}
  \caption{
  A schematic overview of the proposed connection from dark matter to screened Yang-Mills scales which are in association to lepton masses: The averaged mass between Burkert and SNFW of $m_a \sim 0.675\times 10^{-23}\,$eV leads to an explicit symmetry breaking scale of $\Lambda \sim 290$ eV. In the context of an SU(2) theory of the electron this is in the confined phase. The value for the same scale in the deconfined phase as obtained by \cite{Hofmann:2017lmu} is larger by a factor of 15.}
\label{fig:scales}
\end{figure}

Linking the electron scale and the explicit symmetry breaking scale of the axion potential suggest further to marry the whole leptonic sector with three flavours of dark matter. This goes hand in hand with SU(2) fits to the CMB, which found a smaller dark matter density after recombination compared to the $\Lambda$CDM best fit to 2015 Planck data; The ratio between the density parameters of primordial and late-time dark matter ranges between 0.5 and 0.7 \cite{Hahn:2018dih}. This implies at least one event of rapid (on cosmological scales) depercolation of dark matter. Depercolation is the transition of dark energy into dark matter. If the primordial dark matter density in comparison to the current dark matter density in assumed to be roughly 2/1, this is could be a small hint that three axion species are implemented in nature, with corresponding three events of depercolation. As shown in Fig.\,\ref{fig:rH}, the first two events  of depercolation happen during the radiation dominated area, the depecolation of e-lumps happened around z=53 where a clear spike in both matter and dark energy content is visible. The connection from dark matter to screened Yang-Mills scales proposed here  which are in association to lepton masses is shown schematically in Fig.\,\ref{fig:scales}.\nn



\section[Galactic central regions]{Galactic central regions and the dark sector
}\label{Sec5}

The link between the masses of the three species of ultralight axions, whose fuzzy condensates form lumps of typical masses $M_e$, $M_\mu$, and $M_\tau$, with the three lepton families via the Planck-scale originated axial anomaly within confining phases of SU(2) Yang-Mills theories is compelling. In particular, $M_e=M_{200}$ can be determined by fitting of direct observation, as done in section \,\ref{Sec3}, while $M_\mu$ and $M_\tau$ are predicted by an appeal to Eqs.\,(\ref{lumpmassratios}). Such a scenario allows to address two questions: (i) the implication of a given lump's selfgravity for its stability and (ii) the cosmological origin of a given species of isolated lumps.\nn

Before we discuss question (i) we would like to provide a 
thermodynamical argument, based on our knowledge gained about axion and lump masses in terms of Yang-Mills scales and the Planck mass, why Planck-scale axions associated with the lepton families always occur in the form of fuzzy or homogeneous condensates. Namely, the Yang-Mills scales $\Lambda_e$, $\Lambda_\mu=\Lambda_e\, m_\mu/m_e$, $\Lambda_\tau=\Lambda_e\, m_\tau/m_e$ and $\Lambda_\gamma= 2 \pi \,  2.725 \, {\rm K} / 13.87 \sim 10^{-4}$ eV together with Eqs.\,(\ref{maxxaxion}), (\ref{YMeextr}), yield axion masses as 
\begin{align}
\label{axionmassemutau}
\nonumber&m_{a,e}\sim 6.75\times10^{-24}\,\mbox{eV} \,,\\ 
&m_{a,\mu}\sim 2.89\times10^{-19}\,\mbox{eV} \,,\\
\nonumber&m_{a,\tau}\sim 8.17\times 10^{-17}\,\mbox{eV},\\
\nonumber&m_{a,\gamma}\sim 4.12\times 10^{-39}\,\mbox{eV}\,.
\end{align}
The critical temperature $T_c$ for the Bose-Einstein condensation of a quantum gas of free bosons of mass $m_a$ and (mean) number density $n_a\sim M/(m_a \frac43\pi r_B^3)$ is given as 
\eqb
\label{condfrBos}
T_c=\frac{2\pi}{m_a}\left(\frac{n_a}{\zeta(3/2)}\right)^{2/3}\,.
\eqe 
We conclude from Eqs. (\ref{lumpmassratios}),\, (\ref{Bohrradiusfund}),\,(\ref{axionmassemutau}) and (\ref{condfrBos}) that
\begin{align}
\label{critTs}
\nonumber&T_{c,e}\sim 4.16\times 10^{38}\,{\rm eV}\,,\\ 
&T_{c,\mu}\sim 7.61\times 10^{47}\,{\rm eV}\,,\\
\nonumber&T_{c,\tau}\sim 6.09\times 10^{52}\,{\rm eV},\\
\nonumber&T_{c,\gamma}\sim 3.31\times 10^{8}\,{\rm eV}\,.
\end{align} 
All three critical temperatures are comfortably larger than the Planck mass $M_P=1.22\times 10^{19}\,$GeV such that throughout the Universe's expansion history and modulo depercolation, which generates a nonthermal halo of particles correlated on the de Broglie wave length around a condensate core, the Bose-condensed state of e-, $\mu$-, and $\tau$-axions is guaranteed and consistent with $\xi\ll 1$, compare with Eq.\,(\ref{xidef}).\nn

We now turn back to question (i). Explicit lump masses can be obtained from Eqs.\,(\ref{lumpmassratios}) based on the typical mass $M_e=6.3\times 10^{10}\,M_{\odot}$ of an e-lump. One has
\begin{align}
\label{lumpmassemutau}
&M_\mu=1.5\times 10^{6}\,M_\odot\,,\nonumber\\ 
&M_\tau=5.2\times 10^{3}\,M_\odot\,,\nonumber\\ 
&M_\gamma=1.0\times 10^{26}\,M_\odot\,.
\end{align} 

\begin{figure}[h]
  \centering
  \includegraphics[width=11cm]{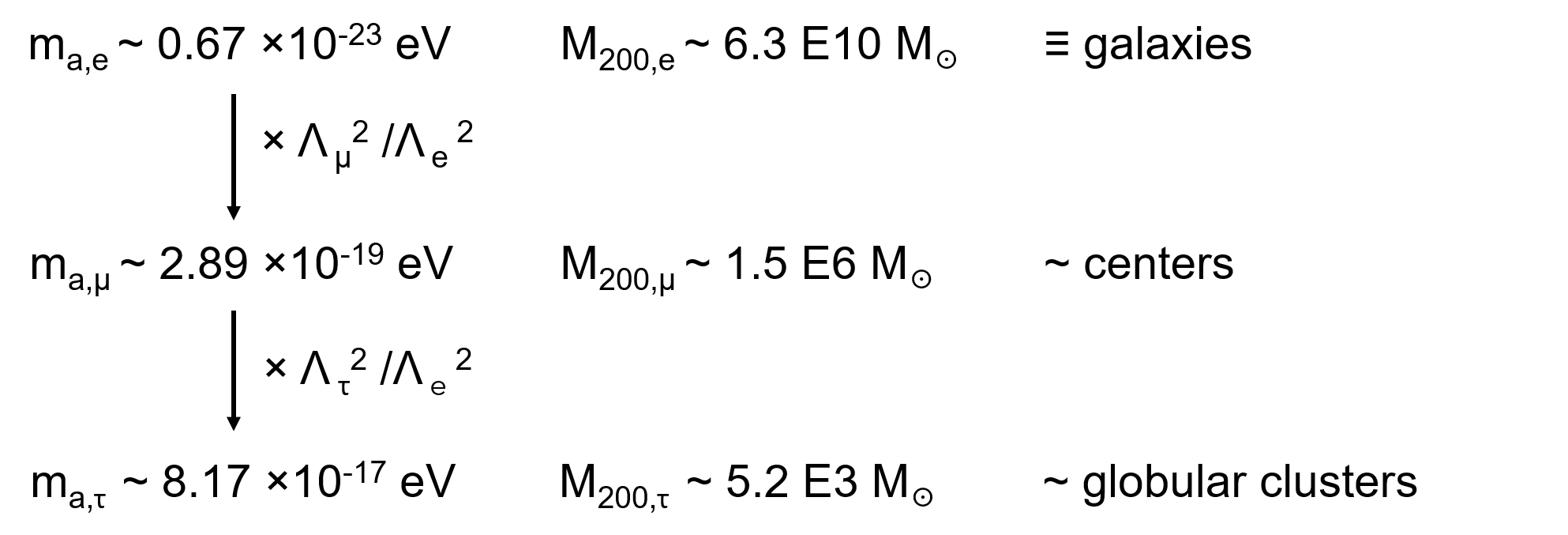}
  \caption{
  An overview of the axion and lump masses proposed here: By definition is the mass of an e-lump associated with a typical galaxy mass, $\mu$-lumps are almost as heavy as galaxy centers and $\tau$-lump masses are similar to the mass of globular clusters.}
\label{fig:scales2}
\end{figure}

A summary of the here obtained masses can be seen in Fig.\,\ref{fig:scales2}.\nn

\begin{figure}[H]\centering
\includegraphics[width=12.5cm]{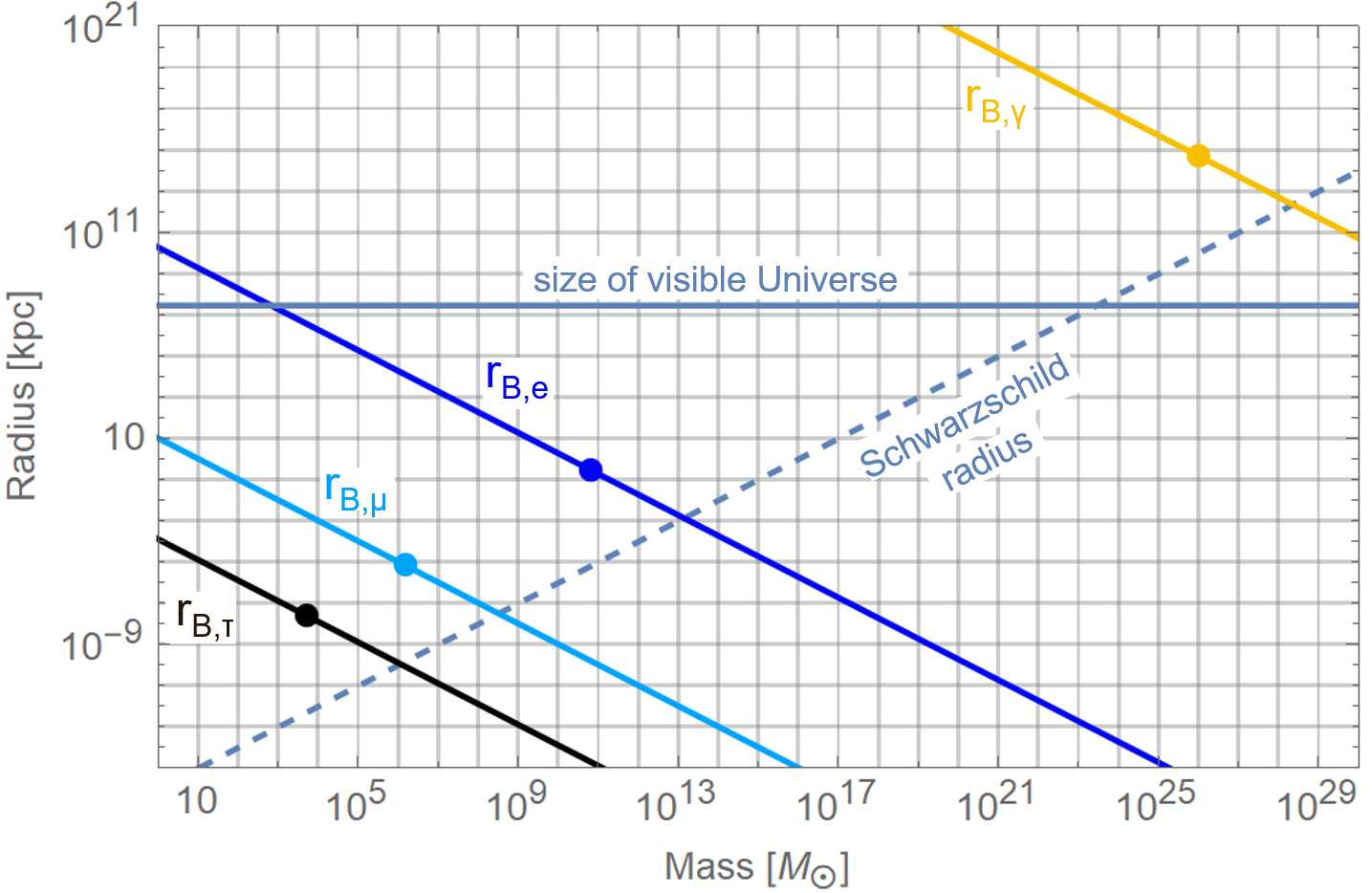}
\caption{Schwarzschild radius (dashed line) and gravitational Bohr radii (solid lines; dark blue: e-lump, turquoise: $\mu$-lump, black: $\tau$-lump and in orange the unpercolated $\gamma$-lump) as functions of lump mass in units of solar mass $M_\odot$. The dots indicate lump masses which derive from the typical e-lump mass $M_e=6.3\times 10^{10}\,M_{\odot}$ suggested by the analysis of the RCs of low-surface-brightness galaxies performed in Sec.\,\ref{Sec3}. Note that because the axion condensate which is in association to the photon scale $\sim 10^{-4}$ eV, the $\gamma$-lump has a typical size which is by far larger than the current observable Universe. It hat no chance to depercolate, that means form lumps / get ripped into lumps. This is why it is spatially homogeneous and we interpret it as dark energy.}
\label{fig:SchwarzschieldradiusVSBohrradii}
\end{figure}

For the computation of the respective gravitational Bohr radii according to Eq.\,(\ref{Bohr}) both quantities, axion mass and lump mass, are required. To judge the gravitational stability of a given lump throughout its entire evolution a comparison between the typical Bohr radius $r_B$ and the typical Schwarzschild radius $r_{\rm SD}$, defined as 
\eqb
\label{Schwarzschildradius}
r_{\rm SD}\equiv\frac{2M}{M_p^2}\,,
\eqe 
is in order. Using $M_e=1.86\times 10^{10}\,M_{\odot}$, Fig.\,\ref{fig:SchwarzschieldradiusVSBohrradii} indicates the implied values of the Bohr radii 
$r_{B,e}$, $r_{B,\mu}$, and $r_{B,\tau}$ by dots on the curves of all possible Bohr radii as functions of their lump masses when keeping the axion masses fixed. Notice that for all three cases, e-lumps, $\mu$-lumps, and $\tau$-lumps, typical Bohr radii are considerably larger than their Schwarzschild radii. Indeed, 
from Eqs.\,(\ref{maxxaxion}), (\ref{MLambda}), and (\ref{Schwarzschildradius}) it follows that
\eqb\label{ratioBSD}
\frac{r_B}{r_{\rm SD}}=\frac12 \kappa^2\,.
\eqe
With $\kappa=314$ we have $r_B/r_{\rm SD}=4.92\times 10^4$. Arguing on an adiabatic basis by following the solid lines in Fig.\,\ref{fig:SchwarzschieldradiusVSBohrradii} down to their intersections with the dashed line, a slow increase in mass by a factor $\sim 2435.73$ of a given lump to reach the critical mass for black-hole formation is unlikely to occur by growth of an axion condensate through collisions with its peers. Moreover, since the mean mass density of a lump scales with the fourth power of the Yang-Mills scale it is excluded even for moderate hierarchies in Yang-Mills scales, such as $\Lambda_\tau/\Lambda_\mu\sim 17$ or $\Lambda_\mu/\Lambda_e\sim 200$, that a lump with a larger scale embedded into a lump with a lower scale can grow substantially by accreting the much less dense condensate of its surroundings. 
What is conceivable though is that lumps of a larger Yang-Mills scale, embedded into a lump of a lower scale, catalyse the latter's gravitational compaction to the point of collapse, see discussion in Sec.\,6.1.    
\nn
With Eq.\,(\ref{lumpmassratios}) we have $M_\mu/M_e \sim 2.3\times 10^{-5}$ such that a dark mass of the selfgravitating dark-matter disk of the Milky Way, exhibiting a radial scale length of $(7.5\cdots 8.85)\,$kpc and a mass of $M_{\rm MW}=(2\cdots 3)\times 10^{11}\,M_{\odot}$ \cite{Kalberla:2007sr},  
implies a $\mu$-lump mass of 
\eqb
\label{compobjMW}
M_\mu=(4.6\cdots 7)\times10^{6}\,M_\odot\,.
\eqe
In \cite{Kalberla:2007sr} the mass of the dark halo of the Milky Way, extending out to 350\,kpc, is determined as $1.8\times 10^{12}\,M_\odot$, and there is a ringlike dark-matter structure within (13\,$\cdots$ 18.5)\,kpc of mass $(2.2\cdots 2.8)\times 10^{10}\,M_\odot$. Since these structures probably are, within the here-discussed framework, due to contaminations 
of a seeding e-lump by the accretion of $\tau$- and $\mu$-lumps we ignore them in what follows. In any case, a dark-matter halo of 350\,kpc radial extent easily accomodates the dark mass ratio $\sim 0.1$ between the selfgravitating dark-matter disk and the dark halo in terms of accreted $\tau$- and $\mu$-lumps. Interestingly, the lower mass bound of Eq.\,(\ref{compobjMW}) is contained in the mass range $(4.5\pm 0.4)\times 10^6\,M_\odot$ \cite{Ghez2008} or $(4.31\pm 0.36)\times 10^6\,M_\odot$ \cite{Gillessen:2008qv} of the central compact object extracted from orbit analysis of S-stars.\nn

Next, we discuss question (ii). Consider a situation where the gravitational Bohr radius $r_B$ exceeds the Hubble radius $r_H(z)=H^{-1}(z)$ at some redshift $z$. Here $H(z)$ defines the Hubble parameter subject to a given cosmological model. In such a situation, the lump acts like a homogeneous energy density (dark energy) within the causally connected region of the Universe roughly spanned by $r_H$. If $r_B$ falls {\sl sizably} below $r_H$ then formerly homogeneous energy density may decay into isolated lumps. In order to predict at which redshift $z_p$ such a depercolation epoch has taken place we rely on the extraction of the epoch  $z_{p,e}=53$ in \cite{Hahn:2018dih} for the depercolation of e-lumps. To extract the depercolation redshifts $z_{p,\mu}$ and $z_{p,\tau}$ we use the cosmological model SU(2)$_{\rm CMB}$ proposed in \cite{Hahn:2018dih} with parameters values given in column 2 of Table 2 of that paper. In Fig.\,\ref{fig:rH} the relative density parameters of the cosmological model SU(2)$_{\rm CMB}$ are depicted as functions of $z$, and the point of e-lump depercolation $z_{p,e}=53$ is marked by the cusps in dark energy and matter. 

The strategy to extract $z_{p,\mu}$ and $z_{p,\tau}$ out of information collected at $z_{p,e}=53$ is to determine the ratio $\alpha_e$ of $r_H=16.4\,$Mpc at $z_{p,e}=53$ and $r_{B,e}=0.26$\,kpc for a typical, isolated, and unmerged e-lump as 
\eqb\label{ratiorHrB}
\left.\alpha_e\equiv\frac{r_H}{r_{B,e}}\right|_{z=z_{p,e}}=55,476\,. 
\eqe
It is plausible that $\alpha_e$ can be promoted to a universal (that is, independent of the Yang-Mills scale and temperature) constant $\alpha$, again, because of the large hierarchy between all Yang-Mills scales to the Planck mass $M_P$. In addition, the ratio of radiation temperature to the Planck mass $M_P$ and to the scales of Yang-Mills theories in their confining phases remains very small within the regime of redshifts considered in typical CMB simulations. Using the cosmological model SU(2)$_{\rm CMB}$, Eq.\,(\ref{Bohrradiusfund}), and demanding $\alpha$ to also set the condition for $\mu$- and $\tau$-lump depercolation, one obtains  
\eqb
\label{zsperc}
z_{p,\mu}=\,40,000,\ \ \ \ z_{p,\tau}=685,000\,.
\eqe

\begin{figure}[H]\centering
\includegraphics[width=11cm]{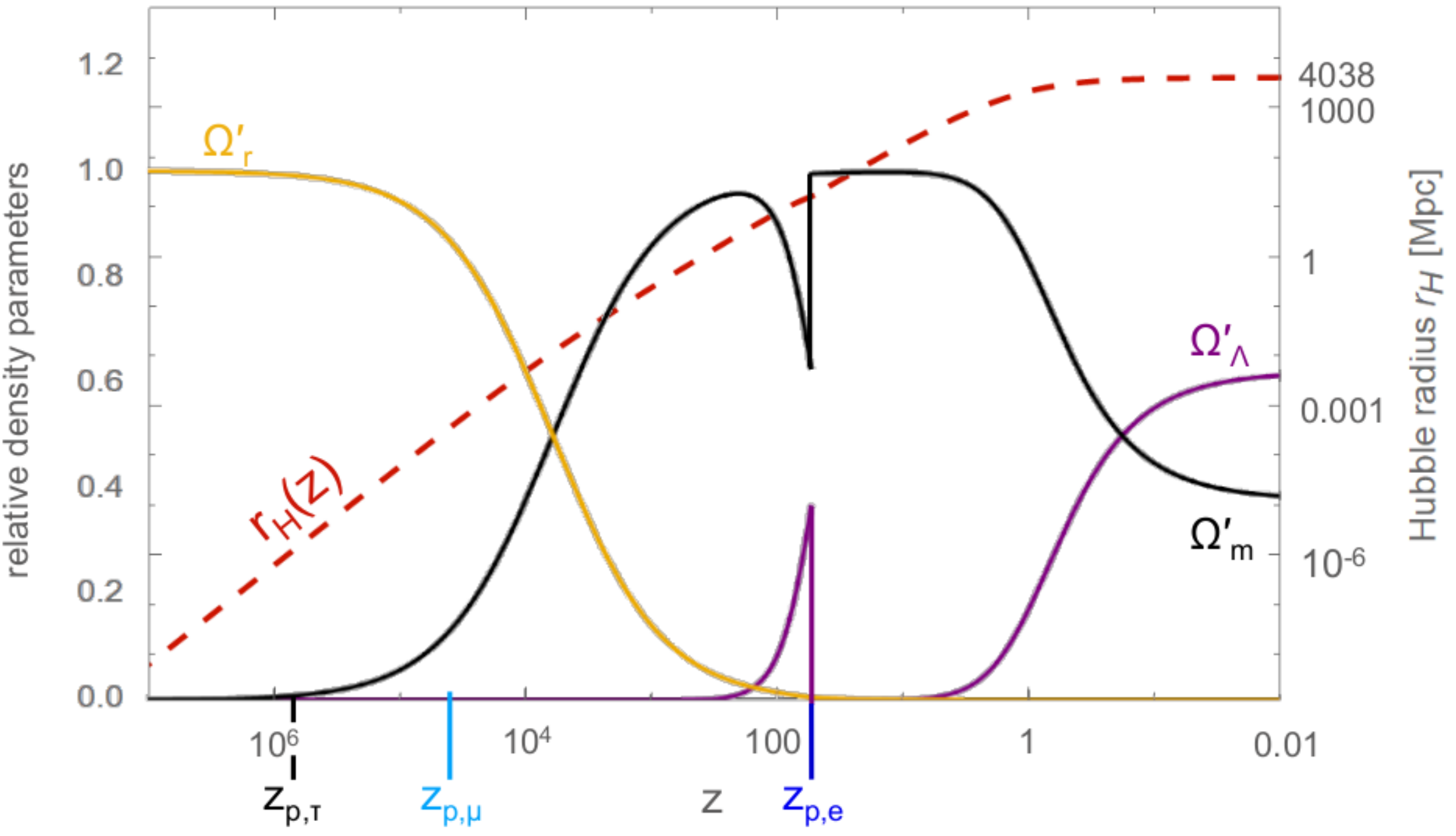}
\caption{Cosmological model SU(2)$_{\CMB}$ of \cite{Hahn:2018dih} (with parameter values fitted to the TT, TE, and EE CMB Planck power spectra and taken from column 2 of Table 2 of that paper) in terms of relative density parameters as functions of redshift $z$. $\Omega^\prime_\Lambda$ stands for dark energy, $\Omega^\prime_{m}$ for total matter (baryonic and dark), and $\Omega^\prime_{r}$ for radiation (three flavours of massless neutrinos and eight relativistic polarisations in a CMB subject to SU(2)$_{\rm CMB}$). The dotted red line represents the Hubble radius of this model. The redshifts of $e$-lump, $\mu$-lump, and $\tau$-lump depercolations are indicated by vertical lines intersecting the $z$-axis. Only e-lump depercolation is taken into account explicitly within the cosmological model SU(2)$_{\CMB}$ since at $z_{p,\mu}=40,000$ and $z_{p,\tau}=685,000$ the Universe is radiation dominated.}
\label{fig:rH2}
\end{figure}

In Fig.\,\ref{fig:rH2} the relative density parameters $\Omega^\prime_\Lambda$ (dark energy), $\Omega^\prime_{m}$ for total matter (baryonic and dark), $\Omega^\prime_{r}$ (total radiation), and the Hubble radius $r_H$ are depicted as functions of $z$. Moreover, the redshifts of $e$-lump, $\mu$-lump, and $\tau$-lump depercolations -- $z_{p,e}$, $z_{p,\mu}$, and $z_{p,\tau}$  --  are indicated by vertical lines intersecting the $z$-axis. 
The depercolation epochs for $\mu$- and $\tau$-lumps at redshifts $z_{p,\mu}=40,000$, and $z_{p,\tau}=685,000$ are not modelled within SU(2)$_{\rm CMB}$ because the Universe then is radiation dominated.\nn

\begin{figure*}\centering
\includegraphics[width=14.7cm]{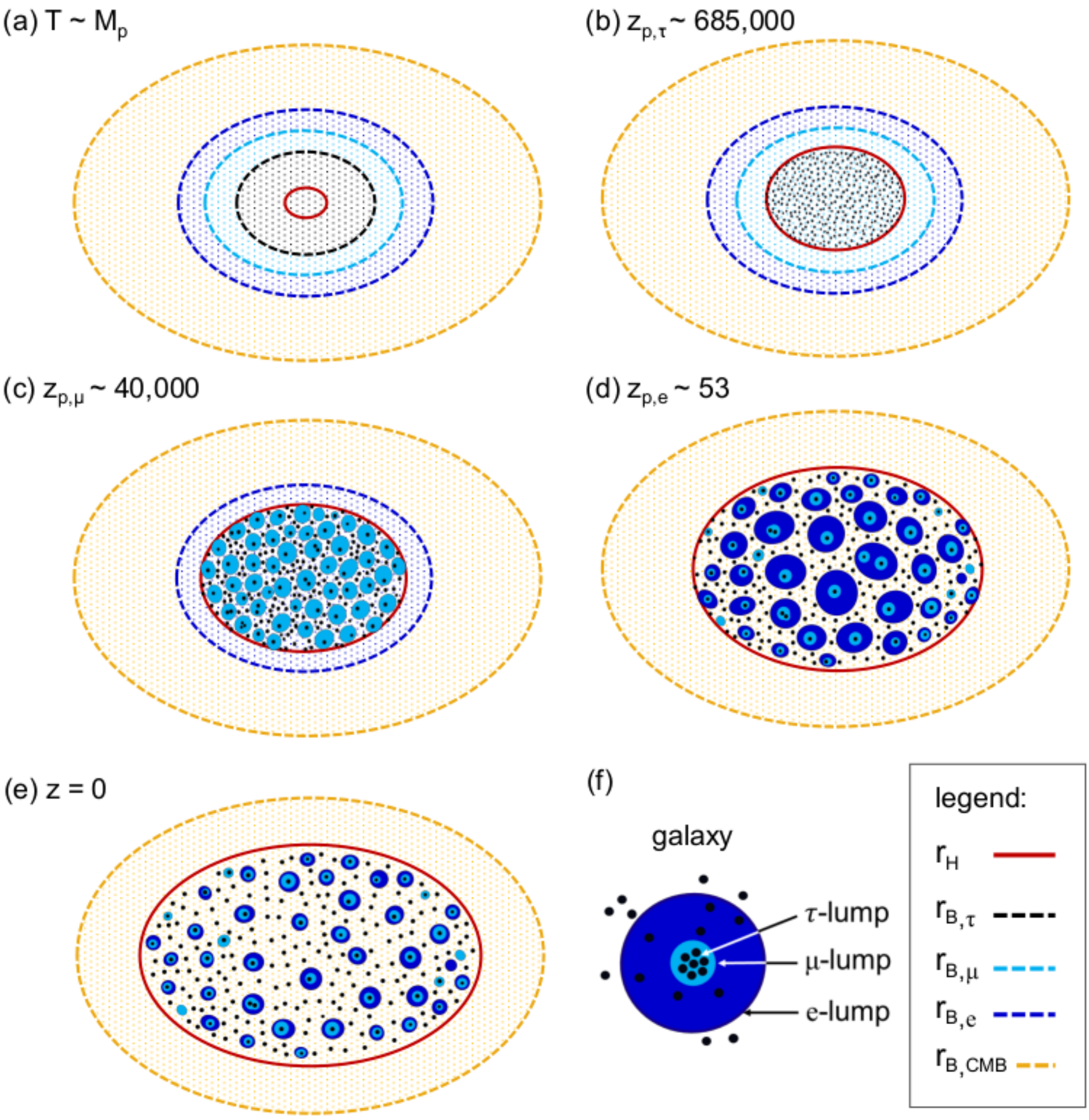}\vspace{-1mm}
\caption{The evolution of the Universe's dark sector according to SU(2) Yang--Mills theories of scales $\Lambda_e=m_e/(15\times 118.6)$, $\Lambda_{\mu}=m_\mu/(15\times 118.6)$, $\Lambda_{\tau}=m_\tau/(15\times 118.6)$ (confining phases, screened), and $\Lambda_{\rm CMB}\sim 10^{-4}\,$eV (deconfining phase, unscreened) invoking Planck-scale induced axial anomalies. The horizon size, set by the Hubble radius $r_H$ at various epochs (\textbf{a} - \textbf{e}), is shown by a red circumference. At epoch (\textbf{a}) gravity induced chiral symmetry breaking at the Planck scale creates a would-be-Goldstone boson which, due to the axial anomaly, gives rise to four ultralight axionic particle species. Their gravitational Bohr radii $r_{B,\tau}$, $r_{B,\mu}$, $r_{B,e}$, and $r_{B,{\rm CMB}}$ are much larger than $r_H$. Therefore, the associated energy densities should be interpreted as dark energy. (\textbf{b}) As the radiation dominated Universe expands the smallest Bohr radius $r_{B,\tau}$ falls below $r_H$. Once the ratio $\alpha\equiv r_H/r_{B,\tau}$ is sufficiently large ($\alpha=$ 55,500) $\tau$-lumps depercolate ($z_{p,\tau}=$ 685,000). (\textbf{c}) As the Universe expands further the Bohr radius $r_{B,\mu}$ falls below $r_H$. When the ratio of $r_H$ and $r_{B,\mu}$ again equals about $\alpha=$ 55,500 $\mu$-lumps derpercolate ($z_{p,\mu}=$ 40,000). The cosmological matter densities of $\tau$ and $\mu$-lumps are comparable \cite{Hahn:2018dih}. Since the mass of an isolated, unmerged $\tau$-lump is by a factor of about $(m_\tau/m_\mu)^2\sim 283$ smaller than the mass of an isolated, unmerged $\mu$-lump it then follows that the number density of $\tau$-lumps is by this factor larger compared to the number density of $\mu$-lumps. (\textbf{d}) Upon continued expansion down to redshift $z_{p,e}=53$ e-lumps depercolate. Their number density is by a factor of $(m_\mu/m_e)^2\sim$ 42,750 smaller than the number density of $\mu$-lumps. (\textbf{e}) The value of $r_{B,{\rm CMB}}$ is vastly larger than $r_H(z=0)$: $r_{B,{\rm CMB}}=2.4\times 10^{10}\,$Mpc vs. $r_H(z=0)=4038\,$Mpc. Therefore, a depercolation of CMB-lumps up to the present is excluded. As a consequence, the condensate of CMB-axions {\sl is} dark energy. (\textbf{f}) Possible dark-matter configuration of a galaxy including $\tau$-lumps and a single $\mu$-lump inside an e-lump.}
\label{fig:DarkHalosComic}
\end{figure*}

In Fig.\,\ref{fig:DarkHalosComic} a schematic evolution of the Universe's dark sector, subject to the SU(2) Yang-Mills theories SU(2)$_{\tau}$, SU(2)$_{\mu}$, SU(2)$_{e}$, and SU(2)$_{\rm CMB}$ invoking Planck-scale induced axial anomalies, is depicted. 
After a possible epoch of inflation and reheating the temperature of the radiation dominated Universe is close to the Planck mass $M_P$, and $r_H\sim M_P^{-1}$. In this situation, the Bohr radii of the various hypothetical lump species (Peccei-Quinn scale $M_P$, SU(2)$_{\tau}$, SU(2)$_{\mu}$, SU(2)$_{e}$, and SU(2)$_{\rm CMB}$ Yang-Mills dynamics) are  much larger than $r_H$, and the (marginal) dark sector of the model then solely contains dark energy. 
Around $z_{p,\tau}=685,000$ (radiation domination) the depercolation of $\tau$-lumps occurs for  $\alpha\equiv r_H/r_{B,\tau}\sim 55,500$. Once released, they evolve like pressureless, non-relativistic particles and, cosmologically seen, represent dark matter.\nn 

As the Universe expands further, the ratio $\alpha\equiv r_H/r_{B,\mu}\sim 55,500$ is reached such that $\mu$-lumps start to depercolate at $z_{p,\mu}=40,000$. Since they contribute to the cosmological dark-matter density roughly the same amount like $\tau$-lumps, see \cite{Hahn:2018dih} for a fit of so-called primordial and emergent dark-matter densities to TT, TE, and EE power spectra of the 2015 Planck data, one concludes from Eq.\,(\ref{lumpmassratios}) that their number density is by a factor  $(m_\tau/m_\mu)^2\sim 283$ smaller than that of $\tau$-lumps if we may, for a first estimate, neglect local gravitational interactions. That is, at $\mu$-lump depercolation there are roughly 300 $\tau$-lumps inside one $\mu$-lump. Each of these $\tau$-lumps possesses a mass of $M_\tau=1.65\times 10^4\,M_\odot$.  This may catalyse the gravitational compaction of the $\mu$-lump in the course of further cosmic evolution, see discussion in Sec.\,6.1.\nn 

At $z_{p,e}=53$ e-lumps depercolate \cite{Hahn:2018dih}. Again, disregarding local 
gravitational binding, we conclude from Eq.\,(\ref{lumpmassratios}) and a nearly equal contribution of each sort of lump to cosmological dark-matter density \cite{Hahn:2018dih} that the number densities of $\mu$- and $\tau$-lumps are by factors of $(m_\mu/m_e)^2\sim 42,750$ and $(m_\tau/m_e)^2\sim 283\times 42,750$, respectively, larger than the number density of e-lumps. At e-lump depercolation we thus have 42,750 $\mu$-lumps and $42,750\times 283\sim 1.2\times 10^7$ $\tau$-lumps within one e-lump.\nn

Again, ignoring local gravitational binding effects, the dilution of $\tau$- and $\mu$-lump densities by cosmological expansion predicts that today we have $42,750/(z_{p,e}+1)^3=0.27$ $\mu$-lumps and $42,750\times 283/(z_{p,e}+1)^3=77$ $\tau$-lumps within one e-lump. Local gravitational binding should correct these numbers to higher values but the orders of magnitude -- O(1) for $\mu$-lumps and O(100) for $\tau$-lumps -- should remain unaffected. 
It is conspicuous that the number of globular clusters within the Milky Way is in the hundreds \cite{GlobularClustersNumbers}, with typical masses between ten to several hundred thousand solar masses \cite{Ghez2008}. With $M_\tau=5.2\times 10^3\,M_\odot$ it is plausible that the 
dark-mass portion of these clusters is constituted by a single or a small number of merged 
$\tau$-lumps. In addition, in the Milky Way there is one central massive and dark object with about $(4.5\pm 0.4)\times 10^6$ \cite{Ghez2008} or $(4.31\pm 0.36)\times 10^6$ solar masses \cite{Gillessen:2008qv}. If, indeed, there is roughly one isolated $\mu$-lump per isolated $e$-lump today then the mass range of the Milky Way's dark-matter disk, interpreted as a merger of few isolated e-lumps, implies the mass range of Eq.\,(\ref{compobjMW}) for the associated $\mu$-lump merger. This range contains the mass of the central massive and dark object determined in \cite{Ghez2008,Gillessen:2008qv}.

\section{Discussion, Summary, and Outlook}


The results of Sec.\,\ref{Sec5} on mass ranges of $\tau$-lumps, $\mu$-lumps, and e-lumps being compatible with typical masses of globular clusters, the mass of the central compact Galactic object \cite{Gillessen:2008qv,Ghez2008}, and the mass of the selfgravitating dark-matter disk of the Milky Way, respectively, is compelling. We expect that similar assignments 
can be made to according structures in other spiral galaxies. \nn

Could the origin of the central compact object in Milky Way be result of $\tau$- and $\mu$-lump mergers?  As Fig.\,\ref{fig:SchwarzschieldradiusVSBohrradii} suggests, a merger of $n\ge 222$ isolated $\tau$- or $\mu$-lumps are required for black hole formation. Since we know that the mass of the central compact object is $\sim 4\times 10^6 M_\odot$ a merger of $n\ge 222$ $\mu$-lumps is excluded for Milky Way. Thus only a merger of $n\ge 222$ $\tau$-lumps, possibly enriched by the consumption of a few $\mu$ lumps, is a viable candidate for black-hole formation in our Galaxy.\nn

The Milky Way's contamination with baryons, its comparably large dark-disk mass vs. the mass of the low-surface-brightness galaxies analysed in Sec.\,3, and possibly tidal shear from the dark ring and the dark halo during its evolution introduce deviations from the simple structure of a typical low-surface-brightness galaxy. Simulations, which take all the here-discussed components into account, could indicate how typical such structures are, rather independently of primordial density perturbations. Isolated $\tau$-, $\mu$-, and e-lumps, which did not accrete sufficiently many baryons to be directly visible, comprise dark-matter galaxies that are interspersed in between visible galaxies. The discovery of such dark galaxies, pinning down their merger-physics, and determinations of their substructure by gravitational microlensing and gravitational-wave astronomy could support the here-proposed scenario of active structure formation on sub-galactic scales.\nn


In this thesis we propose that the dark Universe can be understood in terms of axial anomalies \cite{Adler:1969er,Bell:1969ts,Fujikawa:1979ay} which are invoked by screened Yang-Mills scales in association with the leptonic mass spectrum. This produces three ultra-light axion species. Such pseudo Nambu-Goldstone bosons are assumed to owe their very existence to a gravitationally induced chiral symmetry breaking with a universal Peccei-Quinn scale \cite{Peccei:1977ur} of order Planck mass $M_P=1.22\times10^{19}\,$GeV \cite{Giacosa:2008rw}. We therefore refer to each of these particle species as {\sl Planck-scale} axions. Because of the relation $m_{a,i}=\Lambda_i^2/M_P$ the screened Yang-Mills scale $\Lambda_i$ derives from knowledge of the axion mass $m_{a,i}$.  Empirically, the here-exptracted screened scale $\Lambda_e=287\,$eV points to the first lepton family, compare with \cite{Hofmann:2017lmu}. This enables predictions of typical lump and axion masses in association with two additional SU(2) Yang-Mills theories associating with $\mu$ and $\tau$ leptons.\nn

Even though the emergence of axion mass \cite{Peccei:1977ur} and the existence of lepton families \cite{Hofmann:2017lmu} are governed by the same SU(2) gauge principle, the interactions between these ultra-light pseudo scalars and visible leptonic matter is extremely feeble. Thus the here-proposed relation between visible and dark matter could demystify the dark Universe. An important aspect of Planck-scale axions is their Bose-Einstein, yet non-thermal, condensed state. A selfgravitating, isolated fuzzy condensate (lump) of a given axion species $i=e,\mu,\tau$ is chiefly characterised by the gravitational Bohr radius $r_{B,i}$ \cite{Sin1994} given in terms of the axion mass $m_{a,i}$ and the lump mass $M_i=M_{200,i}$ (virial mass), see Eq.\,(\ref{Bohr}). As it turns out, for $i=e$ the information about the latter two parameters is contained in observable rotation curves of low-surface-brightness galaxies with similar extents. Realistic models for the dark-matter density profiles derive from ground-state solutions of the spherically symmetric Poisson-Schr\"odinger system at zero temperature and for a single axion species. These solutions describe selfgravitating fuzzy axion condensates, compare with \cite{Schive:2014dra}. Two such models, the Soliton-NFW and the Burkert model, were employed in our present extractions of $m_{a,e}$ and $M_e$ under the assumption that the dark-matter density in a typical low-surface brightness galaxy is dominated by a single axion species. Our result $m_{a,e}=0.675\times 10^{-23}\,$eV is consistent with the result of \cite{Bernal2017}: $m_{a,e}=0.554\times 10^{-23}\,$eV. Interestingly, such an axion mass is close to the result $10^{-25}\,$eV\,$\le m_a\le 10^{-24}\,$eV \cite{Hlo_ek_2018} obtained by treating axions as a classical ideal gas of non-relativistic particles -- in stark contrast to the Bose condensed state suggested by Eq.\,\ref{condfrBos} or the gas surrounding it with intrinsic correlations governed by large de-Broglie wavelengths. This value of the axion mass is considerably lower then typical lower bounds obtained in the literature: $m_a>2.9\times 10^{-21}\,$eV \cite{Nadler:2019hjw}, $m_a=2.5^{+3.6}_{-2.0}\times 10^{-21}\,$eV \cite{Maleki:2020sqn}, $m_a>3.8\times 10^{-21}\,$eV \cite{Irsic:2017yje}, and $m_a\sim 8\times 10^{-23}\,$eV in \cite{Schive:2014dra}.
We propose that this discrepancy could be due to the omission of the other two axion species with a mass spectrum given by Eqs.\,(\ref{axionmassemutau}). For example, the dark-matter and thus baryonic density variations along the line of sight probed by a  Lyman-$\alpha$ forest do not refer to gravitationally bound systems and therefore should be influenced by all {\sl three} axion species. Once axions and their lumps are categorised, questions about (i) the cosmological origin of lumps and (ii) their role in the evolution of galactic structure can be asked. Point (i) is addressed by consulting a cosmological model (SU(2)$_{\rm CMB}$ \cite{Hahn:2018dih}) which requires the emergence of dark matter by lump depercolation at defined redshifts, see also \cite{Hofmann:2020wvr}. 
Depercolation of e-lumps at redshift $z_{p,e} = 53$ anchors the depercolations of the two other 
lump species. One obtains $z_{p,\mu}=40,000$ and $z_{p,\tau}=685,000$.\nn

The critical temperature $T_{c,e}$ of SU(2)$_e$ for the deconfining-preconfining phase transition (roughly equal to the temperature of the Hagedorn transition to the confining phase \cite{bookHofmann}) is $T_{c,e}=9.49\,$keV \cite{Hofmann:2017lmu}. A question arises whether this transition could affect observable small-scale angular features of the CMB. In the SU(2)$_{\rm CMB}$ based cosmological model of \cite{Hahn:2018dih} $T_{c,e}=9.49\,$keV corresponds to a redshift of $z_{c,e}=6.4\times 10^7$.
Traversing the preconfining-deconfining phase transition at $z_{c,e}$ an already strongly radiation dominated Universe receives additional radiation density and entropy. However, we expect that the horizon crossing of curvature perturbation at $z>z_{c,e}$, which may influence small-scale matter perturbations, will affect CMB anisotropies on angular scales $l>3000$ only. Therefore, Silk damping would reduce the magnitudes of these multipoles to below the observational errors. Up to the present, lump depercolation does not occur for the Planck-scale axion species associated with SU(2)$_{\rm CMB}$: Here the gravitational Bohr radius of the axion condensate always exceeds the Hubble radius by many orders of magnitude. As for point (ii), the masses and Bohr radii of $\mu$- and $\tau$-lumps seem to be related with the central massive compact object of the Milky Way \cite{Gillessen:2008qv,Ghez2008} and globular clusters \cite{Kalberla:2007sr}, respectively. Within a given galaxy such active components of structure formation possibly originate compact stellar streams through tidal forces acting on $\tau$-lumps. Whether this is supported by observation could be decided by a confrontation of N-body simulations (stars) in the selfgravitating background of the externally deformed lump.\nn  

Apart from cosmological and astrophysical observation, which should increasingly be able to judge the viability of the here-proposed scenario, there are alternative terrestrial experiments which can check the predictions of the underlying SU(2) gauge-theory pattern. Let us quote two examples: First, there is a predicted low-frequency spectral black-body anomaly at low temperatures ($T\sim 5\,$K) \cite{Hofmann:2009yh} which could be searched for with a relatively low instrumental effort. Second, an experimental link to SU(2)$_e$ would be the detection of the Hagedorn transition in a plasma at electron temperature $9.49\,$keV and the stabilisation of a macroscopically large plasma ball at a temperature of $1.3\times 9.49\,$keV \cite{Hofmann:2017lmu}. Such electron temperatures should be attainable by state-of-the-art nuclear-fusion experiments such as ITER or by fusion experiments with inertial plasma confinement.   


\newpage

\section{Multi soliton merger}\label{Multi soliton merger}

\begin{figure}[!htb]
    \centering
    \begin{minipage}{.5\textwidth}
        \centering
          \label{fig:SchwarzschieldradiusVSBohrradiiGenzel}
        \includegraphics[width=7cm]{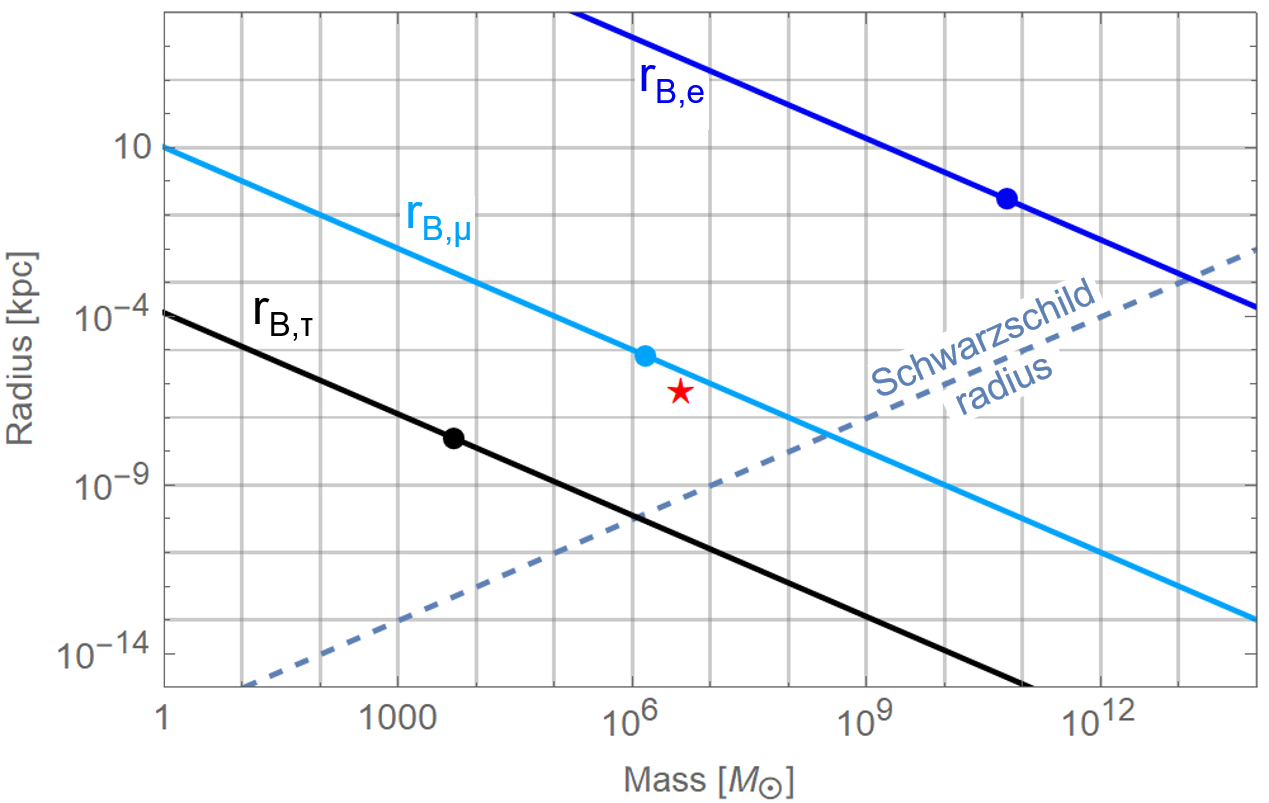}
        \label{fig:prob1_6_2}
    \end{minipage}%
    \begin{minipage}{.6\textwidth}
        \includegraphics[width=7cm]{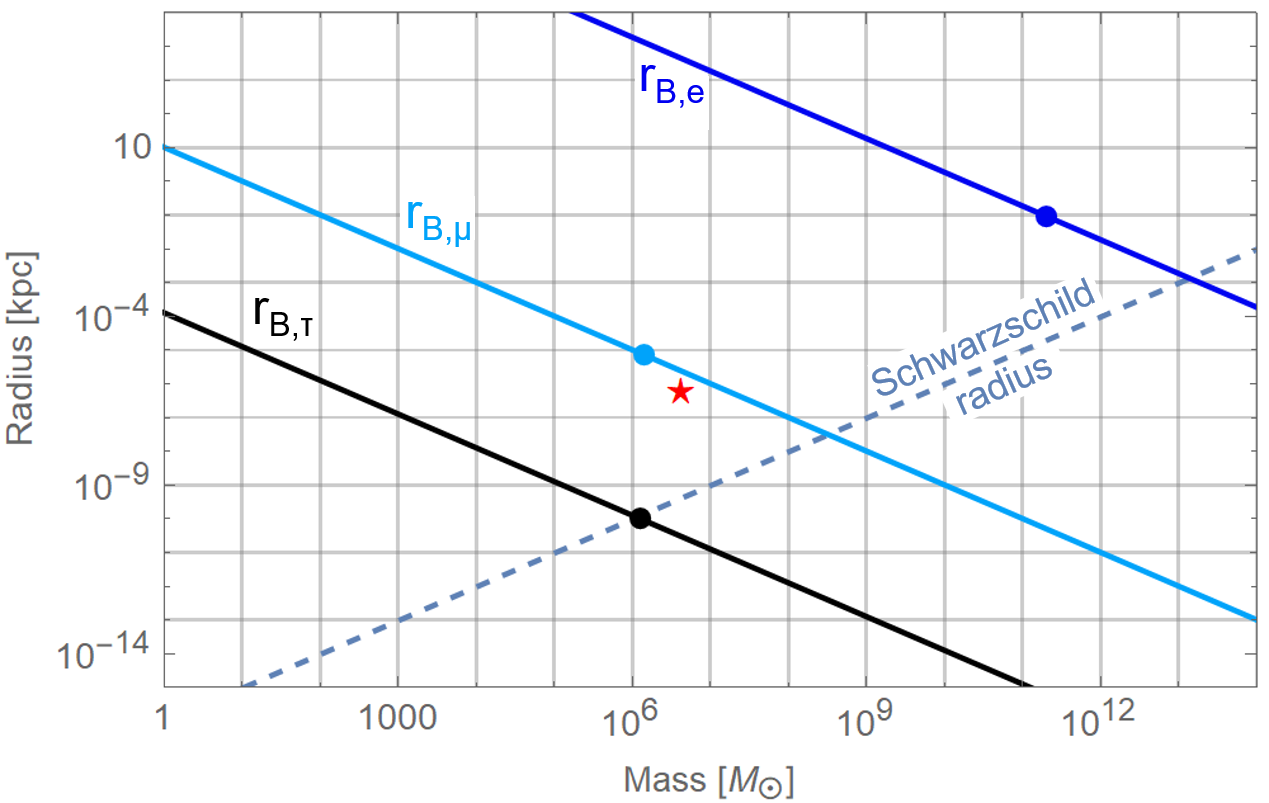}
    \end{minipage}
     \caption{Schwarzschild radius $r_{\rm SD}$ (dashed line) and gravitational Bohr radii (solid lines for an e-lump mass of $2\times 10^{11} M_\odot$ which is the mass of the dark-matter disk within the Milky Way; dark blue: e-lump, turquoise: $\mu$-lump, black: $\tau$-lump and in orange the unpercolated $\gamma$-lump) as functions of lump mass in units of solar mass $M_\odot$. The star-symbol indicates the upper radial bound $r_b$ on Sagittarius A$^*$'s extent and its mass extracted from orbit analysis of S-stars \cite{Gillessen:2008qv,Ghez2008}.}
     \label{fig:lumpscomparison}
\end{figure}

The left-hand side of figure \ref{fig:lumpscomparison} shows that the extracted lump masses which are based on the virial mass $M_{200}$ and lepton scale relations, are gravitationally stable: The typical lump masses which are indicated by dots in the diagram are well above the corresponding Schwarzschild radius. Interestingly, for unscreened leptonic scales, the axion mass for the electron scale would be $m_{a,e}\sim 1.52\times10^{-21}\,\mbox{eV}$ which is allowed by the FDM mass constraints, however, those lumps would lie exactly on the Schwarzschild radius which would make them gravitationally unstable. On subgalactic scales $\mu$-lumps could explain the presence of massive compact objects in galactic centers such as Sagittarius A$^*$ in the Milky Way \cite{Gillessen:2008qv,Ghez2008} while $\tau$-lumps may relate to globular clusters \cite{Kalberla:2007sr}.\nn

On the right-hand side of figure \ref{fig:lumpscomparison} the mass of the dark matter disk of the Milky Way (approximately $2\times10^{11} M_\odot$) was assumed for the e-lump mass. This is equivalent to 3.17 merged e-lumps. For the corresponding $\mu$- and $\tau$-lumps only density estimations have been used based on the idea that each axion type contributes exactly to 1/3 of the total dark matter content in the Universe.\nn
 
Although it is highly interesting to see that $\tau$-lumps could have relatively easily collapsed, consumed $\mu$-lumps and become the  the central object of our galaxy as observed by \cite{Gillessen:2008qv,Ghez2008} (indicated by the red star), we are not able to make reliable statements of such kind without any knowledge of their actual merging behaviour.\nn

\newpage

\begin{figure}[h]
  \centering
  \includegraphics[width=11cm]{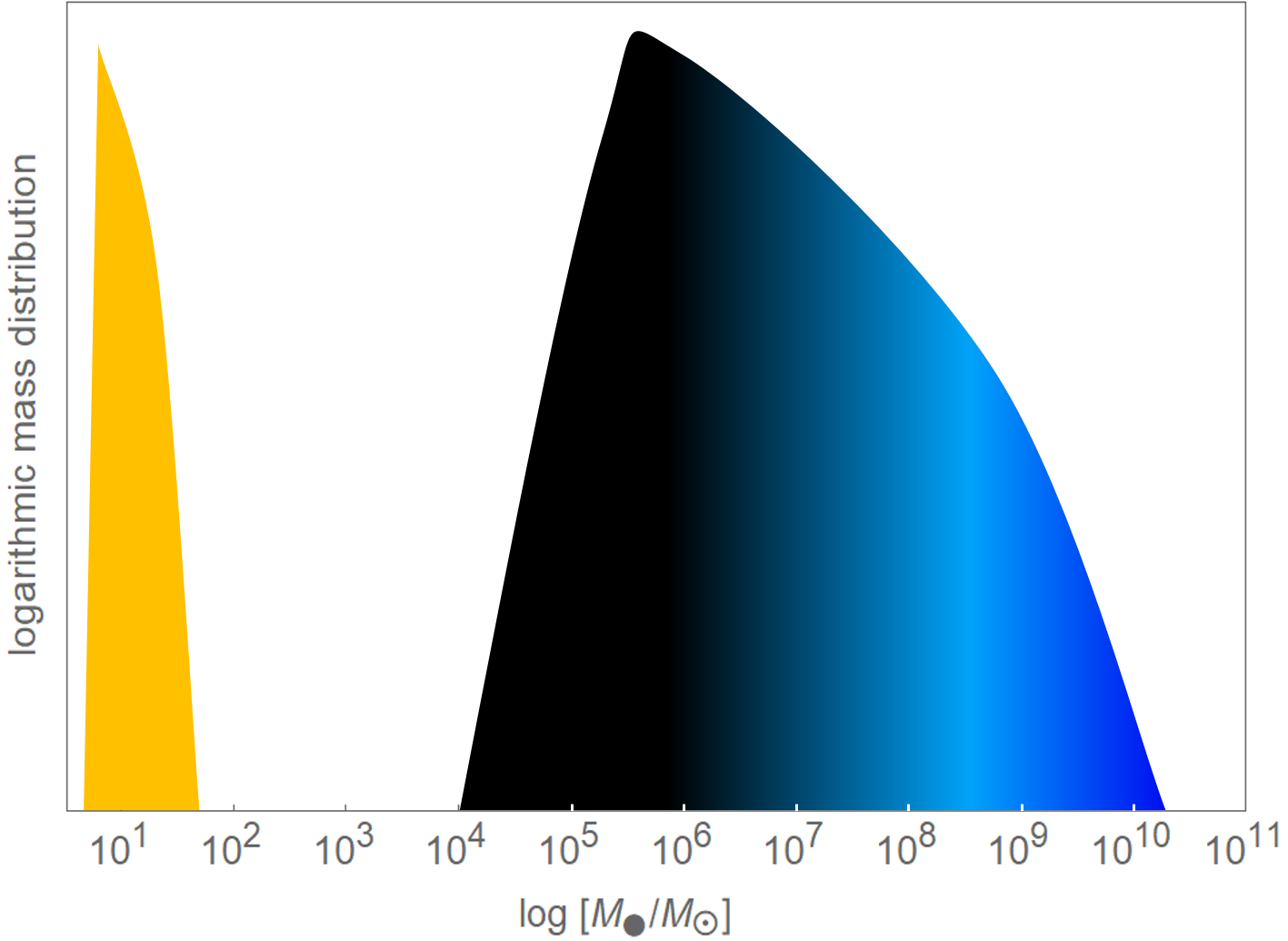}
  \caption{Cartoon illustrating the mass gap between stellar black holes (sBH) and super massive black holes (SMBH). In this graphic the mass of a black hole $M_\bullet$ is shown logarithmic in units of solar masses $M_\odot$. The mass distribution is also logarithmic. It is assumed by many (e.g. by \cite{Barack_2019}) that the mass gap is occupied by BH seeds which grow into SMBHs. In our framework, $\tau$- and  $\mu$-lumps may merge into those SMBHs. The lower bound for a pure $\tau$-lump dark matter merger would be $\sim 10^6 M_\odot$ as seen in Fig.\,\ref{fig:lumpscomparison}. An increasing mass contribution to a SMBH by $\mu$- (turquoise) and e-lumps (dark blue) is indicated by the colour gradients. At $\sim 5 \times 10^8 M_\odot$ a pure $\mu$-lump condensate is most likely to collapse. Note that from our current understanding it is unlikely that any e-lumps would have had enough time to merge into a SMBH. The graphic imitates the original comic by \cite{Barack_2019}.}
\label{fig:massgap}
\end{figure}

This hypothetical origin of the black hole in the center of our galaxy could simultaneously offer an explanation for the mass gap between stellar black holes (sBH) and super massive black holes (SMBH). Following the Bohr radius in Fig.\,\ref{fig:lumpscomparison}, the intersection of $R_{B,\tau}$ and the Schwarzschild radius lies at $\sim 10^6 M_\odot$. This roughly at the beginning of the SMBH distribution as seen in Fig.\,\ref{fig:massgap}. Therefore, we interpret in this framework SMBHs as merged objects
which formed from dark matter lumps, while stellar BHs contain ordinary, baryonic matter.\nn

\chapter{ Summary and Outlook}

\vspace{0.1cm}

\begin{minipage}{0.7\textwidth}
\begin{figure}[H]\centering
\includegraphics[width=10.1cm]{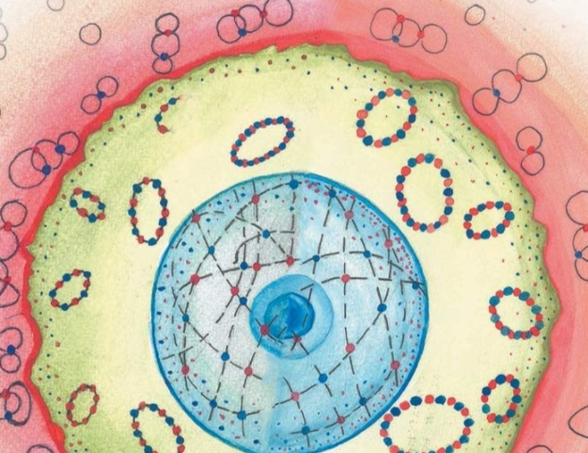}
\caption{Schematic overview of all SU(2)$_e$ phases: The electric deconfining phase (blue), governed by a non-trivial thermal ground state (GS). The magnetic preconfining phase (green) which has a monopole condensate as a GS; and the confining phase (red) which is separated by a non-thermal Hagedorn transition, which has pointlike center vortex loop in the GS.}
\end{figure}
\end{minipage} \hfill
\begin{minipage}{0.3\textwidth}
\begin{itemize}\sffamily

\item[\textcolor{gray!90}{\textbullet}] \textcolor{gray!90}{What is the bigger picture?}

\item[\textcolor{gray!90}{\textbullet}] \textcolor{gray!90}{How could we search for SU(2)$_{\rm CMB}$?}

\item[\textcolor{gray!90}{\textbullet}] \textcolor{gray!90}{How could we search for SU(2)$_e$?}

\item[\textcolor{gray!90}{\textbullet}] \textcolor{gray!90}{What could intergalactic magnetic fields tell us?}
\end{itemize}
\end{minipage}\vspace{1.5cm}

\minitoc

\newpage

\section{Very brief overview}

The purpose of the present work is to extract an axion mass from rotation curves and interpret its implication in the context of an SU(2)\textsubscript{CMB} Cosmology. Such a model has additionally to dark energy $\Omega_{\rm pdm,0}$ a primordial dark matter contribution which existed since shortly after the Big Bang $\Omega_{\rm pdm,0}$ and a dark energy phase transition into dark matter at a redshift of $z \sim 53$, $\Omega_{\rm edm,0}$ \cite{Hahn:2018dih}. We aim at explaining the parameters $\Omega_{\rm edm,0}$ and $\Omega_{\rm pdm,0}$ in terms of axial anomalies subject to a Planck-mass Peccei-Quinn scale and three SU(2) Yang-Mills theories associated with the three lepton families. Thereby we assign the $\mu$- and $\tau$-families to the primordial dark matter and the electron family to the emerging contribution. In addition, an explanation of dark energy is proposed which invokes the SU(2) Yang-Mills theory underlying the CMB. Hence, the explicit gauge-theory content of our model is:

\eqb
\label{SU(2)}
{\rm SU(2)}_{\rm CMB}\,\times\,
\underbrace{{\rm SU(2)}_e\times\,{\rm SU(2)}_{\mu}\times\,{\rm SU(2)}_{\tau}}_\text{{\normalsize \rm SU(3)}}\,. 
\eqe

\begin{figure}[H]
    \centering
        \includegraphics[width=12cm]{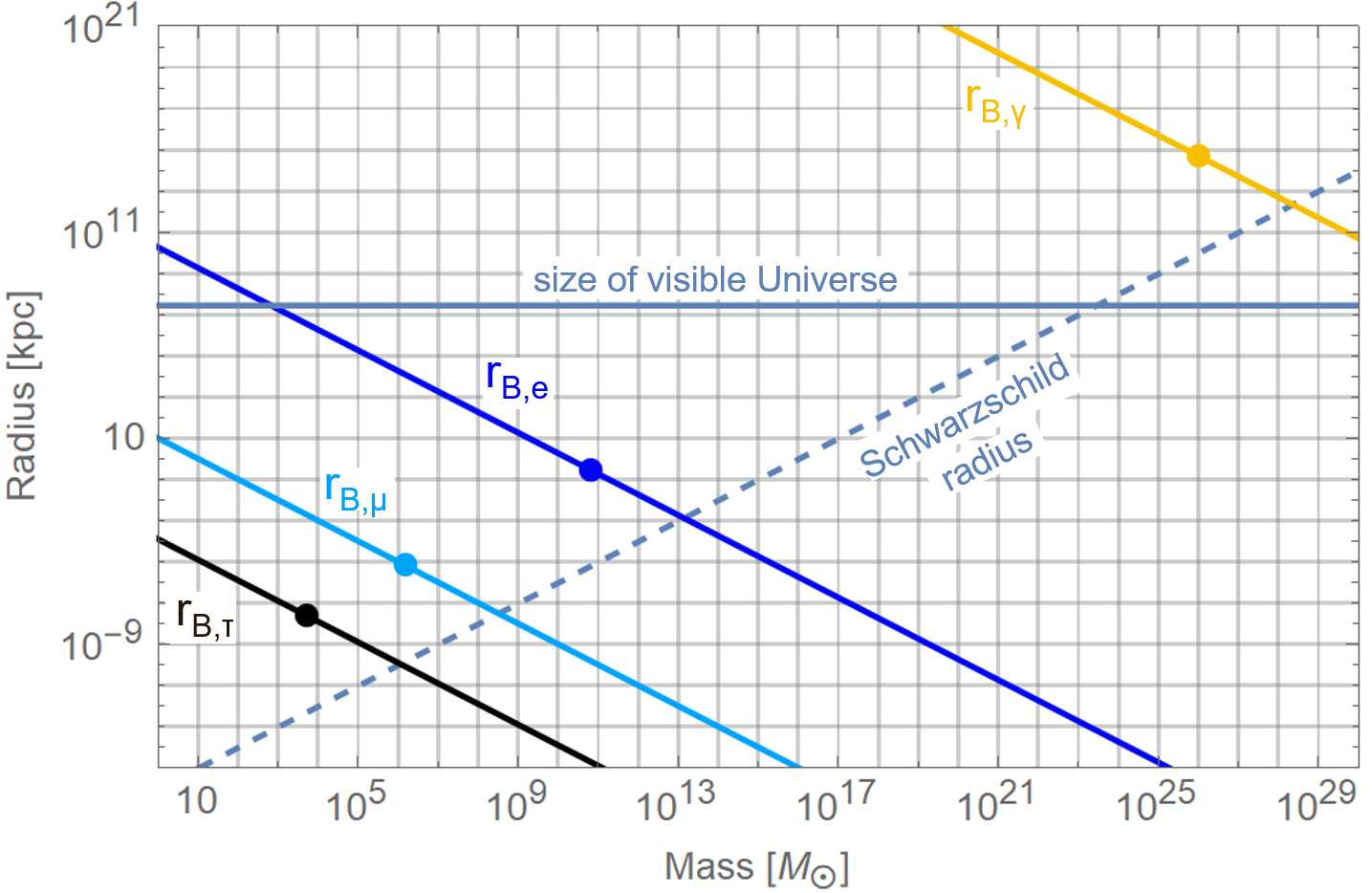}
     \caption{All four SU(2)-theories are coupled to an axion condensate which depercolated for the three leptonic SU(2)-theories, however, not for SU(2)$_{\rm CMB}$. The Schwarzschild radius $r_{\rm SD}$ (dashed line) and gravitational Bohr radii (solid lines for an e-lump mass of $6.3\times 10^{10} M_\odot$; dark blue: e-lump, turquoise: $\mu$-lump, black: $\tau$-lump; and in orange the unpercolated $\gamma$-lump which can be associated to the photon scale $\Lambda \sim 10^{-4}$ eV) as functions of lump mass in units of solar mass $M_\odot$. 
     }
     \label{fig:lumpsPhotons}
\end{figure}

Note that SU(2)$_e\times\,$SU(2)$_{\mu}\times\,$SU(2)$_{\tau}$ is isomorphic to SU(3). This means that not only leptons are assumed to be emergent particles from a (anti)monopole condensate\footnote{Because the theory is dual, these monopoles have an electric charge.}, but indirectly also quarks. For more details on this particular connection, see literature \cite{Hai_Jun_2008} about the fractional quantum-Hall effect. The main result of this thesis is to couple four SU(2)-theories to an axion condensate each which depercolated for the three leptonic SU(2)-theories, however, not for SU(2)$_{\rm CMB}$. Depercolation describes the transition from a spatially homogeneous axion condensate to a spatially inhomogeneous distribution with multiple provisionally bound condensate 'lumps'. The $\gamma$-lump which is attributed to the SU(2)$_{\rm CMB}$ theory is not depercolated yet as it is by many orders of magnitude larger than the current Universe as seen in Fig.\,\ref{fig:lumpsPhotons}.\nn

Based on the typical lump sizes and masses which are obtained solely by scale relations, we speculate that an interplay between $\tau-$ and $\mu-$lumps may be responsible for galactic core formation, e-lumps for the bulk dark matter contribution in galaxies and one single photon-lump may be associated to dark energy. The two biggest assumptions we made where that\nn

\noindent i) all axion fields were created at the Planck scale $\Lambda \sim M_P$,\nn

\noindent ii) the dark matter content of low luminosity galaxies can be approximated to be pure and dominated by a single axion species, the e-lumps.\nn

An additional assumption is that all three leptonic theories contribute equally to today's dark matter content. This is partially motivated by the SU(2)$_{\rm CMB}$ fit to the CMB \cite{Hahn:2018dih}, which suggest a ratio of 0.5 to 0.7 from primordial dark matter ($\tau-$ and $\mu-$lumps) and emergent dark matter (e-lumps). In the hypothetical case that in a galaxy such as our Milky Way enough $\tau-$lumps exist and merge this may create a black hole and motivates the current mass gap between stellar black holes of a few solar masses and super-massive black holes which seem to start at $10^6 M_\odot$. This is exactly the minimal weight of a black hole which is created by a pure $\tau-$lump merger, see the previous section \ref{Multi soliton merger}. Nonetheless, a simultaneous collapse of $\tau-$ and $\mu-$lumps is also feasible in this framework. To some extend this would be a "perfect crime" as the origin of the black hole may obscure it self as the black hole "eats up" all the evidence. The next step would certainly be to match N-body simulation with experimental data.\nn

The nature of this thesis is apart from the rotation curve which coincides with Matos et al. results \cite{Bernal2017} to a considerable extent speculative. Nonetheless, it may be desirable to explore new approaches as $\Lambda$CDM is under pressure \cite{Klypin_1999,Becker_2001,Djorgovski_2001,Bregman:2007ac,Walker_2011,Fixsen_2011,Riess:2016jrr}. Moreover, it can be shown that some of those tensions, e.g. the $H_0$ discrepancy cannot be resolved by a modification of $\Lambda$CDM at low redshift \cite{krishnan2021does}. How can one solidify or disprove aspects of the here applied model, especially if the proposed axions are so light they almost certainly evade any attempt of a direct detection? In the next sections I would like to mention very briefly possible approaches.\nn

\newpage

\section[Multi soliton merger]{Multi soliton merger for three component dark matter model}

\begin{figure}[h]
    \centering
        \includegraphics[width=14cm]{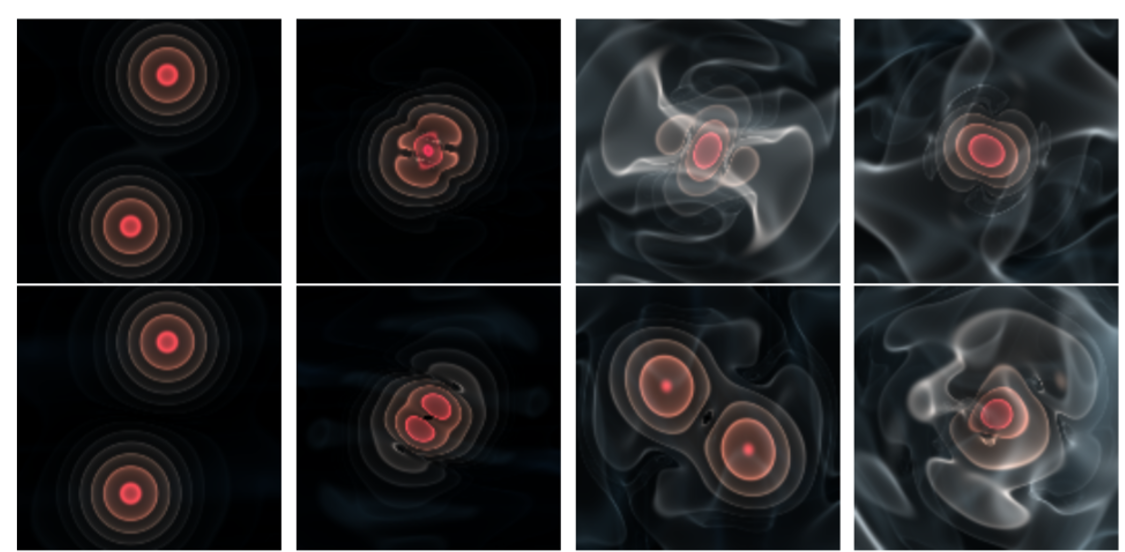}
     \caption{Volume rendered images of two representative binary mergers in phase (top) and with opposite phase (bottom) showing the central region of the computational domain at t = 0.7, t = 0.94, t = 2.0 and t = 7.0 in Mpc/km s
 \cite{Schwabe:2019gil}.}
     \label{fig:mergers}
\end{figure}

Related to the work in this thesis, whether a three component dark matter model is feasible could be researched by simulating 
the dynamics of the three postulated lump types. Therefore, the Schrodinger-Poisson equation needs to be solved numerically. Fig.\,\ref{fig:mergers} shows how such a simulation looks for the merging behaviour of two lumps with the same mass in phase (top) and with opposite phase (bottom) \cite{Schwabe:2019gil}.\nn

As mentioned in the previous chapter, it is highly interesting to see that $\tau$-lumps could have relatively easily collapsed, consumed $\mu$-lumps and become the central object of our galaxy as observed by \cite{Gillessen:2008qv,Ghez2008}. Simulations of galaxy merger could help to decide, whether this or similar situations are plausible or not.\nn

\section{High precision black body search for \texorpdfstring{SU(2)\textsubscript{CMB}}{SU(2)CMB}}

Whether the propagation of photon is described by SU(2)$_{\rm CMB}$ or not can be decided by a search for an SU(2) anomaly, which should be visible in a low temperature search of the black body spectrum. The idea is that the presence of a thermodynamical ground state composed of electric monopoles hinders the propagation of the photon and effectively suppresses frequencies below a critical frequency $\nu_c(T)$.\nn

\begin{figure}[H]\centering
\includegraphics[width=10.1cm]{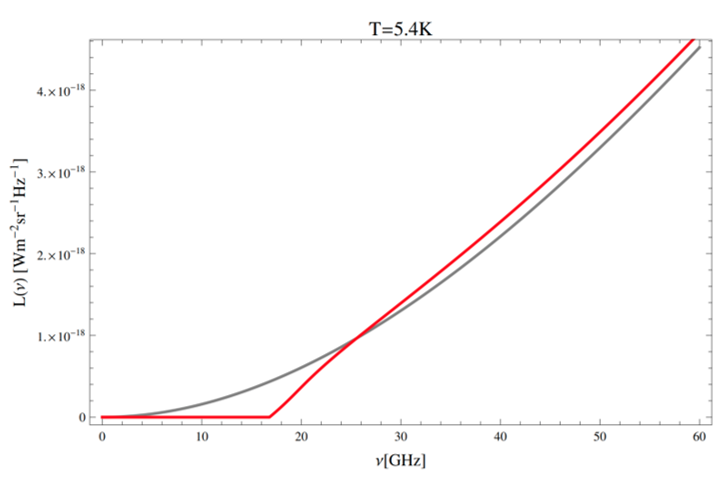}
\caption{ Comparison of the spectral radiance of an SU(2) photon gas (red) and a U(1) gas (grey) at a temperature T=5.4K. Note the spectral gap in the low-frequency region \cite{falquez2011investigation}.}
\label{Falques1}
\end{figure}

See the diploma thesis of Carlos Falquez for more detail, especially for an experimental setup \cite{falquez2011investigation}. See Fig.\,\ref{Falques1} for the expected effect of an SU(2) anomaly in the black body spectrum.
\nn

\section{Asteroseismology in search for \texorpdfstring{SU(2)\textsubscript{e}}{SU(2)e}}

It has been calculated that the critical temperature $T_c$ for the phase transition from the preconfined to the confined phase of SU(2)$_e$ is roughly $T_c = 9.49$\,keV $= 1.1 \times 10^8$\,K \cite{Hofmann:2017lmu}.
This temperature could be reached in the center of our sun. If the critical temperature $T_c$ was reached, a core structure invoked by the preconfined phase of SU(2)$_e$ should be observable by asteroseismology. 

\begin{figure}[H]\centering
\includegraphics[width=10cm]{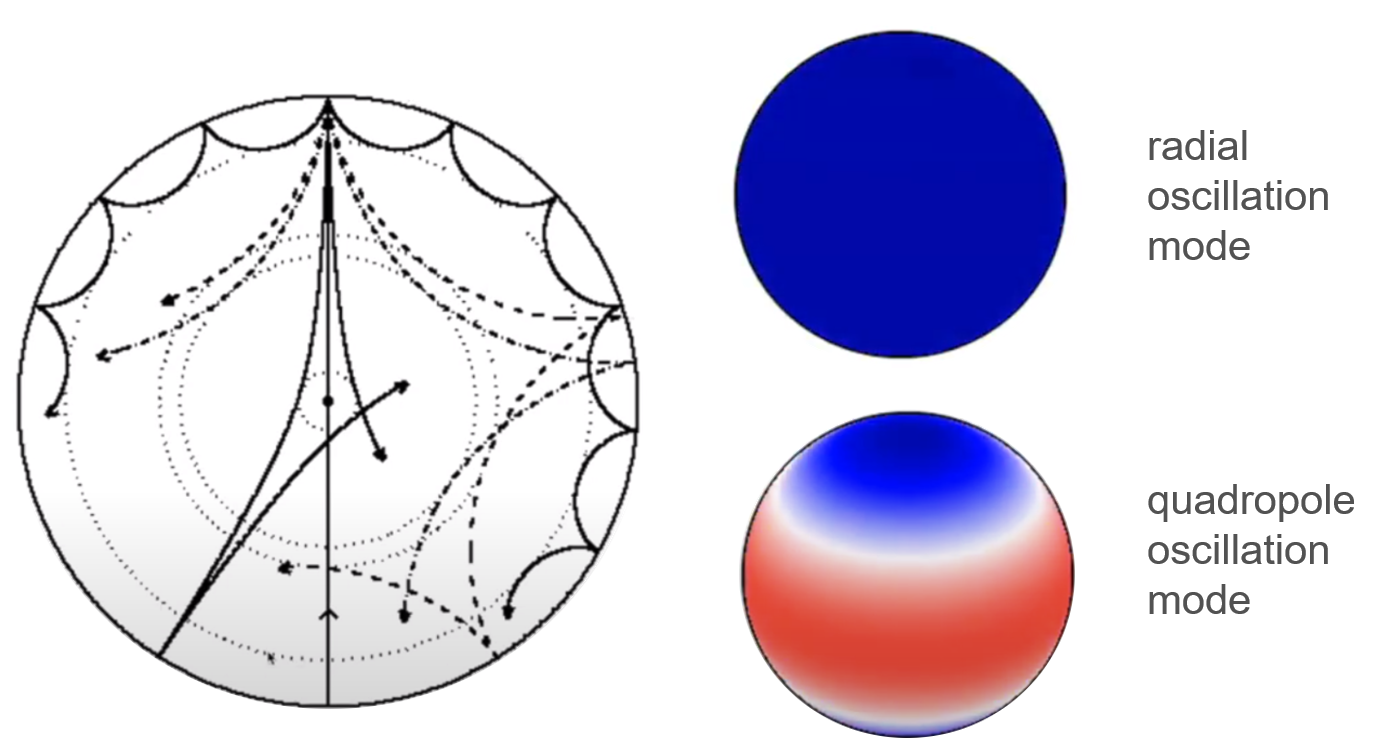}
\caption{ Scheme for the propagation of s- and t- waves inside the sin (left), different modes (right). Credit for the left graphic: Jørgen Christensen-Dalsgaard; Credits for the right graphic: Keaton Bell.}
\end{figure}

\section{Intergalactic magnetic fields in search for magnetic phases of \texorpdfstring{SU(2)$_i$}{SU(2)i}}

\begin{figure}[H]\centering
\includegraphics[width=8cm]{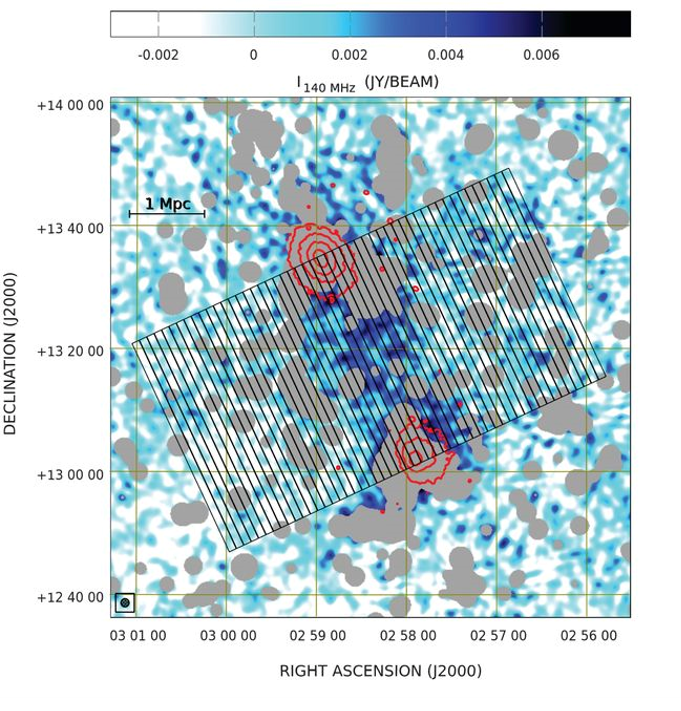}
\caption{Strips used to measure the ridge brightness profile. The x-ray emission from Abell 0399 and Abell 0401 is overlaid in red contours. The color bar represents the LOFAR image in Fig. 1 after masking of sources not related to the radio ridge (gray areas). The width of the strips is 0.108 Mpc (one beam width) and their length is 3 Mpc. \cite{Govoni_2019}.}
\label{spacemagnets}
\end{figure}

The origin of intergalactic magnetic fields in our Universe is still an open problem. A recent study by F. Govoni et al shows that those magnetic fillaments are up to 3 Mpc \cite{Govoni_2019}. One possible origin could be the magnetic phases (preconfined) of the three Yang-Mills theories which are associated to the lepton masses, SU(2)$_i$ with $i \in \{ e, \mu, \tau \}$ which at very high redshift ($z > 10^7$) could have spanned the Universe. In this scenario, the origin is primordial. Determining the latter could be a first hint, but in order to use intergalactic magnetic fields to falsify the SU(2) nature of leptons more severe calculations need to be done. One could try to calculate the magnetic field strengths associated to each Yang-Mills theory. Ultimately, only very sophisticated simulations which should be done on top of a galaxy structure formation could shed light into this mystery. Fig.\,\ref{spacemagnets} shows intergalactic magnetic fields between the glaxy Abell 0399 and Abell 0401 which are as long as 3 Mpc.

\section*{Acknowledgments}

At this point I would like to thank PD Dr. Ralf Hofmann for his supervision, support and encouragement throughout this thesis. I would like to thank my parents for their emotional and financial support; and I am thankful for many inspiring and thought provoking discussions with my friends.\nn


\bibliographystyle{unsrt}
\bibliography{AMainReferences}


\end{document}